\documentclass[fleqn,usenatbib]{mnras}

\usepackage{newtxtext,newtxmath}

\usepackage[T1]{fontenc}

\DeclareRobustCommand{\VAN}[3]{#2}
\let\VANthebibliography\thebibliography
\def\thebibliography{\DeclareRobustCommand{\VAN}[3]{##3}\VANthebibliography}

\usepackage{graphicx}
\usepackage{subfig}
\usepackage{adjustbox}

\newcommand{\hhh}{$^{\mathrm{h}}$}
\newcommand{\mmm}{$^{\mathrm{m}}$}
\newcommand{\sss}{$^{\mathrm{s}}$}
\newcommand{\ddd}{$^{\mathrm{\circ}}$}
\newcommand{\dmm}{$^{\prime}$}

\title[MeerKAT observation of Herschel protoclusters]{MeerKAT observations of Herschel protocluster candidates}

\author[Y. Ding et al.]{
Y. Ding,$^{1}$\thanks{E-mail: yifan.ding19@imperial.ac.uk}
D.\,L. Clements,$^{1}$
L.\,L. Leeuw,$^{2}$
I. Heywood,$^{3,4}$
H. Dannerbauer,$^{5,6}$
A. Parmar,$^{1}$
\newauthor
P. Legodi,$^{4}$
R.\,J.~Ivison,$^{7,8,9,10}$
R. Blake,$^{9}$
C.\,M. Gutiérrez,$^{5,6}$
A. Carnero,$^{5,6}$
W. Sutherland$^{11}$
\\
$^{1}$Astrophysics Group, Imperial College, Blackett Laboratory, Prince Consort
Road, SW7 2BX London, UK\\
$^{2}$Department of Physics and Astronomy, University of the Western Cape, Bellville, 7535, South Africa\\
$^{3}$Astrophysics, Department of Physics, University of Oxford, Keble Road, Oxford OX1 3RH, UK\\
$^{4}$Department of Physics and Electronics, Rhodes University, PO Box 94, Makhanda 6140, South Africa\\
$^{5}$Instituto de Astrofísica de Canarias (IAC), E-38205 La Laguna, Tenerife, Spain\\
$^{6}$Departamento de Astrofísica, Universidad de la Laguna (ULL), E-38206 La Laguna, Tenerife Spain\\
$^{7}$European Southern Observatory (ESO), Karl-Schwarzschild-Strasse~2, D-85748 Garching, Germany\\
$^{8}$School of Cosmic Physics, Dublin Institute for Advanced Studies, 31 Fitzwilliam Place, Dublin D02 XF86, Ireland\\
$^{9}$Institute for Astronomy, University of Edinburgh, Royal Observatory, Blackford Hill, Edinburgh EH9 3HJ, UK\\
$^{10}$ARC Centre of Excellence for All Sky Astrophysics in 3 Dimensions (ASTRO 3D)\\
$^{11}$School of Physical and Chemical Sciences, Queen Mary University of London, 
Mile End Road,  London  E1 4NS, UK
}

\pubyear{2024}

\begin{document}
\label{firstpage}
\pagerange{\pageref{firstpage}--\pageref{lastpage}}
\maketitle

% Abstract of the paper
\begin{abstract}
High-redshift protoclusters consisting of dusty starbursts are thought to play an important role in galaxy evolution. Their dusty nature makes them bright in the FIR/submm but difficult to find in optical/NIR surveys. Radio observations are an excellent way to study these dusty starbursts, as dust is transparent in the radio and there is a tight correlation between the FIR and radio emission of a galaxy. Here, we present MeerKAT 1.28 GHz radio imaging of 3 \textit{Herschel} candidate protoclusters, with a synthesised beam size of $\sim7.5''\times 6.6''$ and a central thermal noise down to $4.35~\mu $Jy/beam. Our source counts are consistent with other radio counts with no evidence of overdensities. Around $95\%$ of the \textit{Herschel} sources have 1.28 GHz IDs. Using the \textit{Herschel} $250~\mu$m primary beam size as the searching radius, we find $54.2\%$ \textit{Herschel} sources have multiple 1.28 GHz IDs. Our average FIR-radio correlation coefficient $q_{250\mu \text{m}}$ is $2.33\pm 0.26$. Adding $q_{250\mu \text{m}}$ as a new constraint, the probability of finding chance-aligned sources is reduced by a factor of $\sim 6$, but with the risk of discarding true identifications of radio-loud/quiet sources. With accurate MeerKAT positions, we cross-match our \textit{Herschel} sources to optical/NIR data followed by photometric redshift estimations. By removing $z<1$ sources, the density contrasts of two of the candidate protoclusters increase, suggestive of them being real protoclusters at $z>1$. There is also potentially a $0.9<z<1.2$ overdensity associated with one candidate protocluster. In summary, photometric redshifts from radio-optical cross-identifications have provided some tentative evidence of overdensities aligning with two of the candidate protoclusters. 
\end{abstract}


%%%%%%%%%%%%%%%%%%%%%%%%%%%%%%%%%%%%%%%%%%%%%%%%%%

%%%%%%%%%%%%%%%%% BODY OF PAPER %%%%%%%%%%%%%%%%%%

	\section{Introduction}\label{sec.intro}
	High density regions at $z>2$ which will become today’s galaxy clusters have been found (e.g., \citealt{steidel1998large}, \citealt{venemans2007protoclusters}), mostly selected in rest-frame UV wavelengths, for example through overdensities of Lyman-break galaxies (LBGs). Recent deep observations using Hyper-Suprime-Cam (HSC) have led to discoveries of $z>4$ protoclusters \citep{toshikawa2018goldrush}. Being galaxy clusters in formation, these high-z protoclusters are important in understanding the evolution of massive galaxies and large scale structure, but given that most of them are selected in the optical wavebands, any dust-obscured galaxies faint in optical are missed. Evidence have been found that a significant number of protoclusters are made up of dusty star-forming galaxies (DSFGs) (e.g., \citealt{daddi2009two,ivison2013herschel,clements2014herschel,casey2015massive,greenslade2018candidate}), it is thus possible that these protoclusters, whose member galaxies undergo simultaneous dusty starbursts, are missed in optical/near-infrared (NIR) surveys such as the LBG searches. Methods of searching for protoclusters in the far-infrared (FIR) are therefore needed. Protoclusters rich in dusty star-forming galaxies ($\sim$5 DSFGs each) do exist but only a few have been found to date (e.g., \citealt{daddi2009two,chapman2009submillimeter,capak2011massive,walter2012intense,dannerbauer2014excess,casey2015massive,oteo2018extreme}). There are some early attempts at searching for FIR/submm protoclusters using high-redshift radio galaxies as signposts of such protoclusters (e.g., \citealt{stevens2003formation,rigby2014searching}). Nonetheless, this method cannot detect most FIR/submm protoclusters as some known FIR protoclusters have no central high-redshift radio galaxies (e.g., \citealt{ivison2013herschel}).\par 
	
	A method of searching for protoclusters of dusty starbursts using \textit{Planck} and \textit{Herschel} imaging data in the FIR and submm (\citealt{herranz2013herschel}, \citealt{clements2014herschel}, \citealt{clements2016h}, \citealt{greenslade2018candidate}) has been proposed and tested. Point sources are first extracted from Planck observations at 857GHz ($350~\mu$m) and 545GHz ($550~\mu$m) at $4.67'$ resolution, then, using \textit{Herschel}-SPIRE (250, 350, 500 $\mu$m) images from large surveys such as H-ATLAS (\citealt{eales2010herschel}) and HerMES (\citealt{oliver2012herschel}), these \textit{Planck} point sources are resolved to $18''-36''$ resolution. Using archival catalogues and visual inspection, \textit{Planck} sources that might originate from overdensities of multiple dusty starbursts while  also being overdense in SPIRE sources at $>3\sigma$ are selected as candidate protoclusters. \citet{greenslade2018candidate} identify 27 candidate protoclusters which all have FIR photometric redshift estimates at $z=2-3$. They also have total estimated star-formation rates (SFRs) of $>10,000~ M_{\odot}/\text{yr}$. With such high SFRs, a significant fraction of their cluster stellar mass is assembled in this phase. Thus these candidate protoclusters likely probe an important epoch of cluster formation when rapid stellar mass assembly takes place. However, with their exact nature unknown, the poor resolution and large beam size of the \textit{Herschel} SPIRE data make it difficult to determine the exact locations (uncertainty $>18''$) of these dusty starburst protocluster galaxies for optical/NIR identification and spectroscopic follow-up.\par 
	
	Radio identifications of FIR/submm objects is a fairly simple and reliable method to obtain their precise locations (e.g., \citealt{ivison2007scuba}, \citealt{chapin2009aztec}, \citealt{casey2012redshift}, \citealt{roseboom2012herschel}, \citealt{hodge2013alma}, \citealt{an2018machine}). Following these previous works, we acquired MeerKAT (\citealt{jonas2009meerkat}) 1.28 GHz radio imaging observations centred on three DSFG protocluster candidates selected from \citet{greenslade2018candidate} to permit their further identification in optical/NIR data. There are several merits in carrying out MeerKAT radio observations on these DSFG protocluster candidates: (a) dust is transparent at radio wavelengths so radio observations will trace the cosmic star formation rate density unbiased by dust emission and absorption; (b) high-resolution radio interferometry observations will greatly help the cross-identification of multi-wavelength counterparts of \textit{Herschel} sources with precise position measurements, as well as investigating the radio morphologies of DSFGs; (c) in optical/NIR images, each \textit{Herschel} SPIRE source (beam size of $\sim18''-36''$) might overlap with multiple optical/NIR sources. This is a difficult situation if we want to do optical/NIR spectroscopy because the counterpart of the FIR source is uncertain. Thanks to the well-known correlation between a galaxy's FIR and radio emission (e.g., \citealt{helou1985thermal}, \citealt{de1985spiral}), we can use MeerKAT to obtain locations down to $\approx 1''$ rms and the optical/near-infrared counterpart is thus unambiguous; (d) Comparing a source's radio flux to its FIR flux can help identify the type of source that contributes to the high FIR flux in our candidate protoclusters, e.g. lensed objects, line-of-sight projections, enhanced starbursts, and radio-loud AGNs. The primary goal of this study is to acquire accurate MeerKAT radio identifications of three \textit{Herschel} candidate protoclusters. This will significantly simplify identifications of \textit{Herschel} sources in the optical/NIR, hence confirming if the candidates are real or not by fitting spectral energy distributions (SEDs) and redshifts to the sources. Furthermore, we will be able to study the stellar mass assembly during their dusty starburst phase and also the impact of environment on the starburst protocluster galaxies (e.g. AGN feedback). More broadly, this work demonstrates how high-quality radio data can help FIR studies, as next-generation radio telescopes such as the Square Kilometre Array (SKA; \citealt{dewdney2009ska}) are coming closer to reality.\par
	 
	This paper is structured as follows: In Section \ref{sec.data} we introduce the data used in this work. Section \ref{sec.extra} describes the source extraction process for our MeerKAT images, and how we account for the extraction completeness and reliability. Source counts extracted from our MeerKAT images are presented in Section \ref{sec.counts}. Section \ref{sec.result} includes the results and discussions. In Section \ref{sec.id} and Section \ref{sec.id.1} we present the cross-identifications (cross-IDs) of \textit{Herschel} sources with MeerKAT sources and investigate the multiplicity of \textit{Herschel} sources seen by MeerKAT. Analysis of the FIR-radio correlation and \textit{Herschel} colours are shown in Section \ref{sec.id.2} and \ref{sec.id.3}. In Section \ref{sec.dis}, we discuss how high-quality radio data can help to improve the robustness of \textit{Herschel} identification in the radio. We show how high quality MeerKAT data help the identification of \textit{Herschel} sources in optical/NIR data in Section \ref{sec.optical}, followed by cross-ID to KiDS and VIKING data which allows photometric redshift estimations in Section \ref{sec.optical.z}. Section \ref{sec.con} concludes the work. Throughout this paper, AB magnitude is used and we assume a standard cosmology, with $H_{0}=67.7~\text{km s}^{-1} \text{Mpc}^{-1}$, $\Omega_{M} = 0.3$, and $\Omega_{\Lambda} = 0.7$.

	\section{Data}\label{sec.data}
	
	\begin{figure*}
		\centering
		\subfloat{\includegraphics[width=0.75\linewidth]{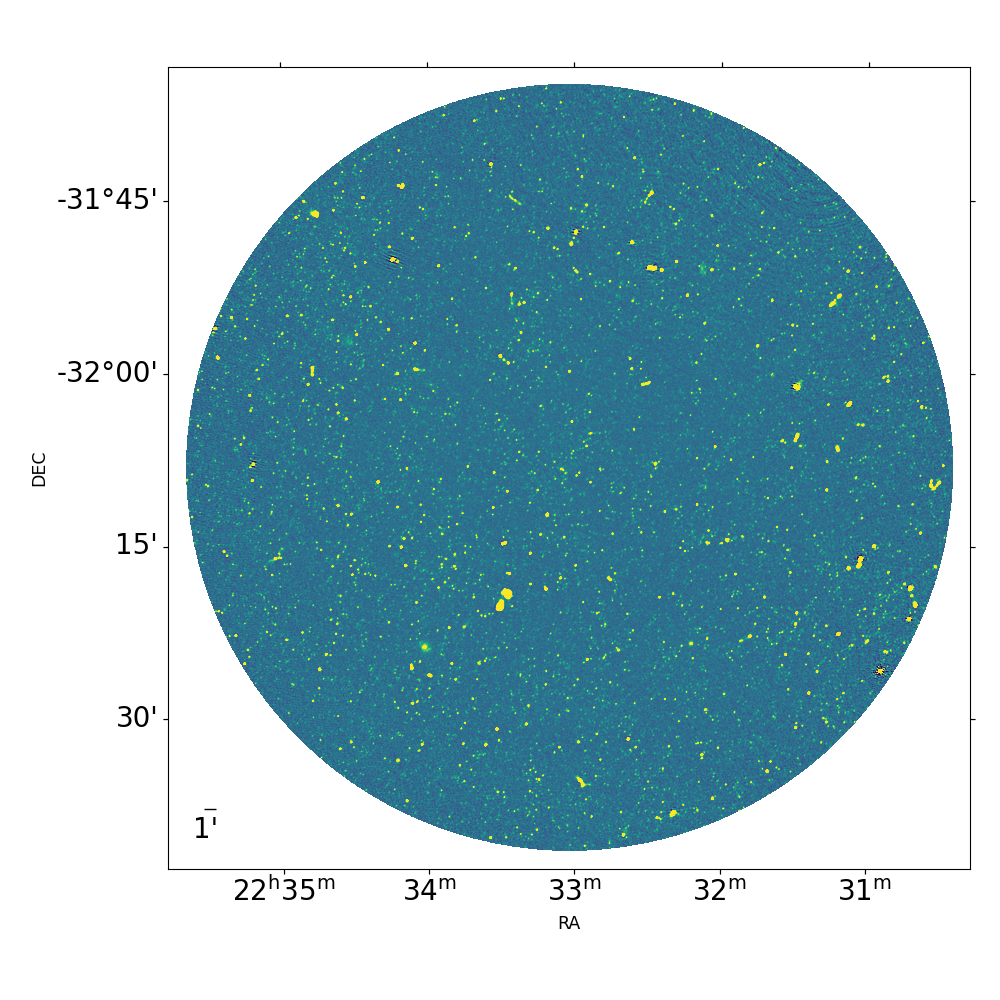}}\\
		\subfloat{\includegraphics[width=0.5\linewidth]{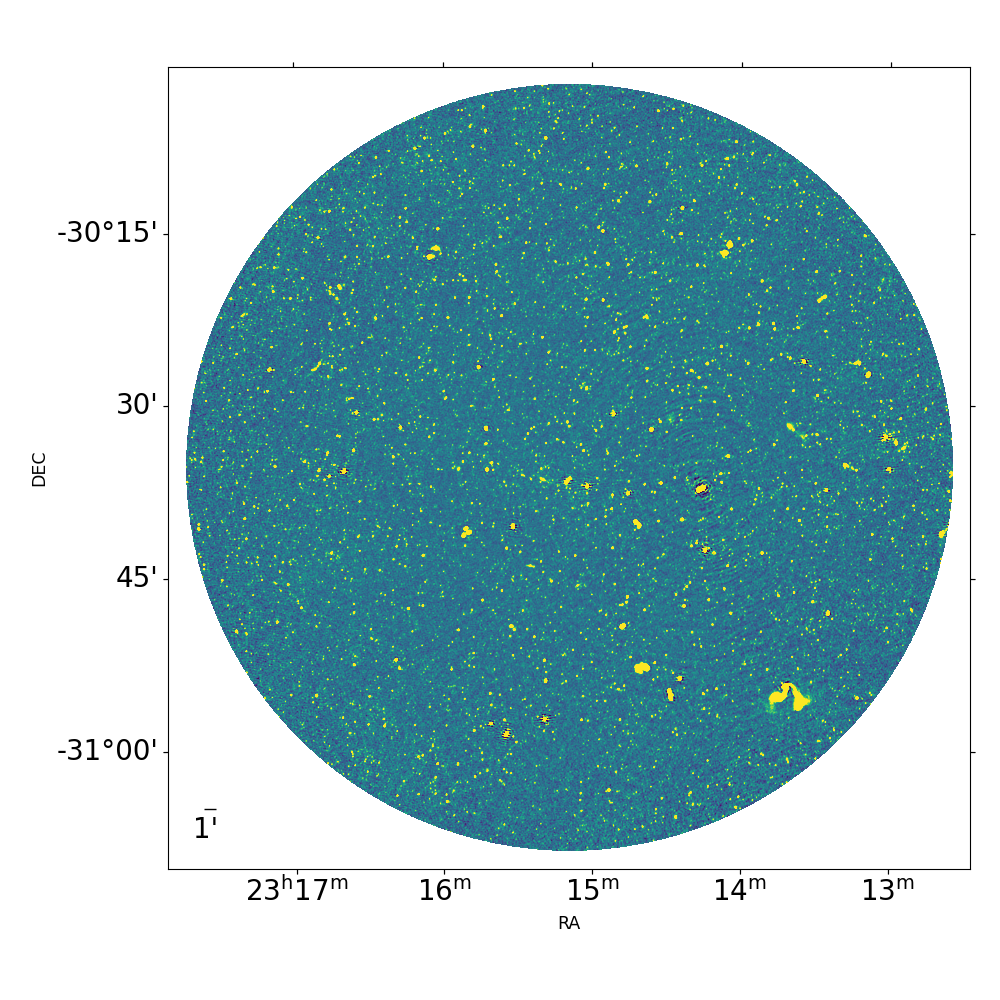}}
		\subfloat{\includegraphics[width=0.5\linewidth]{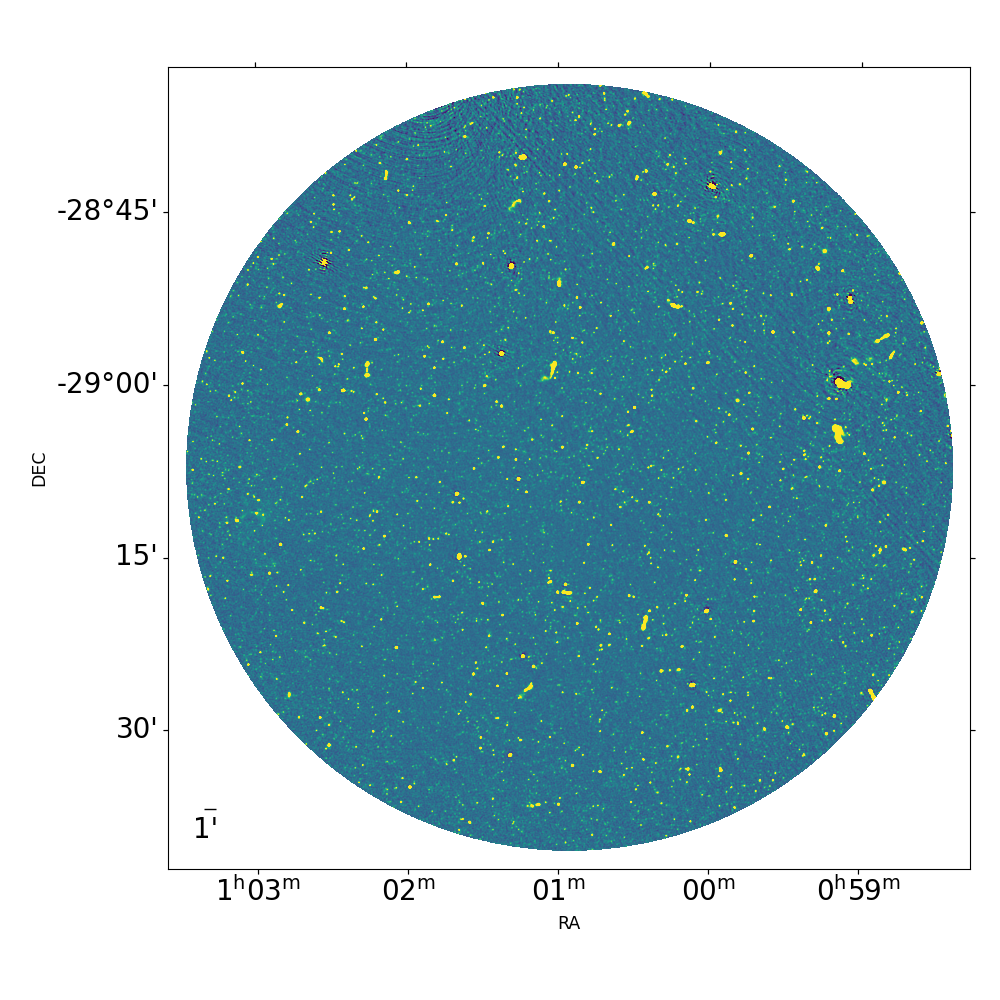}}
		\caption{The MeerKAT 1.28 GHz image of G014 presented in this work with yellow as bright. Top: G014 ; Bottom: G017 (left) and G257 (right). The right ascension (RA) and the declination (DEC) in each image are shown by the x-axis and the y-axis respectively. A scale bar of $1'$ is shown in the bottom left corner of each panel.}
		\label{fig.map}
	\end{figure*}

	\subsection{Target Selection}\label{sec.data.1}
	The three protocluster candidates studied in this paper are chosen from the 27 protocluster candidates selected by \citet{greenslade2018candidate}. We briefly outline their selection method here.\par 
	By cross-matching the \textit{Planck} compact source catalogue with \textit{Herschel} catalogues using the 4.63 arcmin \textit{Planck} beam size at 857GHz as the matching radius, \citet{greenslade2018candidate} identifies 4445 \textit{Herschel} sources matched with 431 \textit{Planck} compact sources. They then exclude \textit{Planck} sources whose counterparts are not potential protoclusters, such as local galaxies, stars or Galactic cirrus. For the remaining \textit{Planck} sources, \citet{greenslade2018candidate} counts the number of 250, 350, and 500 $\mu $m sources with fluxes at $> 25.4$ mJy that lie within 4.63 arcmin of the \textit{Planck} position, with the flux limit chosen to compare to published number counts. Any object with a $3\sigma$ overdensity in any of the three SPIRE bands is classed as a candidate protocluster of galaxies. This results in 27 candidates found.\par 
	Among these 27 candidate protoclusters, the 3 targets selected for this work have the most complete archival multi-band photometric data for further studies on the nature of DSFG protoclusters. The names of the three selected targets are PLCKERC857 G014.99-59.64, PLCKERC857 G017.86-68.67 and PLCKERC857 G257.09-87.10. For convenience, the three observed fields that contain the three candidate protoclusters are referred as G014, G017 and G257 hereafter\par 
	
	\subsection{\textit{Herschel} Data}\label{sec.data.2}
	The candidate protoclusters are selected from the H-ATLAS DR2 (\citealt{smith2017herschel}, \citealt{maddox2018herschel}) South Galactic Pole (SGP) data. The \textit{Herschel} Astrophysical Terahertz Large Area Survey (H-ATLAS) is the largest single key project carried out in open time with the \textit{Herschel Space Observatory} (\citealt{pilbratt2010herschel}). It is a survey of $660~\text{deg}^{2}$ with the PACS and SPIRE cameras in five photometric bands: 100, 160, 250, 350, and 500 $\mu$m. Data release 2 of H-ATLAS covers the SGP fields, and the total $1\sigma$ map sensitivity is 5.7, 6.0, and 7.3 mJy at 250, 350, and 500 $\mu$m, respectively. The SGP field, which includes our MeerKAT maps, is centred approximately at right ascension 0\hhh6\mmm and declination $-32^{\circ}44'$(J2000) and has an area of $317.6~\text{deg}^{2}$. The catalogues of FIR and submillimeter sources for the SGP field contains 193,527 sources detected at more than $4\sigma$ in any of the 250, 350 and 500 $\mu$m bands. The source detection process is based on the 250 $\mu$m map.\par 
	
	\subsection{Ancillary Data}\label{sec.data.3}
	The SGP field is covered by many large surveys which provide extensive data for further studies. These include the 2-Degree-Field Galaxy Redshift Survey (2dF, \citealt{colless20012df}), the VST Kilo-Degree Survey (KIDS, \citealt{de2013kilo}) which is an optical imaging survey in $u,g,r,i$, and the VISTA Kilo-Degree Infrared Galaxy Survey (VIKING, \citealt{edge2013vista}) which is a NIR imaging survey in $Z,Y,J,H$ and $K_{s}$. The H-ATLAS DR2 also has some overlap with the Dark Energy Survey (DES, \citealt{abbott2018dark}) which is a deep optical survey in $u,g,r,i,z,Y$, as well as with the recently-released ESO public survey Southern H-ATLAS Regions Ks-band Survey (SHARKS, \citealt{dannerbauer2022sharks}).\par 
	
	\subsection{MeerKAT 1.28 GHz Data}\label{sec.data.4}
	
	The three protocluster candidate fields were observed\footnote{Project code SCI-20190418-LL-01} with MeerKAT's L-band receivers (856--1712~MHz) between 12 and 15 June 2019. Details of the observations are provided in Table \ref{tab:targets}. The FWHM primary beam size is $\sim 34'$ at 1.28 GHz, covering an area of $\sim 0.971~\text{deg}^{2}$ for each image. The pixel scale of each image is about $1.1''\times1.1''$. Figure \ref{fig.map} displays our MeerKAT observations of the three fields. \par 
	
	The data were retrieved from the MeerKAT archive\footnote{\url{https://archive.sarao.ac.za}} in Measurement Set format, averaged down in frequency to 1024 channels prior to transfer. Initial flagging of the data was followed by the determination of bandpass, delay and time-dependent complex gain solutions from scans of the primary (PKS~B1934-638 in all cases) and secondary calibrators, using the {\sc casa} package {casa2022}. The resulting gain solutions were applied to the target scans, which were then flagged using {\sc tricolour}. This implements a MeerKAT-optimised version of the {\sc sumthreshold} algorithm \citep{offringa2010}, accelerated using {\sc dask-ms} \citep{perkins2022}.\par 
	
	A single round of phase and delay self-calibration is performed, by imaging using {\sc wsclean} \citep{offringa2014} with the deconvolution constrained to regions of genuine emission, and deriving the self-calibration solutions using {\sc cubical} \citep{kenyon2018}. Following this the data were subjected to a round of direction-dependent calibration, with imaging performed using {\sc ddfacet} \citep{tasse2018,tasse2023a}, and directional gain solutions derived for $\sim$10 user-defined directions using {\sc killms} \citep{smirnov2015,tasse2023b}. The final image was primary beam corrected by dividing it by an azimuthally-averaged model of the Stokes I beam derived using the {\sc katbeam}\footnote{\url{https://github.com/ska-sa/katbeam}} package. Full details of the reduction process are provided by \citet{heywood2022}. The scripts that performed the reduction are available online for general use \citep{heywood2020}. All processing was done using the \emph{ilifu} cluster\footnote{\url{https://www.ilifu.ac.za/}} in Cape Town, and all of the packages mentioned above were containerised using {\sc singularity} \citep{kurtzer2017}.

	\begin{table*}
		\centering
		\normalsize
		\caption{Names, positions, associated secondary calibrators, and total on-source time for each of the protocluster candidate target fields.}
		\begin{tabular}{lllllll} 
			\hline
			Name            & RA                    & Dec                       & Secondary     & Time on-source  & Synthesised beam size & Central thermal noise\\
							& (J2000)               & (J2000)                   & calibrator    & (h) &  &  ($\mu \text{Jy/beam}$)          \\ \hline
			G014.99$-$59.64 & 22\hhh33\mmm02\sss.16 & $-$32\ddd08\dmm16\farcs8  & J2152$-$2828  & 3.75  & $7.5'' \times 6.6''$ &  4.35       \\
			G257.09$-$87.10 & 01\hhh00\mmm55\sss.68 & $-$29\ddd07\dmm19\farcs2  & J0025$-$2602  & 4 &  $7.6'' \times 7.0''$  &  5.04        \\
			G017.86$-$68.67 & 23\hhh15\mmm09\sss.36 & $-$30\ddd35\dmm27\farcs6  & J2341$-$3506  & 3.5  & $7.5'' \times 6.3''$ &  5.62       \\ \hline 
		\end{tabular}
		\label{tab:targets}
	\end{table*}

	\section{Source Extraction}\label{sec.extra}
	\subsection{Extraction Software}\label{sec.extra.1}
	To improve the robustness of the final catalogues as well as comparing different source extraction tools, sources are extracted from the MeerKAT maps with both the radio source extraction tool PyBDSF (the Python Blob Detector and Source Finder, \citealt{mohan2015pybdsf}) and the SExtractor (Source Extractor, \citealt{bertin1996sextractor}).It should be noted that SExtractor is not designed for radio images and it may find unphysical sources smaller than the synthesised beam.\par
	PyBDSF first calculates the background rms and mean. Regions where fluxes are $3 \sigma$ above the mean flux of the image and peak fluxes $5\sigma$ above the mean are fitted with Gaussians then extracted as valid sources. Visual inspection finds that a single run of PyBDSF extraction on our images leaves a significant number of faint sources undetected. Therefore, to increase the extraction completeness, we have two PyBDSF runs for each map: the original MeerKAT image is used for the first run, then a second extraction run is carried out on the residual map of the first run. The default flagging options are used to exclude residuals from the first run being detected as sources. Visual inspection of the second extraction run does not find any false detection originating from first run residuals. The two resulting catalogues are then concatenated into one final catalogue. The final PyBDSF catalogues of the G014, G017 and G257 MeerKAT images contain 5925, 5415 and 5795 sources respectively.\par 
	
	SExtractor uses a different approach. Two passes are made through the data. During the first pass, a model of the sky background is built, and several global statistics are computed. During the second pass, image pixels are background-subtracted, filtered and segmented on-the-fly. Detections are then deblended, pruned (or, “CLEANed”), and enter the measurement phase. Finally, the measured quantities which are defined by the user are written to the output catalogue, after cross-matching with an optional input list. We set the single-pixel detection threshold to be $\sim 1.8\sigma$ with the minimum number of connected pixels above the threshold set to 5, which results in a $\geq 4 \sigma$ final detection limit for a valid source. The default settings are used for the other parameters. The final SExtractor catalogues of the G014, G017 and G257 MeerKAT maps contain 16403, 16557 and 16339 sources respectively. For the measured total flux densities of the detected sources, we use FLUX\_BEST from the SExtractor catalogue. In most cases, FLUX\_BEST uses the fluxes measured by integrating pixel values within an adaptively scaled aperture, but it will use the corrected isophotal fluxes if the contribution of other nearby sources exceeds $10\%$. The catalogued fluxes are in units of the image pixel which is Jy/beam. We have converted these to total fluxes in $Jy$ by:
	\begin{equation}\label{eq.0}
		\text{Total~flux(Jy)=Flux(JY/beam)} \times \frac{\text{Pixel~area}(\text{deg}^{2})}{\text{Synthesised~beam~area}(\text{deg}^{2})}
	\end{equation}
\par 

	\subsection{Extraction Completeness \& Reliability}\label{sec.extra.2}
\begin{figure*}
		\centering
		\subfloat[Source extraction completeness]{\includegraphics[width=0.5\linewidth]{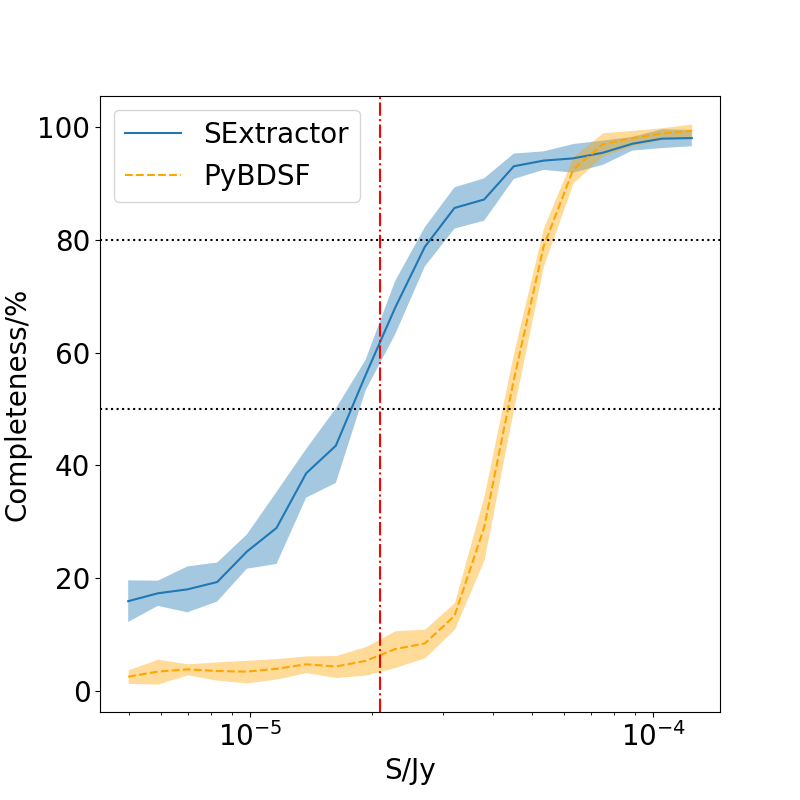}}
		\subfloat[Reliability]{\includegraphics[width=0.5\linewidth]{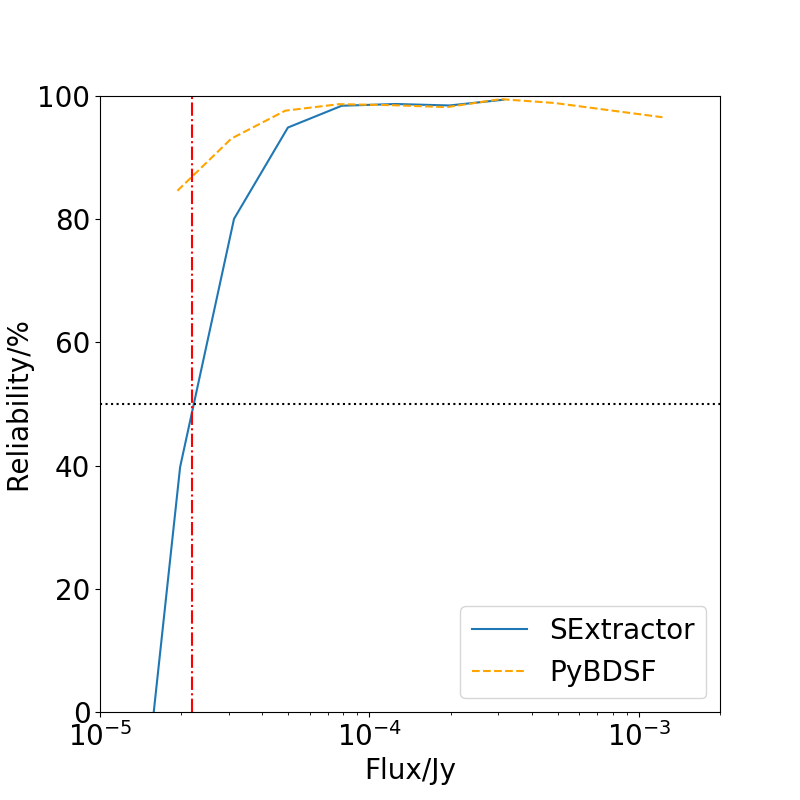}} \\		
		\subfloat[Flux distributions]{\includegraphics[width=0.5\linewidth]{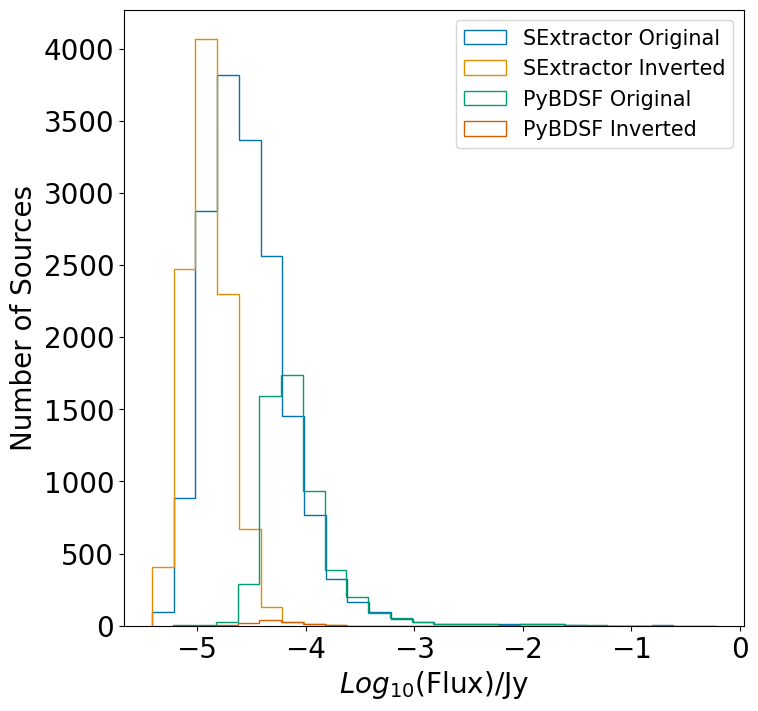}}
		\subfloat[SExtractor-PyBDSF flux comparison]{\includegraphics[width=0.5\linewidth]{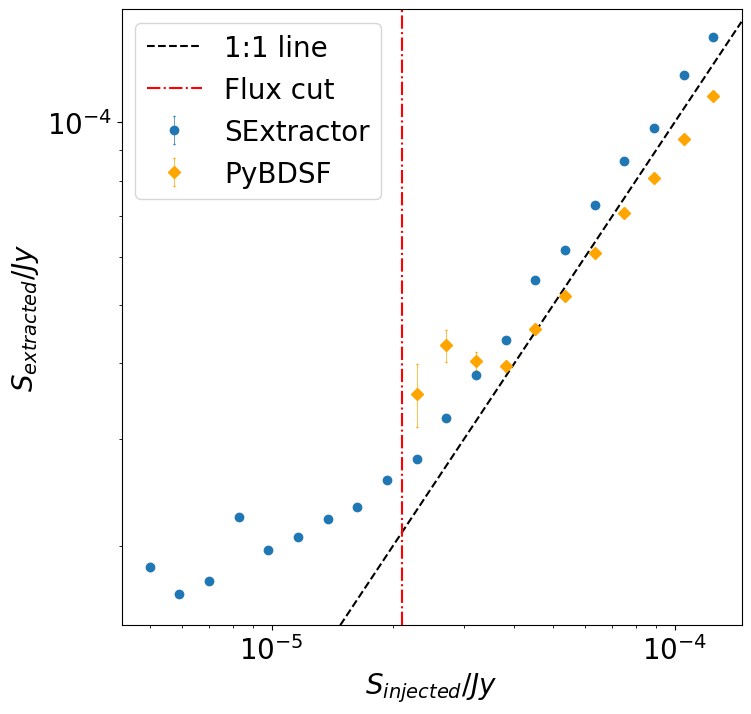}}
		\caption{ (a) Extraction completeness as a function of 1.28 GHz flux. The shaded areas show the respective standard deviation, the black dotted lines indicate 50\% and 80\% completeness, and the red dashed-dotted line shows the $S=0.021$ mJy flux cut. It can be seen that SExtractor (blue solid line) has better completeness at fainter fluxes. SExtractor reaches a 50\% completeness at $S\approx 0.017$ mJy while PyBDSF (orange dashed line) reaches a 50\% completeness at $S\approx 0.035$ mJy with smaller uncertainties. (b) The reliability of the source extraction with SExtractor and PyBDSF. The blue solid line shows the result from SExtractor and the orange dashed line shows the reliability of PyBDSF. The black dotted line indicates 50\% reliability and the red dashed-dotted line shows the $S=0.021$ mJy flux cut. PyBDSF barely picks up any spurious objects, whereas over half of sources detected by SExtractor are likely spurious at fluxes lower than $\sim 0.021$ mJy. (c) Flux distributions of sources extracted from both the original maps and the flux-inverted maps. (d) Average total flux of injected sources recovered by SExtractor and PyBDSF vs. average total flux of sources injected into the map in Jy. The blue dots are SExtractor fluxes and the yellow diamonds are PyBDSF fluxes. The black dashed line is the 1:1 line, the red dashed-dotted line is the $S=0.021$ mJy flux cut. At $S>0.021$ mJy SExtractor slightly overestimates the flux density of a source, while PyBDSF only have reliable flux density measurements at $S>0.030$ mJy and it increasingly underestimates source flux densities as the flux level goes higher. At $S>0.021$ mJy, a 1.42 linear proportionality is found between PyBDSF flux measurements and SExtractor flux measurements.}
		\label{fig.extract}
	\end{figure*}

	By injecting fake sources into the original MeerKAT maps and re-extracting sources following the same extraction procedures described in the last section, we quantify the extraction completeness of both PyBDSF and SExtractor as shown in Figure \ref{fig.extract}a. The fake sources are compact 2-D Gaussians which are randomly distributed on the maps, the size of the Gaussians is comparable to the general size of real sources, most of which have FWHMs of $\sim 6''-12''$. 100 sources with the same total flux are injected and re-extracted, then the same procedure is repeated with increasing total flux until the completeness reaches 100\%. Injecting more than 100 fake objects into our images will lead to simulated sources confused with real sources, therefore at each flux level we iterate the injection process 10 times, then take the average completeness as the final number to produce a robust measure of completeness. To avoid overlapping between injected sources and real MeerKAT sources, all simulated sources are set to be at least $6''$ away from any real source.\par 
	
\begin{figure*}
		\centering
		\subfloat[MeerKAT]{\includegraphics[width=0.5\linewidth]{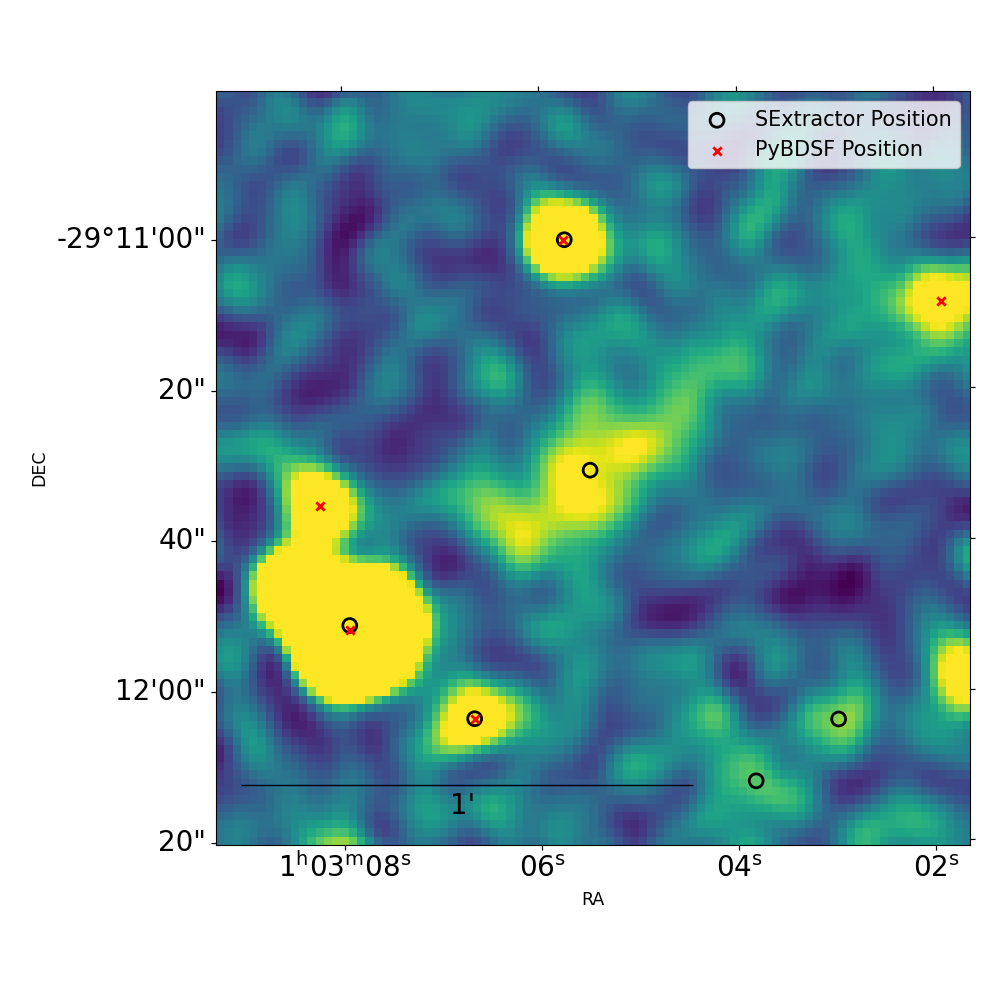}}
		\subfloat[SHARKS]{\includegraphics[width=0.5\linewidth]{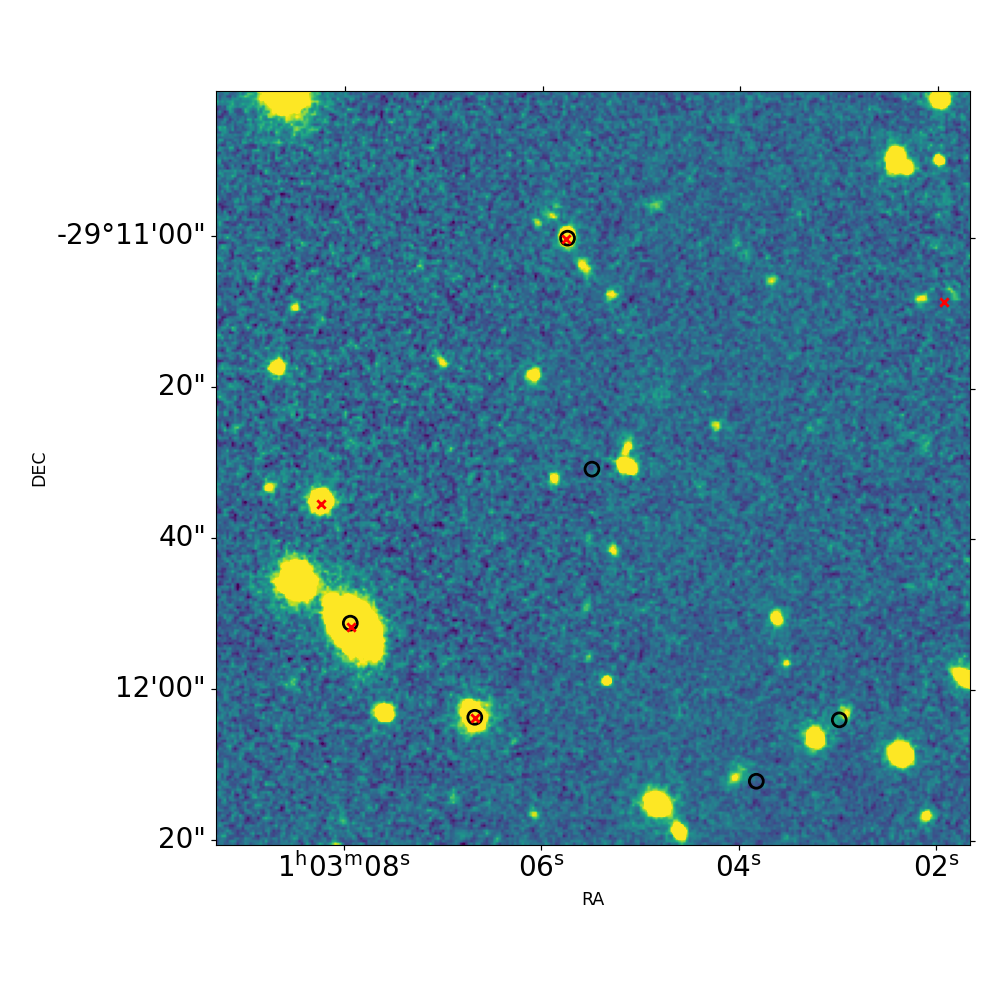}}
		\caption{An example of a faint source detected by SExtractor but not PyBDSF. MeerKAT 1.28 GHz image (a) and SHARKS Ks image (b) are positionally matched and shown, with RA as x-axis and DEC as y-axis. The black circles indicate sources detected by SExtractor, the red crosses show the PyBDSF-detected sources, a scale bar of $1'$ is shown in the bottom left of the left panel. As can be seen from the SHARKS Ks image, there is a faint source right at the centre of the images. However, this faint source is only detected by Source Extractor while ignored by PyBDSF. PyBDSF only detects the much brighter sources in this region.}
		\label{fig.sex_better}
	\end{figure*}

	It can be seen that SExtractor has higher completeness at fainter fluxes, as it reaches  50\% and 80\% completeness at $S\approx 0.018$ mJy and $S\approx 0.027$ mJy while PyBDSF has the same completenesses at $S\approx 0.036$ mJy and $S\approx 0.048$~mJy, although PyBDSF's completeness has smaller variations. This indicates that SExtractor can extract sources at $0.018\leq S \leq 0.036$~mJy which are difficult for PyBDSF to detect. An example is shown in Figure \ref{fig.sex_better}. \par 
	
	\begin{figure}
		\centering
		\includegraphics[width=\columnwidth]{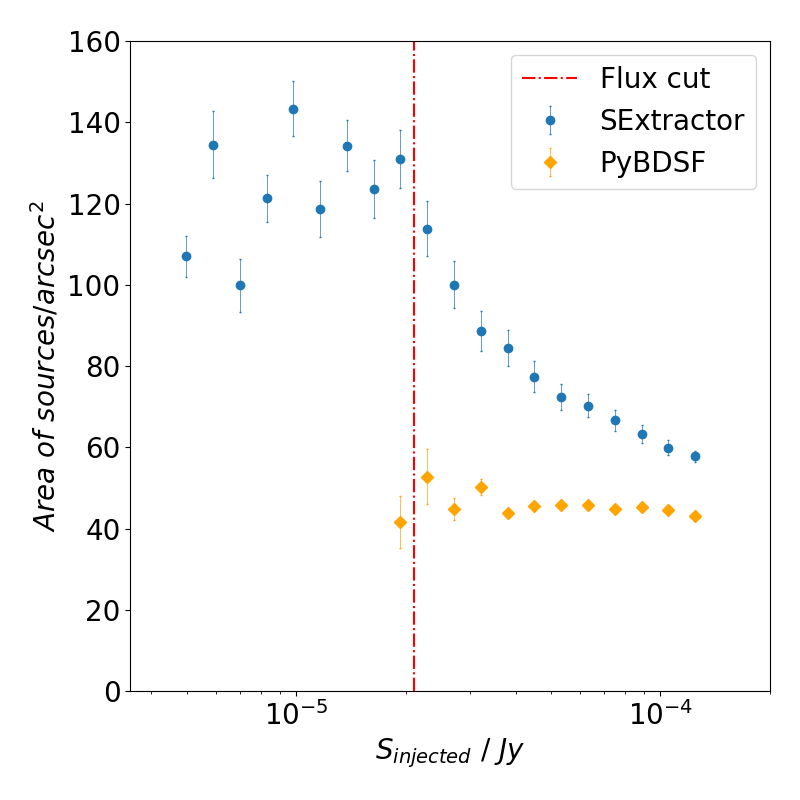}
		\caption{The average area of injected sources (in $\text{arcsecond}^{2}$) extracted by SExtractor (blue dots) and PyBDSF (orange diamonds) as a function of flux density. The red dashed-dotted line shows the $S=0.021$~mJy cut. The average area of SExtractor source decreases as the flux density increases, while the area of PyBDSF detections is roughly a constant after the flux cut. This agrees well with the flux density offset between the two source finders seen in Figure \ref{fig.extract}d, supporting the hypothesis that SExtractor is measuring a slightly larger area per source than PyBDSF does.}
		\label{fig.size}
	\end{figure}

	\begin{figure*}
		\centering
		\subfloat[]{\includegraphics[width=0.333\linewidth]{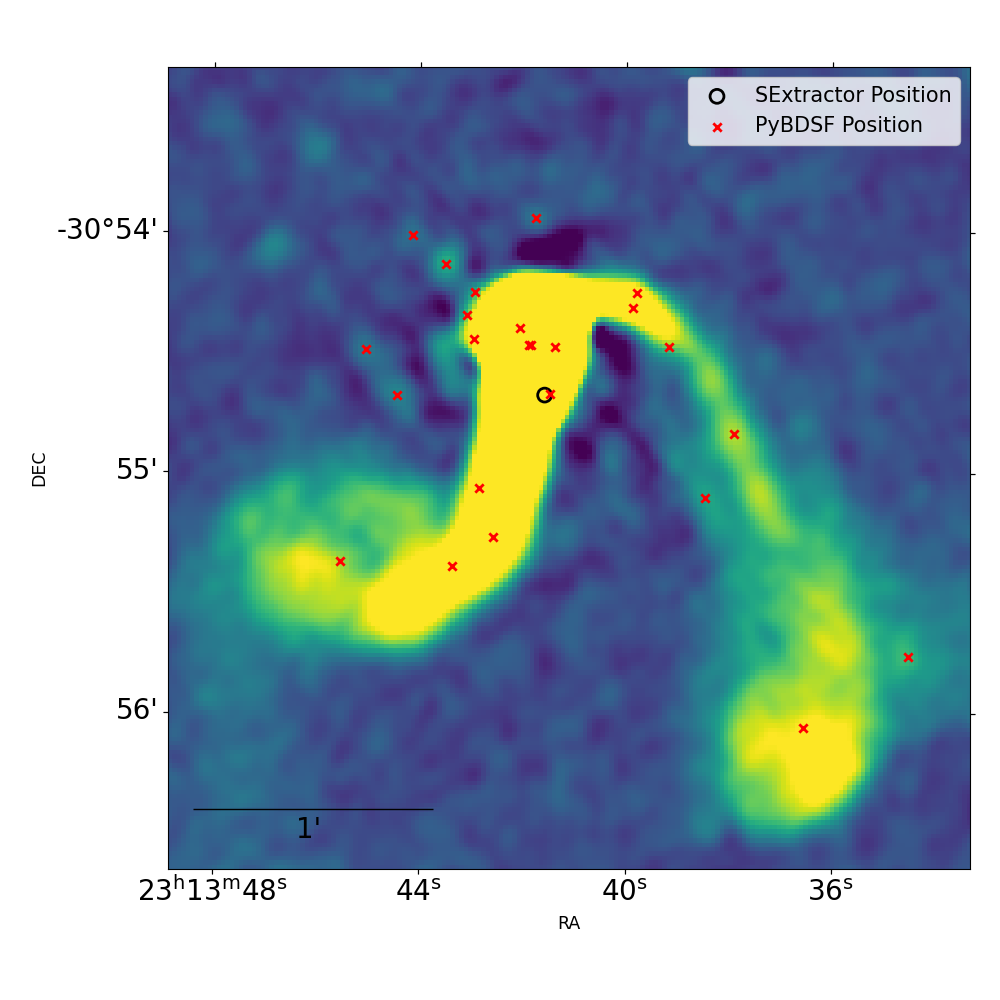}}
		\subfloat[]{\includegraphics[width=0.333\linewidth]{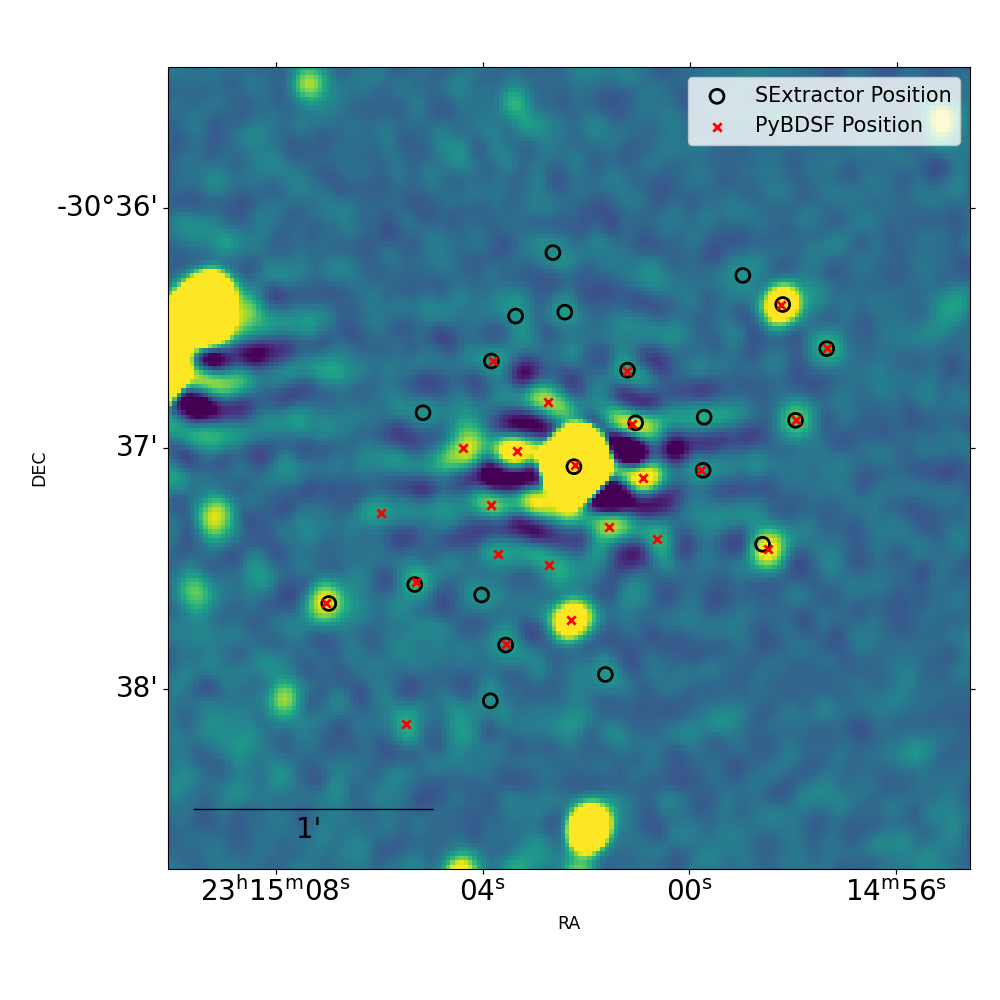}}
		\subfloat[]{\includegraphics[width=0.333\linewidth]{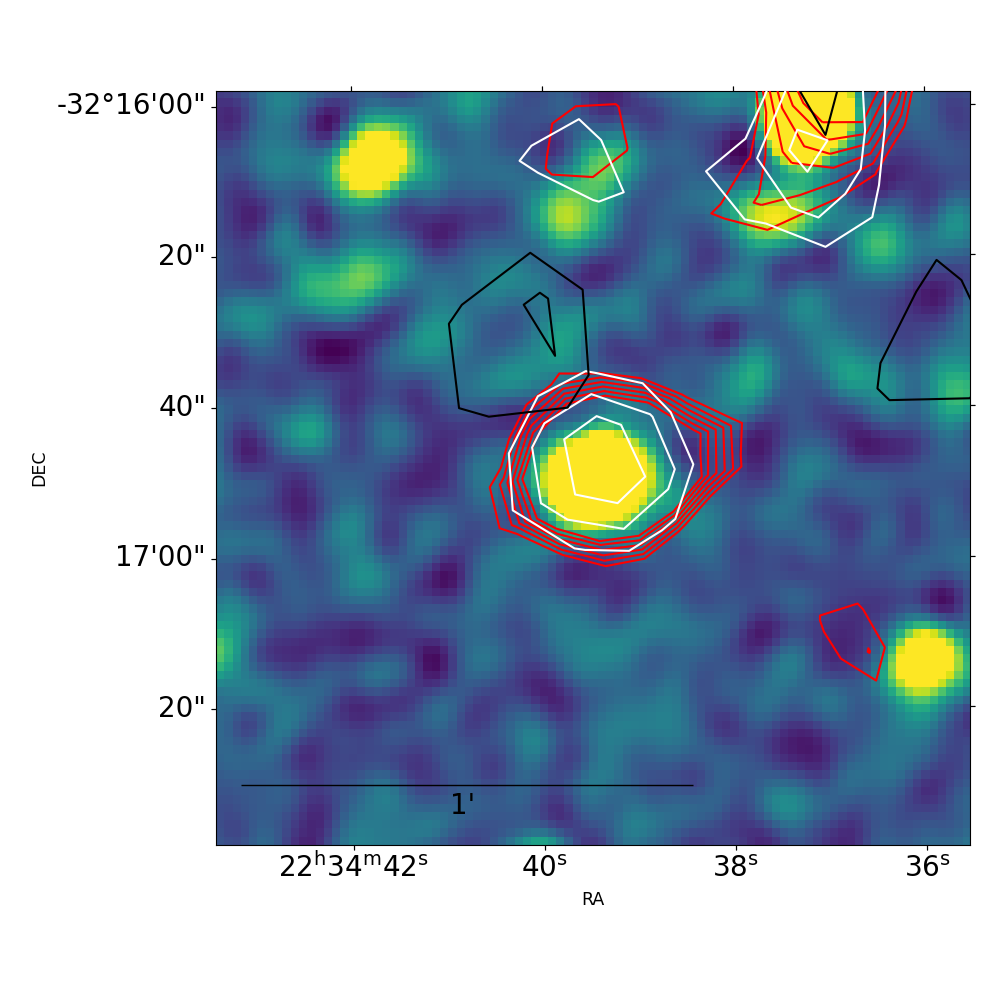}} \\
		\subfloat[]{\includegraphics[width=0.333\linewidth]{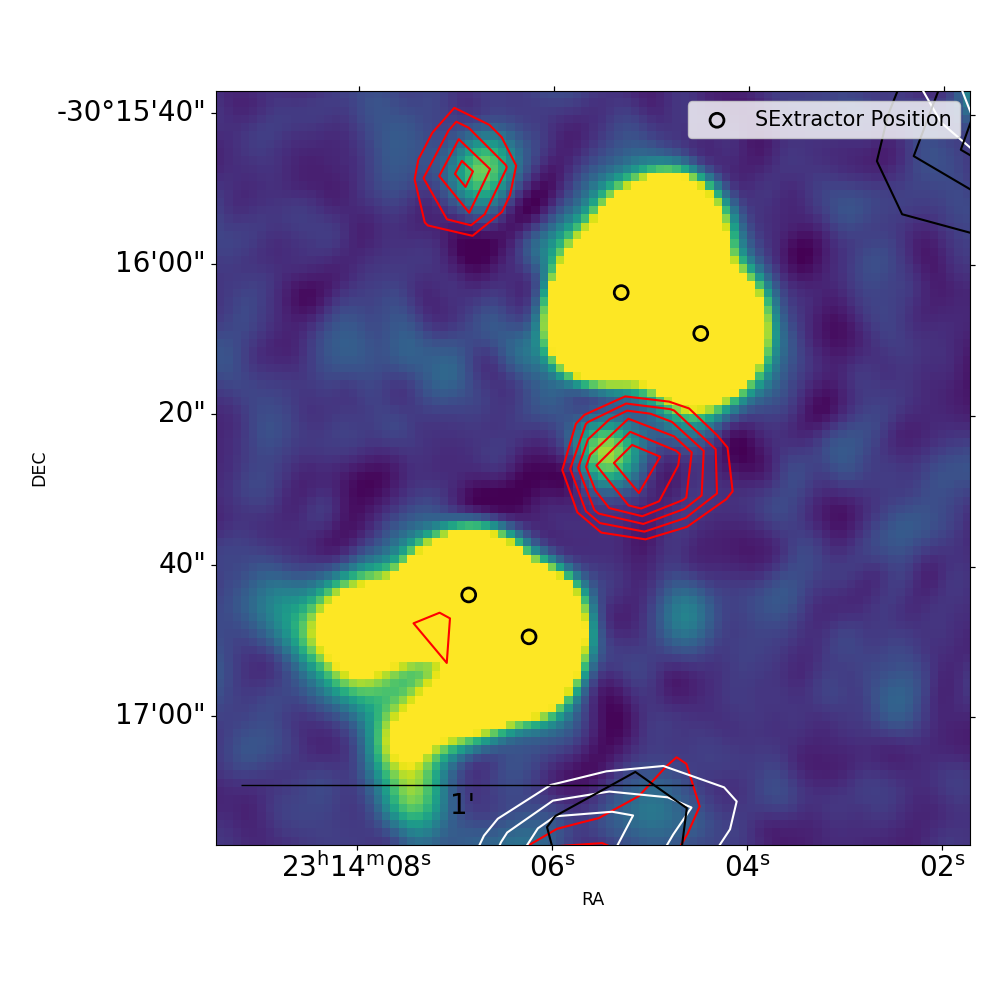}}
		\subfloat[]{\includegraphics[width=0.333\linewidth]{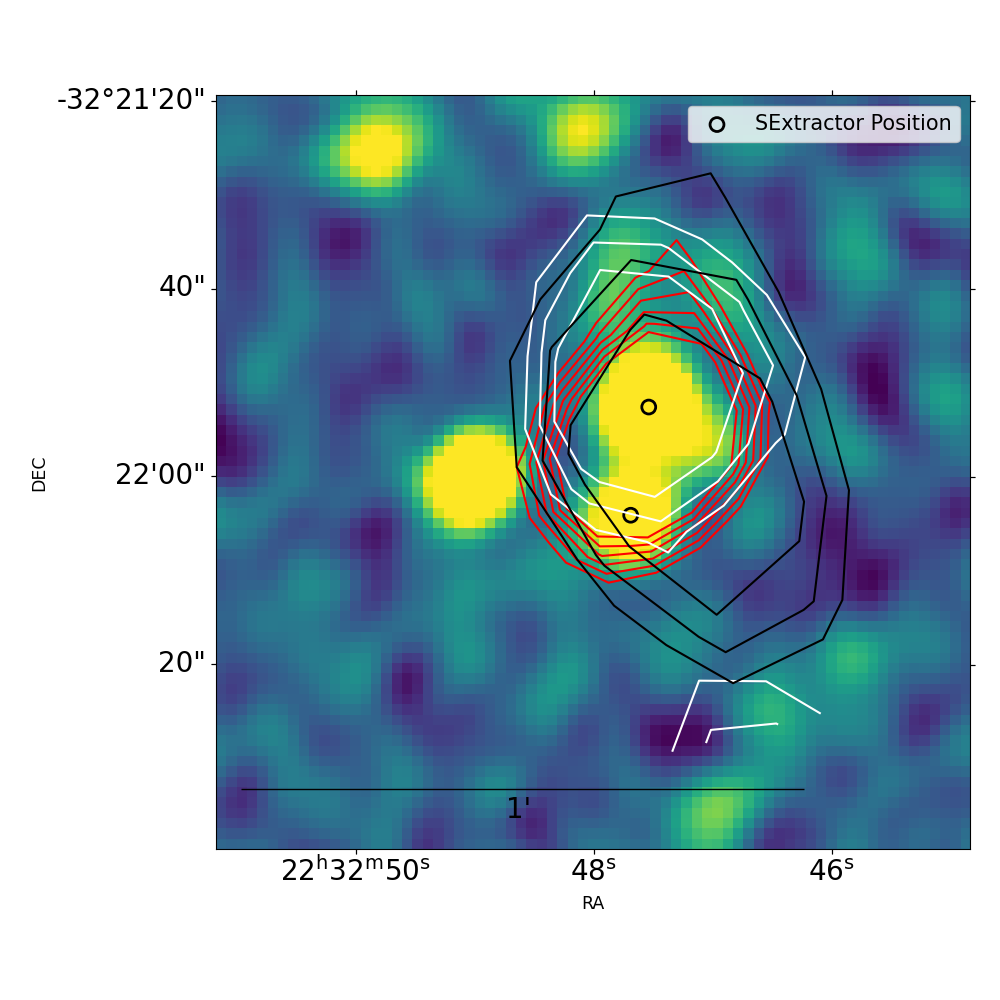}}
		\subfloat[]{\includegraphics[width=0.333\linewidth]{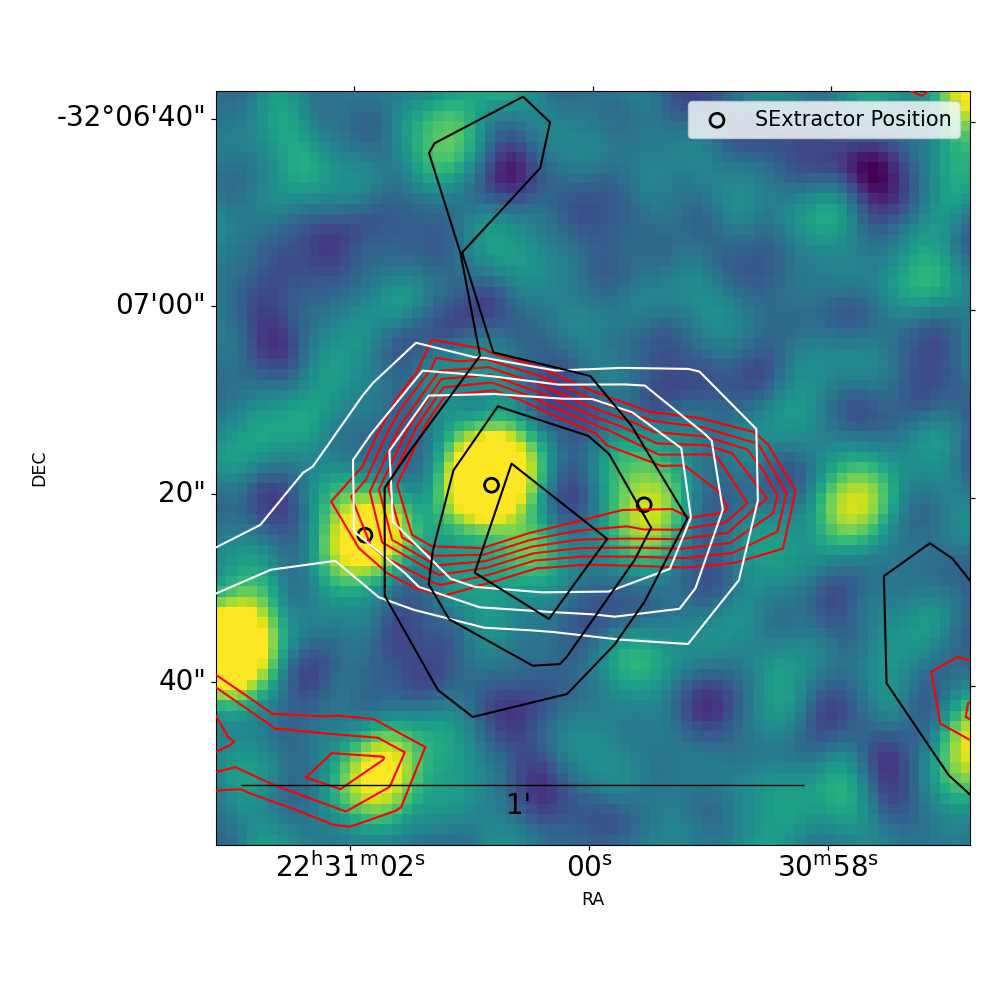}} \\
		\subfloat[]{\includegraphics[width=0.333\linewidth]{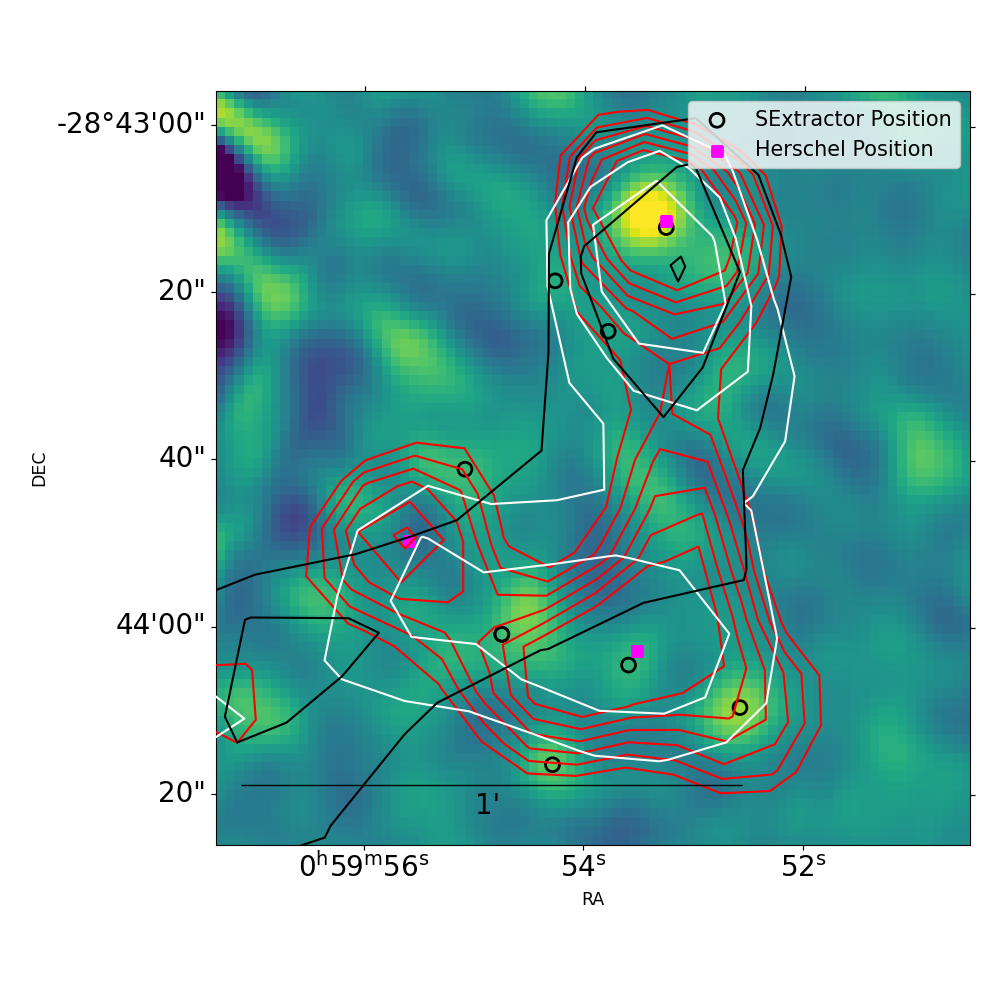}}
		\subfloat[]{\includegraphics[width=0.333\linewidth]{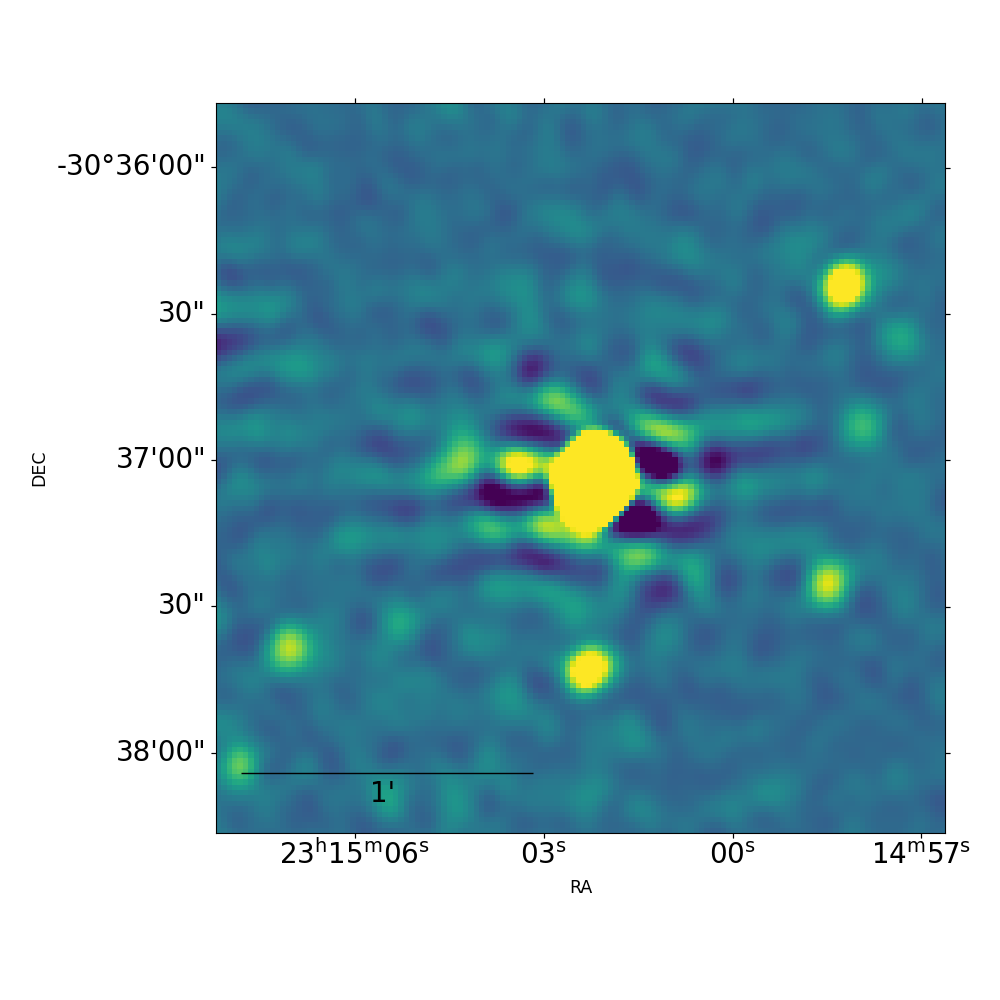}}	
		\subfloat[]{\includegraphics[width=0.333\linewidth]{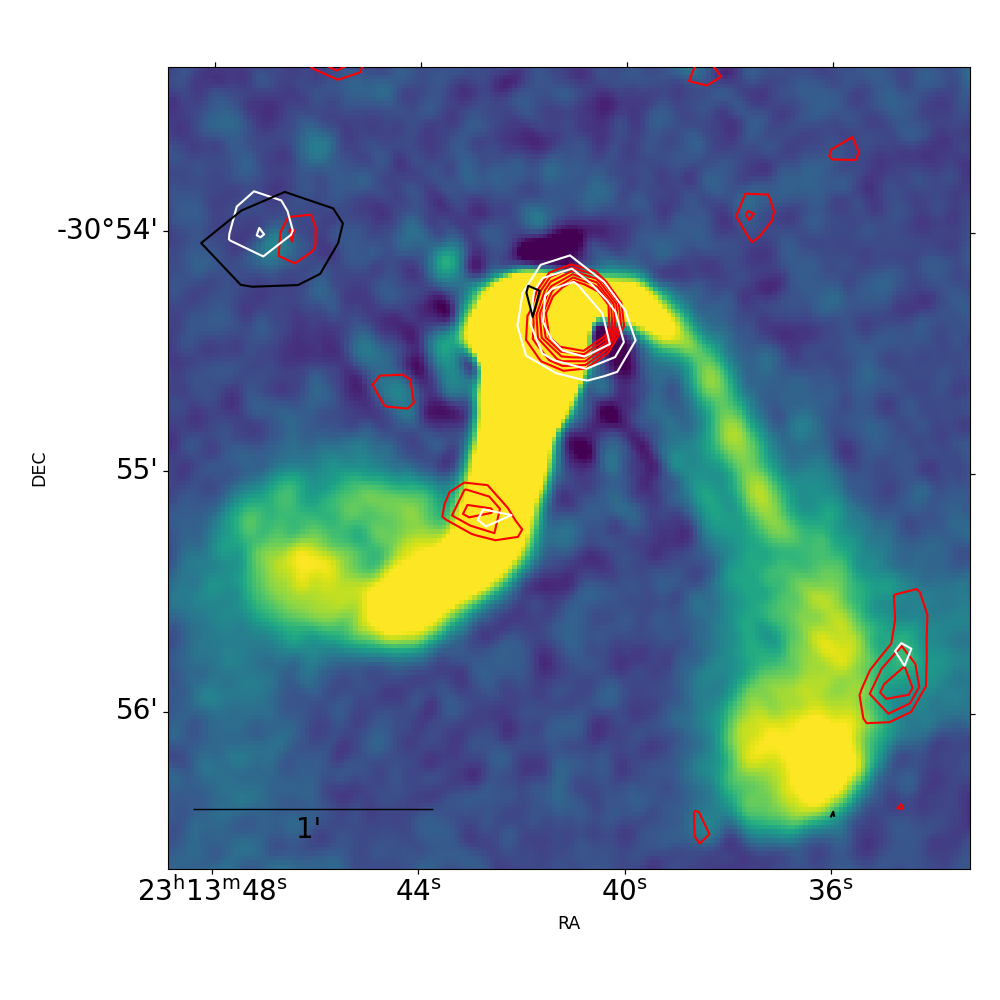}}				
		\caption{MeerKAT image cutouts. The background images are the 1.28 GHz MeerKAT observations, the red contour illustrates the $5-10\sigma$ \textit{Herschel} 250 $\mu$m flux density, the white and black contours are the $3-5\sigma$ \textit{Herschel} 350 and 500 $\mu$m flux density respectively. A scale bar of $1'$ is shown in every cutout. Positions of SExtractor detected MeerKAT sources are shown in black circles, PyBDSF positions are shown in red crosses and \textit{Herschel} positions are shown in magenta squares. (a) A bright source with extended morphology. (b) A bright source with strong beam artefacts. (c) An example of the \textit{Herschel}-MeerKAT cross-identification. (d) A true \textit{Herschel}-MeerKAT cross-identification not being matched due to this source's extended morphology. (e) A \textit{Herschel} source with two MeerKAT counterparts. (f) A \textit{Herschel} source with three MeerKAT counterparts. (g) A \textit{Herschel} source with apparent MeerKAT multiplicity of 7. There are actually 3 \textit{Herschel} sources close to each other, overlapping with 7 faint MeerKAT sources. The true number of counterparts of each \textit{Herschel} sources will be $1-4$. There is an additional faint MeerKAT source to the left of the \textit{Herschel} source at the top. It could be a spurious detection and is not matched to any \textit{Herschel} source. (h) The MeerKAT sources with the lowest $q_{250\mu \text{m}}$, it is a very bright radio sources with beam artefacts around it. (i) A radio-loud source with low $q_{250\mu \text{m}}$. The radio source is associated with two jets orthogonal to each other. A \textit{Herschel} source coincides with the centre of this extended radio structure, suggestive of AGN activity.}
		\label{fig.cutout}
	\end{figure*}

	To assess the reliability of the two extractors, in other words what fraction, as a function of flux, of the detected sources are real objects instead of noise, we invert our images by multiplying the original flux maps by -1, then re-extract catalogues with SExtractor and PyBDSF using the same parameters. The resulting flux distributions are then compared to the original distributions, as shown by Figure \ref{fig.extract}c. Objects detected in the flux-inverted images are mostly spurious sources induced by noise or random fluctuations. Therefore measuring the number of objects in the flux-inverted images will provide an estimate of how many spurious sources our source extractors detect at different flux levels in real images. We define the reliability as:
	\begin{equation}\label{eq.6}
		\text{Reliability}=\frac{N_{\text{real}}-N_{\text{inverted}}}{N_{\text{real}}}
	\end{equation}
	where $N_{\text{real}}$ and $N_{\text{inverted}}$ are the number of sources extracted from real images and flux-inverted images respectively. Figure \ref{fig.extract}b shows the reliabilities of the two extractors as a function of flux. For PyBDSF, very few spurious objects are detected as real sources, showing a very high source extraction reliability. For SExtractor, spurious objects contribute over half the number counts at fluxes lower than $\sim 0.021$~mJy as indicated by the red dashed line, which means sources extracted by SExtractor with $S \leq 0.021$~mJy cannot be considered reliable. Comparing Figure \ref{fig.extract}a\&b, while PyBDSF has better extraction reliability than SExtractor at any flux, SExtractor has much better completeness at fainter fluxes. SExtractor can detect sources down to $S=0.021$~mJy with reasonable reliability and a completeness of $\sim 60\%$, meanwhile PyBDSF can only reach a completeness of $\sim 7\%$ at this flux. Therefore, at this flux level, the fraction of real detected objects is far higher in the SExtractor based catalogue, even with the lower reliability.  \par 
	To evaluate how well the two codes measure flux densities, we compare the average total fluxes of injected fake sources recovered by SExtractor and PyBDSF, as a function of injected flux. Figure \ref{fig.extract}d shows the results. It can be seen that SExtractor will slightly overestimate source flux densities but its flux density measurements are consistently reasonable at $S>0.021$~mJy, while PyBDSF have better flux recovery at $S>0.030$~mJy but it will increasingly underestimate source flux densities as the flux level goes higher. Fitting a linear regression model to the SExtractor and PyBDSF extracted fluxes, we find a 1.42 proportionality at $S>0.021$~mJy, meaning that the flux density of a source measured by SExtractor is 1.42 times higher than that measured by PyBDSF on average. This offset between the measured flux densities is likely to be a result of different detection algorithms implemented by the software. PyBDSF fits Gaussians to regions that are $3 \sigma$ above the mean flux of the image with peaks above $5 \sigma$, then measures the total fluxes within the fitted Gaussians, while SExtractor identifies regions which contain at least 5 connected pixels that are above $1.8 \sigma$ significance then estimates the total fluxes by integrating pixel values within an adaptively scaled elliptical aperture. Therefore, a possible explanation of why SExtractor measures a higher total flux is that for each detected source, the elliptical aperture used by SExtractor will include more pixels than the Gaussians fitted by PyBDSF, leading to a higher measured flux. Figure \ref{fig.size} shows the average sizes of simulated sources extracted by SExtractor and PyBDSF as a function of flux density. The average size of SExtractor sources decreases as the flux density increases, while the average size of PyBDSF detections remains roughly constant after the flux cut. This agrees well with the trend seen in Figure \ref{fig.extract}d, which is supportive of the hypothesis that SExtractor is measuring a larger area per source. Furthermore, \citet{hale2019radio} finds that PyBDSF has an excess of positive residuals, suggesting that the Gaussian models of PyBDSF are under-fitting the sources in the field and hence slightly underestimating the true fluxes. After visual examination, we find that for bright sources with extended morphologies or obvious artefacts, SExtractor generally tends to blend them with nearby smaller sources or noise, boosting the measured fluxes, whereas PyBDSF is more likely to pick up small flux variations inside or around extended objects as individual sources. Figure \ref{fig.cutout}a\&b illustrates two example bright sources and the positions of objects identified by SExtractor and PyBDSF around the bright sources. As can be seen in Figure \ref{fig.cutout}a, for a bright source with extended morphology, PyBDSF has identified numerous objects inside and around it, whereas SExtractor has taken the entire area as one large source. For a bright source with beam artefacts as in Figure \ref{fig.cutout}b, PyBDSF has picked up several small blobs induced by the synthesised beam, while SExtractor has only identified one or two such artefacts as sources. In short, the origin of the flux measurement offset between the two source finders is still unclear and requires further investigations to be determined. For the rest of this paper, we will apply the empirically found correction factor of 1.42 to our SExtractor-measure flux densities for comparisons to results produced by PyBDSF. \par
	
	In conclusion, we decide to use the SExtractor catalogue as our final catalogue, with a flux cut at $0.021$~mJy to minimise spurious objects in the catalogue while achieving a higher completeness. After this flux cut, there are 11016, 11509 and 13332 sources in the G014 G107 and G257 maps respectively.

	\subsection{MeerKAT 1.28 GHz Source counts}\label{sec.counts}
	One of the main goals of our MeerKAT observations is to aid the identification of DSFG protocluster candidates selected from \textit{Herschel} data. To look for traces of protoclusters in the observed area, we calculate the brightness-weighted differential number counts $S^{2}n(S)$ (i.e. Euclidean normalised number counts) for our extracted sources with a flux range between $0.021$~mJy $\leq S < 1$~Jy, where S is the flux density of the source and n(S) is the the number of sources per steradian with flux densities between $S$ and $S + dS$. 
	\begin{equation}\label{eq.7}
		n(s)=\frac{N}{A \times dS}
	\end{equation}
	where $N$ is the number of sources between $S$ and $S+dS$ and $A$ is the total area which is $\sim 0.971~\text{deg}^{2}$ for each image. We count the extracted sources in bins of width 0.2 dex, centred on 1.28 GHz flux densities i.e. $\log_{10}[S(\text{Jy})]=-4.6,-4.4,-4.2,-4.0,-3.8,...,0.0$, then compare our results with source counts at similar frequencies in the literature. Corrections based on the completeness and reliability obtained in Section \ref{sec.extra.2} are applied to our number counts. To compare with literature results which use PyBDSF, we also apply the 1.42 SExtractor-PyBDSF flux correction to our number counts to make them comparable. Table \ref{tab.1} shows our 1.28 GHz source counts. The uncertainties in the differential counts are the quadratic sum of the Poisson uncertainties in samples of size $N$ in each bin.\par 
	
	\begin{table*}	
		\centering
		\normalsize
		\caption{The 1.28 GHz source counts in this work, corresponding to Figure \ref{fig.count}. For each flux density bin of width 0.2 dex, the bin centre $\log_{10}[S]$, raw source counts $N$ in the bin, the integral counts $N_{\text{integral}}$ (i.e. cumulative raw source counts) and the uncorrected brightness-weighted counts $S^{2}n(s)$ in the bin is shown. The brightness-weighted counts corrected for completeness, reliability and PyBDSF-SExtractor flux ratio $S^{2}n(s)_{\text{corr}}$ is also displayed.}
		\label{tab.1}
		\begin{adjustbox}{width=12cm}	
			\begin{tabular}{ c c c c c } 
				\hline
				\hline
				$log_{10}[S/\text{Jy}]$ & $N$ & $N_{\text{integral}}$ & $S^{2}n(s)/\text{Jysr}^{-1}$ & $S^{2}n(s)_{\text{corr}}/\text{Jysr}^{-1}$\\
				\hline
				\hline
				-4.65 & 7031 & 7031 & $381.9\pm 4.6$ & $575.7\pm 5.6$\\
				-4.45 & 10136 & 17167 & $872.6\pm 8.7$ & $853.2\pm 8.6$\\
				-4.25 & 8150 & 25317 & $1112.0\pm 12.3$ & $964.3\pm 11.5$\\
				-4.05 & 5196 & 30513 & $1123.6\pm 15.6$ & $908.3\pm 14.0$\\
				-3.85 & 2716 & 33229 & $930.8\pm17.9$ & $699.6\pm 15.5$\\
				-3.65 & 1220 & 34449 & $662.7\pm 19.0$ & $509.7\pm 16.6$\\
				-3.45 & 584 & 35033 & $502.8\pm 20.8$ & $382.9\pm 18.2$\\
				-3.25 & 303 & 35336 & $413.4\pm 23.8$ & $325.8\pm 21.1$\\
				-3.05 & 163 & 35499 & $352.5\pm 27.6$ & $280.3\pm 24.6$\\
				-2.85 & 111 & 35610 & $380.4\pm 36.1$ & $331.5\pm 33.7$\\
				-2.65 & 62 & 35672 & $336.8\pm 42.8$ & $335.8\pm 42.7$\\
				-2.45 & 52 & 35724 & $447.7\pm 62.1$ & $352.0\pm 55.0$\\
				-2.25 & 33 & 35757 & $450.3\pm 78.4$ & $353.8\pm 69.5$\\
				-2.05 & 25 & 35782 & $540.6\pm 108.1$ & $496.0\pm 103.6$\\
				-1.85 & 25 & 35807 & $856.8\pm 171.4$ & $820.2\pm 167.7$\\
				-1.65 & 16 & 35823 & $869.1\pm 217.3$ & $758.3\pm 203.0$\\
				-1.45 & 9 & 35832 & $774.8\pm 258.3$ & $600.9\pm 227.5$\\
				-1.25 & 4 & 35836 & $545.8\pm 272.9$ & $1088.5\pm 385.4$\\
				-1.05 & 6 & 35842 & $1297.5\pm 529.7$ & $215.6\pm 215.9$\\
				-0.85 & 2 & 35844 & $685.4\pm 484.7$ & $1025.3\pm 592.8$\\
				-0.65 & 2 & 35846 & $1086.4\pm 768.2$ &  $541.7\pm 542.4$\\
				-0.45 & 0 & 35846 & $0.0\pm 0.0$ & $0.0\pm 0.0$\\
				-0.25 & 1 & 35847 & $1395.3\pm 1395.3$ & $1360.6\pm 1362.5$\\
				\hline
				\hline
			\end{tabular}
		\end{adjustbox}
	\end{table*}
	
\begin{figure*}
		\centering
		\includegraphics[width=\linewidth]{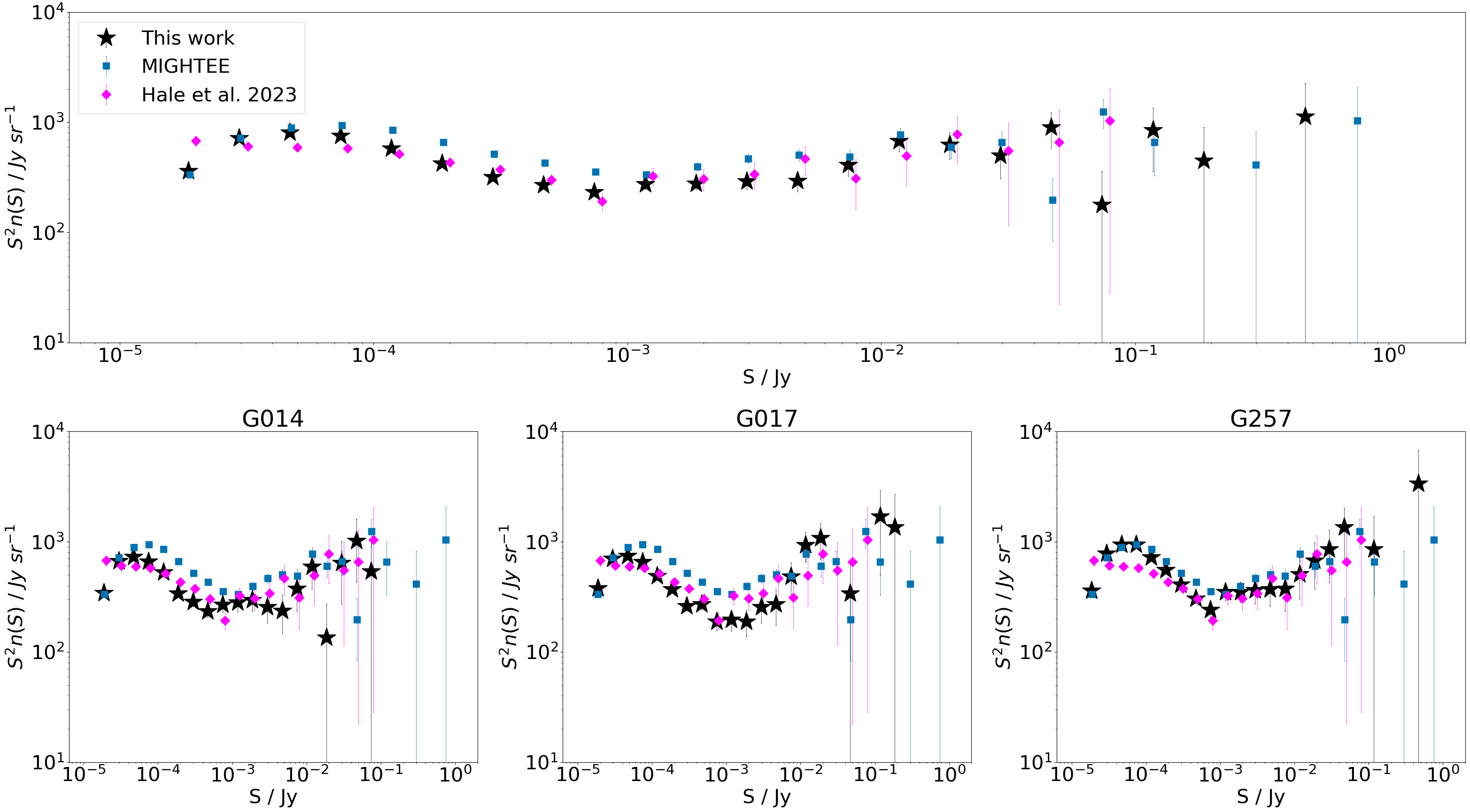}
		\caption{Brightness-weighted differential source counts as a function of flux. The black stars are our MeerKAT 1.28 GHz counts, the blue squares are number counts extracted from MIGHTEE (\citealt{heywood2022mightee}) COSMOS and XMM-LSS images following our extraction processes and the magenta diamonds are the MIGHTEE source counts in \citet{hale2023mightee}. The top panel shows our total counts by integrating counts from the 3 maps, while the bottom 3 panels show the source counts from individual maps, G014, G017 and G257 from left to right. Comparing our number counts to those in \citet{hale2023mightee}, we can see an excess at the flux of $\sim 0.1$~mJy in G257, suggesting the possibility of real overdensities associated with the candidate protoclusters. However, if we compare our number counts to the counts extracted from MIGHTEE data using the same extraction procedures as us, the excess at $S\approx0.1$~mJy now disappears, showing that it is probably a systematic effect caused by different data processing methods.}
		\label{fig.count}
	\end{figure*}

\begin{figure}
		\includegraphics[width=\columnwidth]{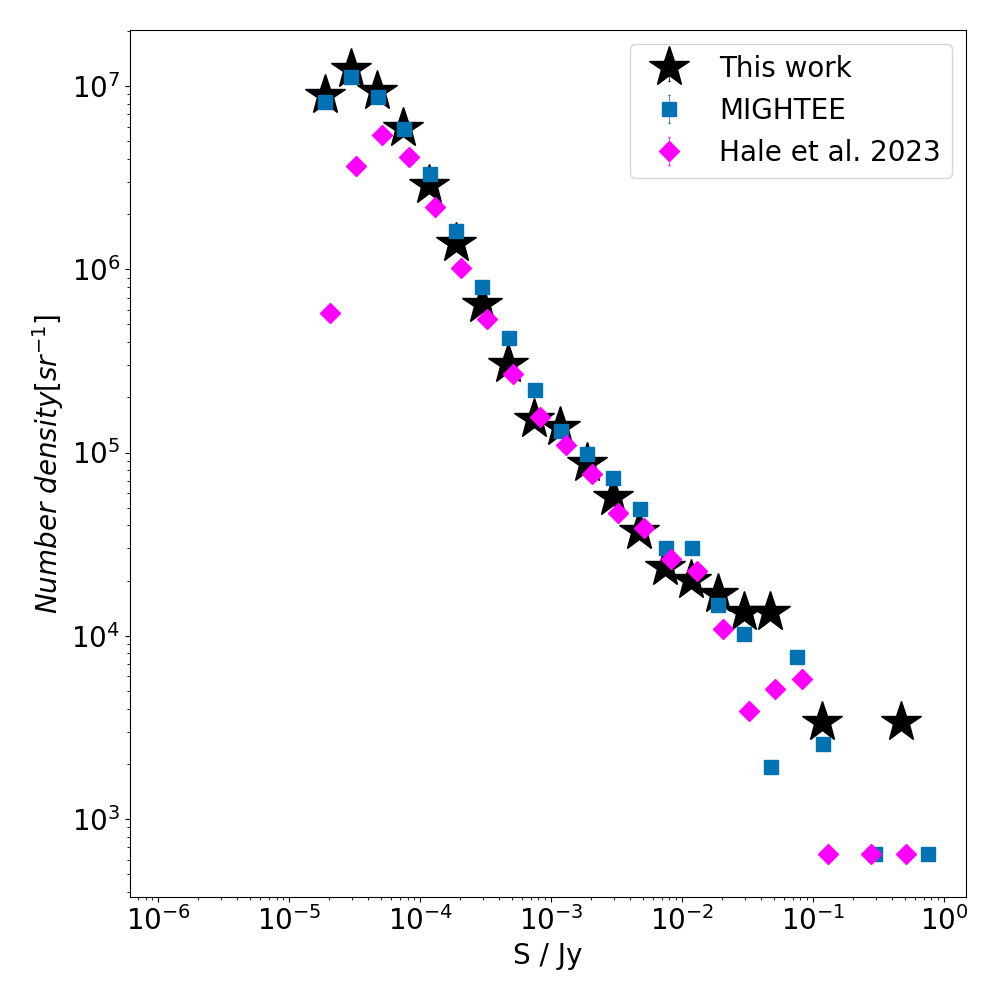}
		\caption{Number of sources per steradian as a function of flux. The black stars are our MeerKAT 1.28 GHz counts in G257, the blue squares are number counts extracted from MIGHTEE (\citealt{heywood2022mightee}) COSMOS and XMM-LSS images following our extraction processes and the magenta diamonds are the MIGHTEE source counts in \citet{hale2023mightee}. The error bars are too small to be seen. The excess at $S \approx 0.1$~mJy seen in Figure \ref{fig.count} is no longer visible in this case, proving it just being a result of the differences in data processing.}
		\label{fig.count.nd}
	\end{figure}

	To choose data for comparison, it is important to select data as similar to ours as possible. This makes MIGHTEE data (\citealt{heywood2022mightee}) a very good choice, as MIGHTEE uses the same instrument, MeerKAT, to survey the COSMOS and XMM-LSS fields, and its data are reduced in a similar manner using the same software, CASA (\citealt{mcmullin2007asp}). The official source counts of MIGHTEE are presented in \citet{hale2023mightee}. We also download the MIGHTEE robust $0.0$ science images in both COSMOS and XMM-LSS fields, then use SExtractor to extract sources from the two images following the same extraction steps described in Section \ref{sec.extra.1}. We combine the number counts in both COSMOS and XMM-LSS fields, then compare these number counts to our source counts. Figure \ref{fig.count} shows our 1.28 GHz counts together with other 1.4GHz radio source counts from \citet{hale2023mightee}, and the number counts extracted from MIGHTEE (\citealt{2016mks..confE...6J,heywood2022mightee}) COSMOS and XMM-LSS images following the same extraction processes used in this work. It can be seen that compared to the number counts in \citet{hale2023mightee}, the 1.28 GHz source counts of this work have an excess at $S\approx0.1$~mJy in G257. At this flux both the completeness and reliability of SExtractor are $> 95\%$, therefore this excess of source counts is not likely caused by over correction. In addition, the PyBDSF number counts also show the same excess at the same flux of $\sim 0.1$~mJy. Since our images are centred at 3 DSFG protocluster candidates, this excess in number counts suggest the possibility of a real overdensity in G257. However, comparing to the re-extracted MIGHTEE counts, this excess is no longer seen. We additionally compare the raw number counts per steradian in G257 to the raw counts of \citet{hale2023mightee} and MIGHTEE re-extraction, as the raw number counts are less sensitive to uncertainties. Figure \ref{fig.count.nd} shows the results. It can be seen that in this case the excess at $S \approx 0.1$~mJy seen in our G257 brightness-weighted differential number counts compared to that in \citet{hale2023mightee} is hardly visible. This means that this excess is probably a result of the differences in data processing such as extraction completeness or extraction parameters used, instead of being evidence of real overdensity.\par 
		
\begin{figure*}
		\centering
		\includegraphics[width=\linewidth]{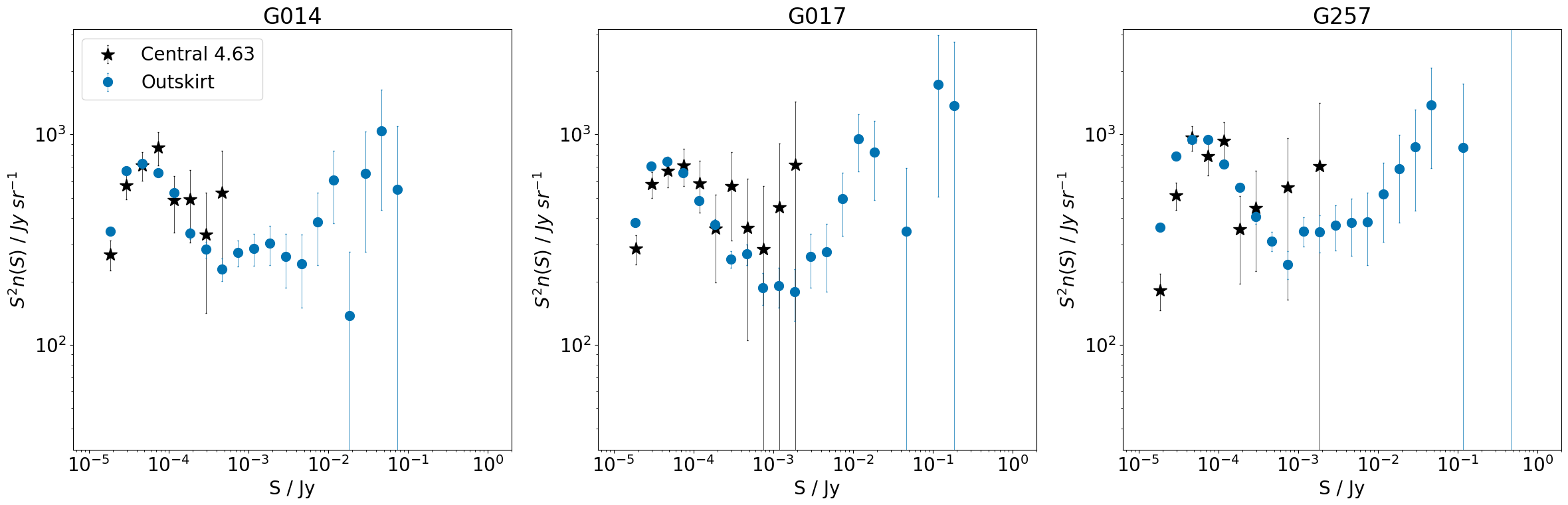}
		\caption{Brightness-weighted differential 1.28 GHz source counts as a function of flux.The black stars are source counts in the central $4.63'$ region, while the blue circles are counts outside the central region.}
		\label{fig.count.local}
	\end{figure*}

	We then compare the source counts in the central $4.63'$ \textit{Planck} beam, where the candidate protoclusters are located, to the source counts outside the \textit{Planck} beam which gives the local field counts. Figure \ref{fig.count.local} shows the resulting counts. There is no excess seen at the faint flux end. At higher fluxes the number count errors become very large due to the very small number of bright sources in the field centres. By calculating the density contrast as follows:
	\begin{equation}\label{eq.1}
		\delta  = \frac{n_{\text{centre}}}{n_{\text{field}}}-1
	\end{equation}
	where $n_{\text{centre}}$ is the number density in the centre and $n_{\text{field}}$ is the number density in the entire field, we have $\delta = -0.08\pm 0.07$, $\delta = -0.14\pm 0.06$ and $\delta = -0.25\pm 0.05$ for G014, G017 and G257 respectively. This means that there is no radio source overdensity in the central regions compared to other areas.
	
	In summary, the source counts of the MeerKAT images presented in this work are consistent with other radio counts at similar frequencies. There is no direct sign of overdensity in our source counts. 

	\section{Results \& Discussion}\label{sec.result}
	\subsection{\textit{Herschel} Identifications}\label{sec.id}
	Due to the finite size of the telescope, \textit{Herschel} has large beam sizes ($\geq 18''$). It is therefore hard to obtain accurate positions of \textit{Herschel} sources without the help of high-resolution observations at other wavelengths. However, FIR-bright sources are generally heavily dust-obscured objects which are difficult to observe with high-resolution optical/UV instruments as dust strongly absorbs at these wavelengths. In this scenario, high-resolution radio observations become an appropriate solution, as: (1) dust is transparent at radio wavelength; (2) radio interferometers such as MeerKAT usually have good angular resolution; (3) there is a known correlation between star-forming galaxies' FIR emission and their radio emission (e.g. \citealt{de1985spiral}, see Section \ref{sec.id.2} for more discussion), which will help rule out random radio sources overlapping with \textit{Herschel} sources by chance. With more accurate positions of \textit{Herschel} sources determined by radio observations, it is much easier to find their cross-identifications in other wavelengths and hence further multi-wavelength studies.\par

\begin{figure}
		\centering
		\includegraphics[width=\columnwidth]{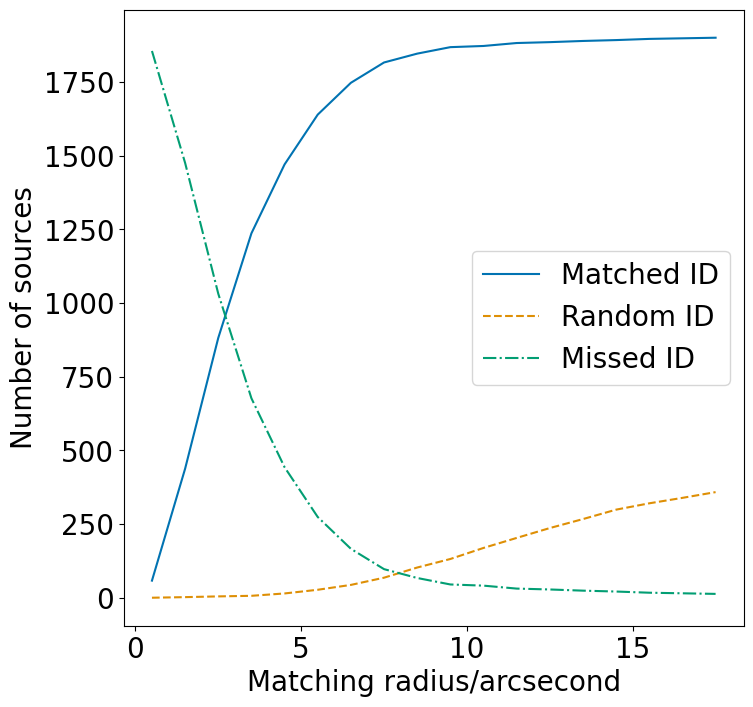}
		
		\caption{Number of \textit{Herschel} sources with real MeerKAT counterparts (blue solid line) / missed MeerKAT counterparts (green dashed-dotted line) / random identification (yellow dashed line) as a function of matching radius. At $\sim 7.5''$ the gradient of the increase in number of matched sources significantly reduces while the number of random IDs starts to rise, this means that using matching radii smaller than $7.5''$ will miss a large part of true cross-identifications, while using matching radii larger than $7.5''$ will include more background sources aligned by chance. Using a matching radius of $7.5''$ will maximise the number of true IDs while minimising the number of missed IDs and random IDs.}
		\label{fig.radius}
	\end{figure}

	One of the most important factors in the identification is the matching radius applied. A matching radius too small will result in missing many true counterparts, while a radius too large will include irrelevant background sources. In both cases the reliability of the cross-identification will be reduced. To determine the best matching radius, we investigate the number of \textit{Herschel} sources which have 1.28 GHz counterparts as a function of matching radius. The results are illustrated in Figure \ref{fig.radius}. Also plotted in Figure \ref{fig.radius} is the number of \textit{Herschel} sources with no counterpart. Both numbers have a sharp gradient at matching radius $\leq \sim 7.5''$ that levels out at a much steadier rate. This indicates that using matching radii smaller than $7.5''$ will miss a large fraction of true cross-identifications. This is further supported by a sky simulation reproducing the 1.28 GHz number counts. The simulated catalogue has the same flux density distribution as the real catalogue, but the source positions are completely random (see Section \ref{sec.dis} for a similar simulation with more details). The number of \textit{Herschel} sources matched to the simulated sources as a function of searching radius is also shown in Figure \ref{fig.radius}. This shows that the number of random identifications starts to rise rapidly beyond searching radius $\sim 7''$. Therefore we select $7.5''$ as our final matching radius and carry out cross-identification.  \par 
	There are 1914 \textit{Herschel} sources in total which are inside our MeerKAT observed fields, 1817 of them have 1.28 GHz counterparts, resulting in a matching percentage of $\sim 94.9\%$. The number of matched sources in G014, G017 and G257 is 419, 599 and 799 respectively. Figure \ref{fig.cutout}c shows an example of a cross-matched source. It should be noted that only $18.3\%$ of our matched \textit{Herschel} sources have $>3\sigma$ detections in all 3 SPIRE bands, with $99.2\%$, $60.8\%$ and $19.5\%$ of the \textit{Herschel} sources having $>3\sigma$ detections in 250, 350 and 500 $\mu$m band respectively. \par 
	
	Visual inspection of the \textit{Herschel} sources without MeerKAT counterparts reveals that some of them do overlap with 1.28 GHz objects, but their 1.28 GHz counterparts are either close to other bright sources, and hence ignored or blended with nearby bright sources by SExtractor, or they have extended morphologies and the separations between the central positions assigned by SExtractor and \textit{Herschel} are larger than the matching radius. Figure \ref{fig.cutout}d shows an example. The centre of this 1.28 GHz source with two jet-like structures clearly coincides with the \textit{Herschel} contour, but SExtractor does not recognise its centre as an individual source, probably because it is too close to one of the jets which are much brighter, hence being ignored or identified as parts of the bright jets.\par

	To further evaluate the robustness of our 1.28 GHz identifications of \textit{Herschel} sources, we estimate the Poisson probability of an identification being a chance alignment (e.g., \citealt{downes1986parkes}, \citealt{biggs2011laboca}). Assuming random distribution of 1.28 GHz sources, the expected number of sources within a radius $r$ is:
	\begin{equation}\label{eq.2}
		\mu = \pi r^{2} n
	\end{equation}
	where $n$ is the blank field number density of sources brighter than $S$. Then the Poisson probability of one random object within $r$ is:
	\begin{equation}\label{eq.3}
		P=1-e^{-\mu}
	\end{equation}
	Using the number density brighter than $S=0.021$~mJy from \citet{hale2023mightee} and our search radius of $7.5''$ for $r$, the probability that one 1.28 GHz source lines up with a \textit{Herschel} source by coincidence is $7.33\%$, in other words the probability of it being a good identification is $(1-P)=92.67\%$. Given that we have a matching percentage of $\sim 94.9\%$ with some genuine cross-IDs not counted due to their extended morphologies, this leads to the conclusion that $\gtrapprox 90\%$ of \textit{Herschel} sources will have one or more true 1.28 GHz counterparts at $S>0.021$~mJy.

	\subsubsection{Multiplicity of \textit{Herschel} Sources}\label{sec.id.1}
	Since the \textit{Herschel}-SPIRE beam sizes are much larger than the beam size of our MeerKAT images, there is a chance that a \textit{Herschel} source has multiple corresponding MeerKAT sources. It is known that most single dish DSFGs are blends of multiple submillimeter sources (e.g., \citealt{ivison2007scuba}, \citealt{wang2010sma}, \citealt{karim2013alma}, \citealt{stach2018alma}, \citealt{greenslade2020nature}). One of the most popular hypotheses to explain this multiplicity of DSFGs is that most DSFGs are ongoing/late-stage mergers (e.g. \citealt{hopkins2008cosmological}, \citealt{wardlow2018alma}), but  whether or not  the multiple components of any given DSFG are truly physically associated is unclear. CO observations of 12 submillimetre galaxies (SMGs) (\citealt{tacconi2008submillimeter}; \citealt{engel2010most}) suggests that all the galaxies in the sample are either ongoing mergers or compact dense galaxies which are considered as late stage mergers, meaning that SMGs are likely to be high-redshift analogues to local (U)LIRGs. To the contrary, models from \citet{hayward2013spatially} and \citet{cowley2015simulated} indicate that multiple sources resolved from single-dish detected SMGs are generally not associated, hence are not merger companions. In conclusion, for now it is only clear that most single-dish detected DSFGs (or more broadly speaking SMGs) are blends of multiple sources. Here, we utilise our 1.28 GHz data to investigate whether the multiplicity of DSFGs can still be seen at this wavelength, and if it can provide any insight into  the nature of DSFGs.\par
\begin{figure}
		\includegraphics[width=\linewidth]{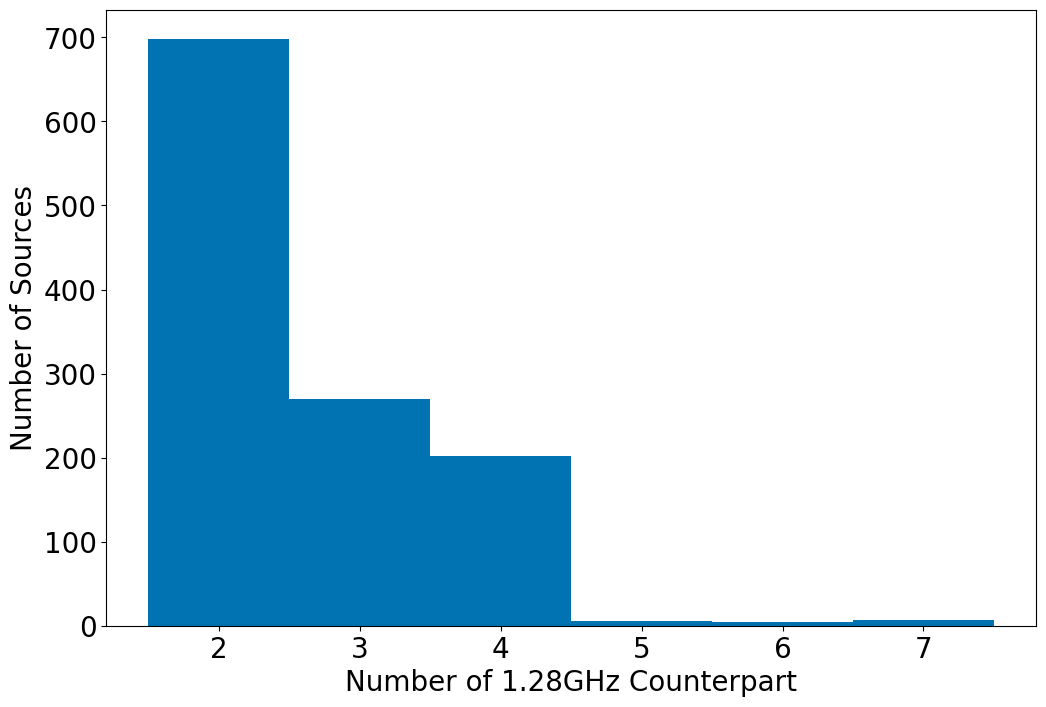}
		\caption{The distribution of the number of MeerKAT matched to \textit{Herschel} sources. Most \textit{Herschel} matched sources have $2-4$ MeerKAT counterparts, while the highest number of radio components of a single source is 7. Visual inspection suggests all \textit{Herschel} sources with more than 4 radio components are in fact several \textit{Herschel} sources lying close to each other. The true number of MeerKAT counterparts should be 1-4, see Figure \ref{fig.cutout}g}
		\label{fig.counterpart}
	\end{figure}

 
	We re-match our MeerKAT sources to the H-ATLAS DR2 catalogue, using a matching radius of $17.6''$ which is the primary beam size of the \textit{Herschel} SPIRE 250 $\mu$m channel. This matching radius is used to search for all potential MeerKAT counterparts of \textit{Herschel} sources, especially for unresolved sources. However, using a matching radius as large as $17.6''$ will boost the risk of random 1.28 GHz sources being mis-identified to $34.2\%$. We will call this new catalogue the multiplicity catalogue for convenience in the following sections. Figure \ref{fig.cutout}e\&f illustrates some \textit{Herschel} sources matched with multiple MeerKAT counterparts. We find that 1885 \textit{Herschel} sources have MeerKAT counterparts at this matching radius and 1038 out of the 1885 \textit{Herschel} sources have 2 or more radio counterparts. This results in $54.2\%$ of \textit{Herschel} sources with multiple 1.28 GHz IDs. The majority of the 1038 \textit{Herschel} matched sources have $2-4$ components with 698, 270 and 59 of them having two, three and four components respectively, while the maximum number of counterparts of a single \textit{Herschel} source is 7, as shown by Figure \ref{fig.counterpart}. Visual examination of \textit{Herschel} sources with more than 4 MeerKAT counterparts reveals that all of them are actually two or more close \textit{Herschel} sources acting as one large object during cross-matching, leading to their apparent multiplicities larger than 4. The actual number of counterparts of these sources will be 1-4. An example of this is shown in Figure \ref{fig.cutout}g. 
	
	To further investigate the nature of \textit{Herschel} sources with 1.28 GHz multiplicity, we carry out two-sample Kolmogorov–Smirnov tests (hereafter, K-S tests) to examine whether or not the flux density distribution of \textit{Herschel} sources with multiple 1.28 GHz components is intrinsically different to sources without such multiplicity. The two-sample K-S test is used to test whether two underlying one-dimensional probability distributions differ. If there is a true difference between the flux density distributions of the two populations, it implies that multiple 1.28 GHz components of a \textit{Herschel} sources are likely to be associated instead of being mere chance-alignments. For this analysis, \textit{Herschel} detections below $3\sigma$ are removed for each SPIRE band.\par
\begin{figure*}
		\centering
		\includegraphics[width=\linewidth]{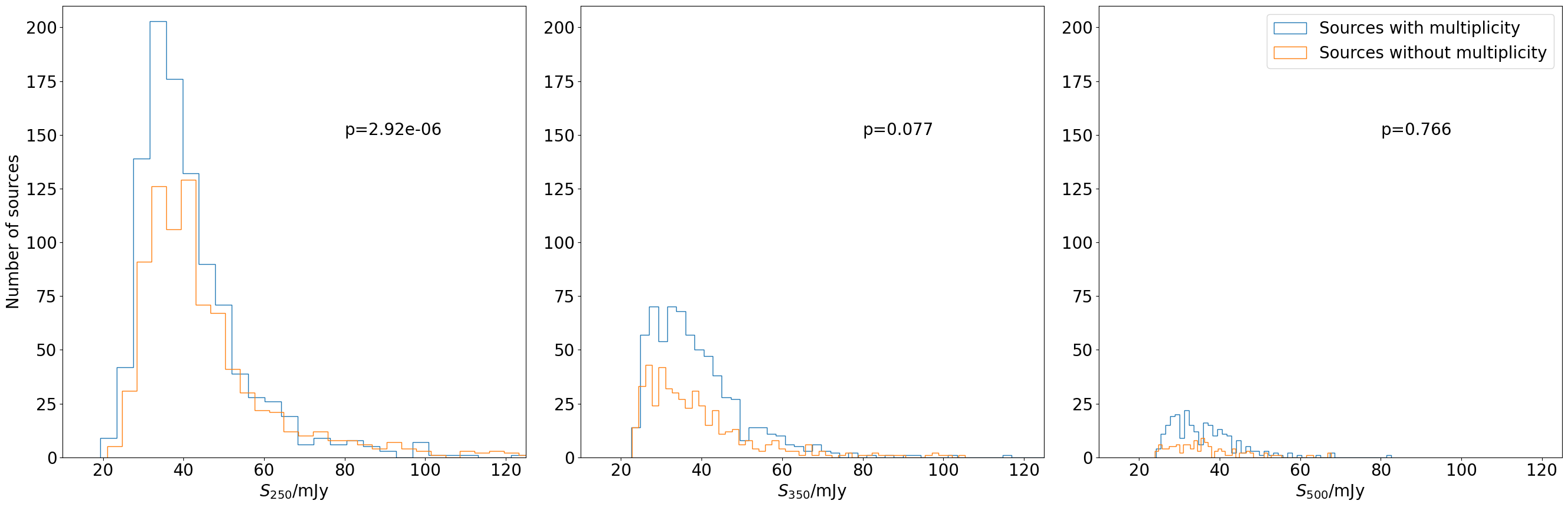}
		\caption{Flux density distributions of \textit{Herschel} sources with/without 1.28 GHz MeerKAT counterparts. From left to right are the distributions in $250$, $350$ and $500~\mu$m. The resulting p-values of the two-sample K-S tests are shown. For a confidence level above 95\%, only the $250~\mu$m distributions of the two populations are statistically different.}
		\label{fig.ks}
	\end{figure*}

 
	Figure \ref{fig.ks} illustrates the flux density distributions of \textit{Herschel} sources with/without 1.28 GHz MeerKAT multiplicity, with the associated p-values of the K-S tests shown in the corresponding panels. Statistically, we can be over 99.9\% confident that \textit{Herschel} sources with/without MeerKAT multiplicity have different flux distributions at $250~\mu$m, but in the other two SPIRE bands the confidence level that the two populations have different flux distributions drops to $\sim92.3\%$ and $\sim33.4\%$. In other words, the two-sample K-S test results of our sample suggest that the difference between the flux density distributions of \textit{Herschel} sources with and without multiple 1.28 GHz counterparts is only significant at $250~\mu$m. The average $250~\mu$m fluxes and standard deviations of \textit{Herschel} sources with/without MeerKAT multiplicity are $42.2\pm 16.1$ mJy and $47.5\pm 23.1$ mJy respectively. One possible explanation of why the $250~\mu$m distributions of the two populations are different is that there are too few \textit{Herschel} sources without multiplicity detected at $350$ and $500~\mu$m, which can also be seen from Figure \ref{fig.ks}. Since the $250~\mu$m band has the best sensitivity and resolution of all 3 SPIRE bands while the $500~\mu$m band has the worst, there will be more faint sources without multiple components detected in the $250~\mu$m band compared to the other two bands. While in the other two bands, bright sources which are more likely to be blends of multiple fainter sources will be dominant and faint sources with no multiplicity will be too few to reflect their genuine flux distribution. This also explains the increasing trend of p-value from 250 to $500~\mu$m, as the sensitivity of the band decreases. Another possible explanation is that, since we are using a larger matching radius here, hence higher percentage of chance alignments, and there are far more faint $250~\mu$m sources than bright ones, there will be more chances for MeerKAT sources to randomly match a faint rather than bright SPIRE sources, leading to a fainter $250~\mu$m flux distribution of sources with multiple MeerKAT IDs. \par 
	Overall, our results indicate that \textit{Herschel} sources with multiple 1.28 GHz MeerKAT components have a different $250~\mu$m flux density distribution to \textit{Herschel} sources with just one MeerKAT counterpart, while at $350$ and $500~\mu$m there are insufficient \textit{Herschel} sources to make a statistically significant comparison. This provides further evidence that the multiple 1.28 GHz counterparts of a \textit{Herschel} source are associated instead of being a chance-alignment. However, this could also just be a bias due to lack of data or inaccurate determinations of multiple radio counterparts to \textit{Herschel} sources at large matching radius.  We need wider area radio surveys to cover more $350$ and $500~\mu$m sources and other techniques to achieve better cross-IDs at larger searching radius to draw further conclusions.

	\subsubsection{The Far-IR Radio Correlation}\label{sec.id.2}
	
	A surprisingly tight correlation between galaxies' FIR emission and their radio emission was discovered in the late 1970s, first proposed by \citet{van1971observations}. The FIR radiation is believed to come from thermal re-radiation of dust-enshrouded regions while the radio radiation is thought to be synchrotron emission from cosmic ray electrons accelerated in supernova remnants. This naturally explains the observed correlation as the massive stars that eventually end as supernovae are those stars which substantially heat up their surrounding gas and dust (\citealt{harwit1975infrared}). Numerous studies have been carried out on the Far-Infrared Radio Correlation (FIRC; e.g., \citealt{helou1985thermal}; \citealt{de1985spiral}; \citealt{yun2002radio}; \citealt{jarvis2010herschel}; \citealt{molnar2021non}). \par
	The FIRC has many useful applications. It can be used to estimate $L_{\text{IR}}$ with radio data when no IR data are available, or it can be used to complement or constrain the FIR SEDs when FIR data are inadequate (e.g., \citealt{yun2002radio}). Likewise, the FIRC can also be used to study the relation between obscured star formation and galaxies' radio emission, as $\text{SFR}\propto L_{\text{IR}}\propto L_{\text{1.4GHz}}$. \par 
	
\begin{figure}
		\includegraphics[width=\columnwidth]{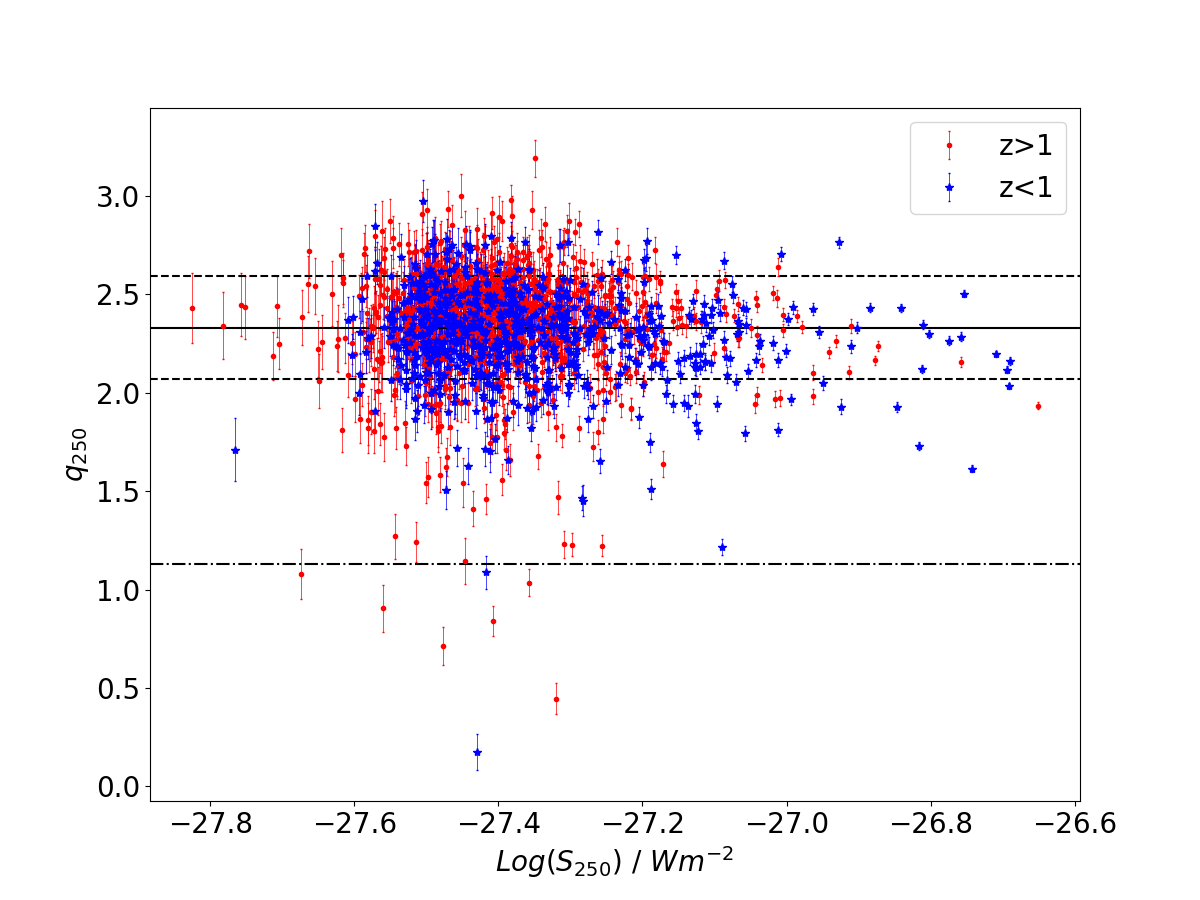}
		\centering
		\caption{$q_{250\mu \text{m}}$ against the 250 $\mu$m flux density for all sources in the three maps. The red dots are sources with {\sc EAZY} photometric redshift larger than one or with no photometric redshift assigned, the blue stars are sources with $z_{\text{EAZY}} < 1$. The solid line is the mean value of $q_{250\mu \text{m}}$ with dashed lines being its $1\sigma$ standard deviation. The dotted-dash line is the lower $5\sigma$ clipping boundary. The mean $q_{250\mu \text{m}}$ value is $2.33\pm 0.26$ for all sources, it is comparable to the results in \citet{ivison2010blast} which is $\sim 2.26\pm 0.35$ with relatively small dispersion.}
		\label{fig.firc}
	\end{figure}

	Following \citet{ivison2010blast}, we calculate the monochromatic $q_{250\mu \text{m}}$ to investigate the FIRC as:
	\begin{equation}\label{eq.4}
		q_{250\mu \text{m}}=\log_{10}\frac{S_{250\mu \text{m}}}{S_{\text{1.28 GHz}}}
	\end{equation}
	where $S_{250\mu \text{m}}$ is the $250~\mu$m flux density of a source and $S_{\text{1.28 GHz}}$ is its 1.28 GHz flux density. Using the \textit{Herschel}-MeerKAT matched catalogue with $7.5''$ matching radius, the $q_{250\mu \text{m}}$ as a function of 250 $\mu$m flux is shown in Figure \ref{fig.firc}. It should be noted that the $q_{250\mu \text{m}}$ calculated here is derived from observed frame quantities due to the lack of accurate redshift information for most sources. Therefore our results are expected to have a redshift dependency which will not be seen in samples with K-corrections applied (e.g., \citealt{ivison2010blast}). The mean value of our $q_{250\mu \text{m}}$ is $2.33\pm 0.26$, which is consistent with the mean value of $2.26\pm 0.35$ in \citet{ivison2010blast}. This further indicates that our \textit{Herschel}-MeerKAT cross-identifications are of good reliability. Moreover, using the {\sc EAZY} photometric redshifts estimated in Section \ref{sec.optical.z}, we divide our sample into two subsamples. One of the two subsamples includes sources with {\sc EAZY} photometric redshifts larger than one or with no photometric redshift assigned, and the other includes sources with {\sc EAZY} redshift lower than one. The mean $q_{250\mu \text{m}}$ are $2.34\pm 0.28$ and $2.29\pm 0.22$ for $z_{\text{EAZY}} > 1$ and $z_{\text{EAZY}} < 1$ sources respectively. There is no significant difference between the two subsamples. We additionally compare the $q_{250\mu \text{m}}$ of all sources within the $4.63'$ radius circles in the field centres to the other sources, as they are potential members of the candidate protoclusters. The average $q_{250\mu \text{m}}$ is $2.31\pm 0.21$ which is consistent with the overall value. It should be noted that the source with the lowest $q_{250\mu \text{m}}$ is in the centre of G017, it is a very bright radio source with some beam artefacts around it, see Figure \ref{fig.cutout}h.\par 
	
\begin{figure}
		\centering
		\includegraphics[width=\columnwidth]{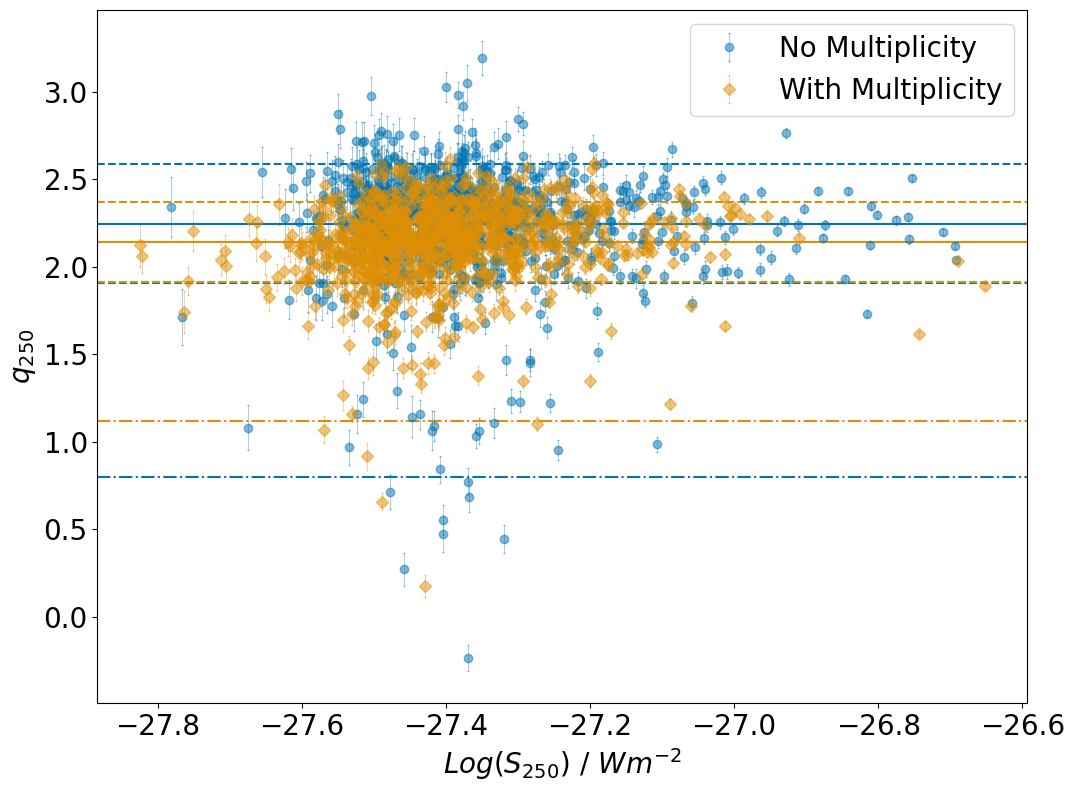}
		\caption{$q_{250\mu \text{m}}$ against the 250 $\mu$m flux density for \textit{Herschel} sources with only one 1.28 GHz counterparts (blue dots) and \textit{Herschel} sources with multiple 1.28 GHz counterparts (orange diamonds). The solid lines are the mean values of $q_{250\mu \text{m}}$ of the two populations, with dashed lines being the corresponding $1\sigma$ standard deviations. The dotted-dash lines are the respective lower $5\sigma$ clipping boundaries.The average $q_{250\mu \text{m}}$ values for the two populations are $2.14\pm 0.22$ and $2.24\pm 0.34$ for \textit{Herschel} sources with and without multiplicity respectively. }
		\label{fig.firc.multi}
	\end{figure}

	We also compare the FIR-radio correlation between \textit{Herschel} sources with and without 1.28 GHz multiplicity utilising our multiplicity catalogue. Since we cannot determine the FIR flux contributions of each radio component of a given \textit{Herschel} source, the $q_{250\mu \text{m}}$ of a \textit{Herschel} source with multiple counterparts is calculated by integrating the radio flux densities of all its 1.28 GHz components. The results are shown in Figure \ref{fig.firc.multi}. The average $q_{250\mu \text{m}}$ values are $2.14\pm 0.22$ and $2.24\pm 0.34$ for \textit{Herschel} sources with and without multiplicity respectively. Again, we perform a K-S test to examine whether there is a statistically significant difference between the two populations. The p-value of this K-S test is $2.86\times 10^{-34}$ which means it is very likely that the $q_{250\mu \text{m}}$ distributions of the two populations are different. \textit{Herschel} sources with 1.28 GHz multiplicity show a lower $q_{250\mu \text{m}}$ on average with smaller scatter. A possible cause of this difference in $q_{250\mu \text{m}}$ is that we derive the $q_{250\mu \text{m}}$ of \textit{Herschel} sources with multiple radio components by adding up all the radio counterparts' flux densities, which may lead to higher 1.28 GHz flux densities and hence lower $q_{250\mu \text{m}}$. As already mentioned in Section \ref{sec.id.1}, this could also be a consequence of the fact that the probability of random alignments is higher in the multiplicity catalogue and there are more faint $250~\mu$m sources than bright ones. In this case if a considerable fraction of matched sources are just chance alignments and these radio sources are not significant FIR emitters, it would explain the lower $q_{250\mu \text{m}}$ of \textit{Herschel} sources with multiple counterparts seen here. Whether this difference is real or just a bias can only be investigated by further studies with accurate determinations of each 1.28 GHz components' contributions to the \textit{Herschel} flux densities.\par 
	
	Sources with low $q_{250\mu \text{m}}$ have much larger radio fluxes than expected from their FIR fluxes. In other words they are radio-loud and radio-loudness is a typical feature of AGN activity. We flag sources with very low $q_{250\mu \text{m}}$ by performing sigma-clipping to the $q_{250\mu \text{m}}$ in all catalogues. Sources with $q_{250\mu \text{m}}$ $5\sigma$ smaller than the mean are excluded, then the average $q_{250\mu \text{m}}$ and the standard deviation are recalculated. We iterate this process to the point where no further source is excluded. The resulting $q_{250\mu \text{m}}$ for the $7.5''$ catalogue is $2.33\pm 0.24$, with 8 sources being flagged out. For the multiplicity catalogue, the updated $q_{250\mu \text{m}}$ are $2.15\pm 0.21$ and $2.26\pm 0.29$ for \textit{Herschel} sources with and without multiplicity with 5 and 8 sources being flagged out respectively. After the sigma-clipping, the average values of $q_{250\mu \text{m}}$ for all catalogues do not change significantly, while the $1\sigma$ standard deviations become smaller in all cases as an expected consequence of sigma-clipping. In both Figure \ref{fig.firc} and Figure \ref{fig.firc.multi} we can see there are sources lying outside the lower 5$\sigma$ boundaries. Visual examination of these sources finds that some of them are sources with interesting morphologies such as head-tail sources or jets which are more likely to be true identifications with potential AGN activity. Figure \ref{fig.cutout}i shows an example with extended radio structure coinciding with a \textit{Herschel} source, suggestive of a possible AGN with radio jets heavily enshrouded by dust. This demonstrates that $q_{250\mu \text{m}}$ could be an effective selection criterion for radio-loud AGN searches in future studies. Additionally, an interesting observation is that in Figure \ref{fig.firc}, the majority of radio-loud sources have photometric redshift larger than one or have no optical/NIR IDs to provide the photometry needed for redshift estimations. This further suggests that these sources are heavily dust-enshrouded AGN as they are faint in the optical/NIR.

	\subsubsection{Herschel Colours}\label{sec.id.3}
	
	\begin{figure*}
		\centering
		\subfloat[]{\includegraphics[width=0.5\linewidth]{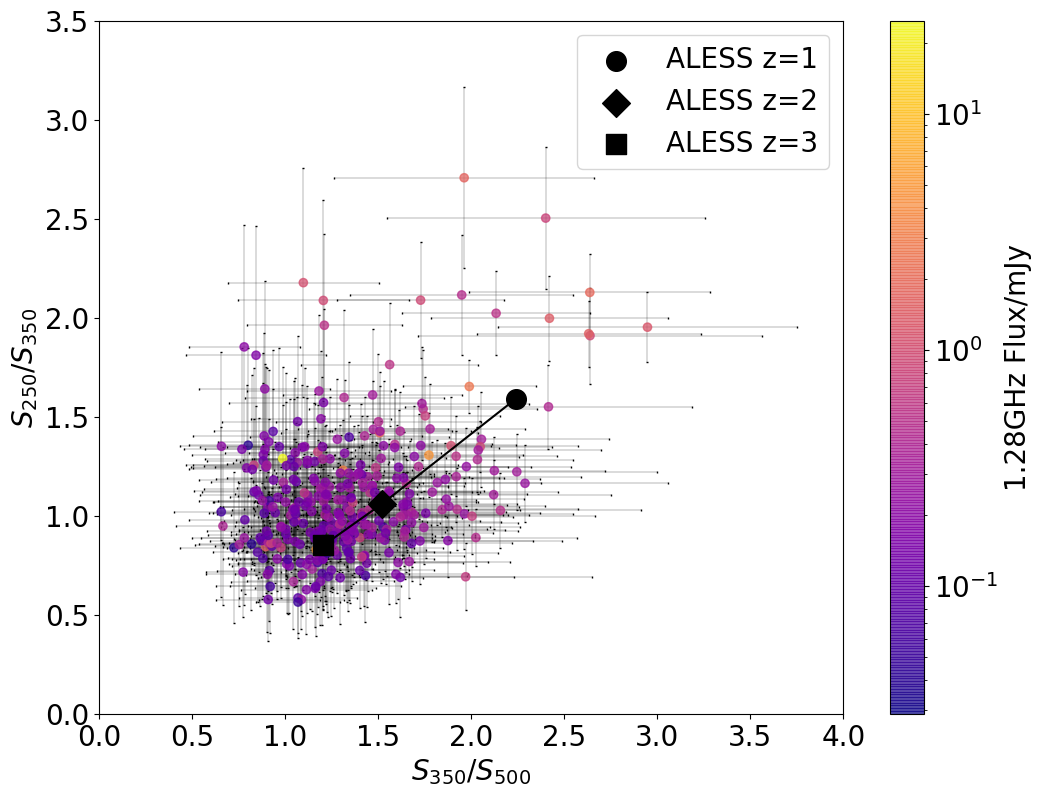}}
		\subfloat[]{\includegraphics[width=0.5\linewidth]{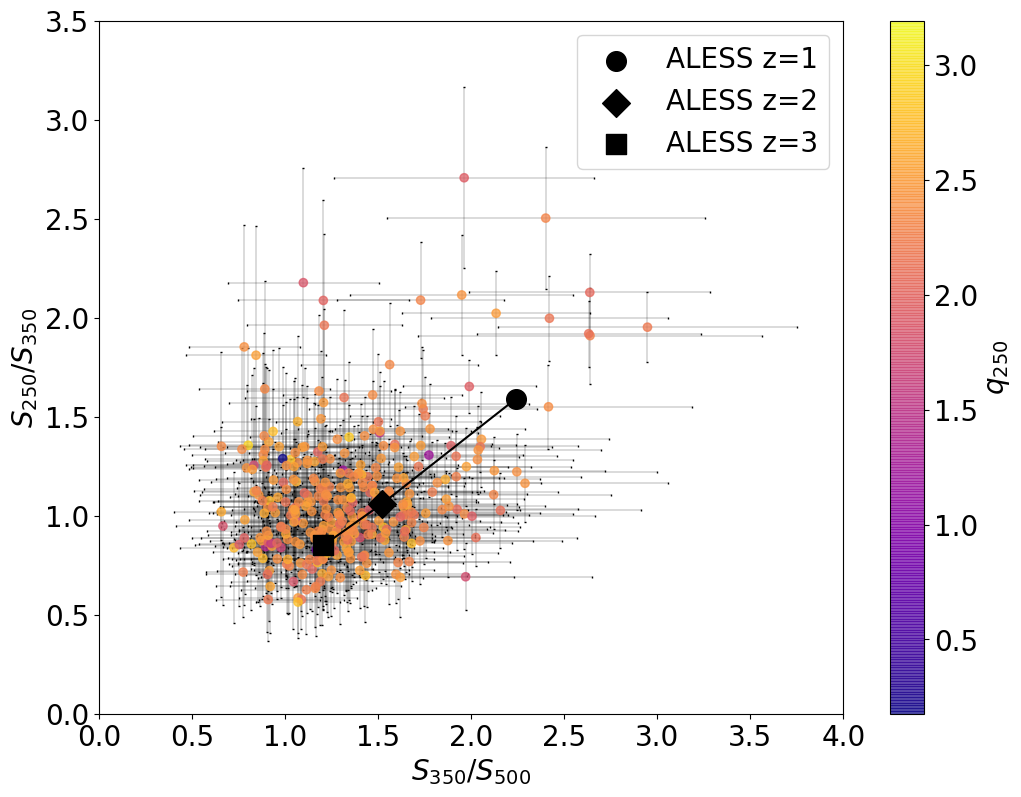}}
		
		\caption{\textit{Herschel} $S_{250}/S_{350}$ and $S_{350}/S_{500}$ colours of our MeerKAT sources. The \textit{Herschel} colours of the ALESS template SED (\citealt{da2015alma}) at redshift of 1,2 and 3 are also shown as black dot, diamond and square respectively. Most sources reside in regions indicated as redshift 2-3. (a) \textit{Herschel} colour-colour diagram, colour-coded by their 1.28 GHz MeerKAT fluxes. No obvious correlation between the 1.28 GHz flux and the \textit{Herschel} colours can be observed. (b) The same diagram, colour-coded by their FIRC factor $q_{250\mu \text{m}}$, there is also no apparent correlation between the $q_{250\mu \text{m}}$ and the \textit{Herschel} colours.}
		\label{fig.colour}
		
	\end{figure*}
	
	\begin{figure*}
		\centering
		\subfloat[]{\includegraphics[width=0.32\linewidth]{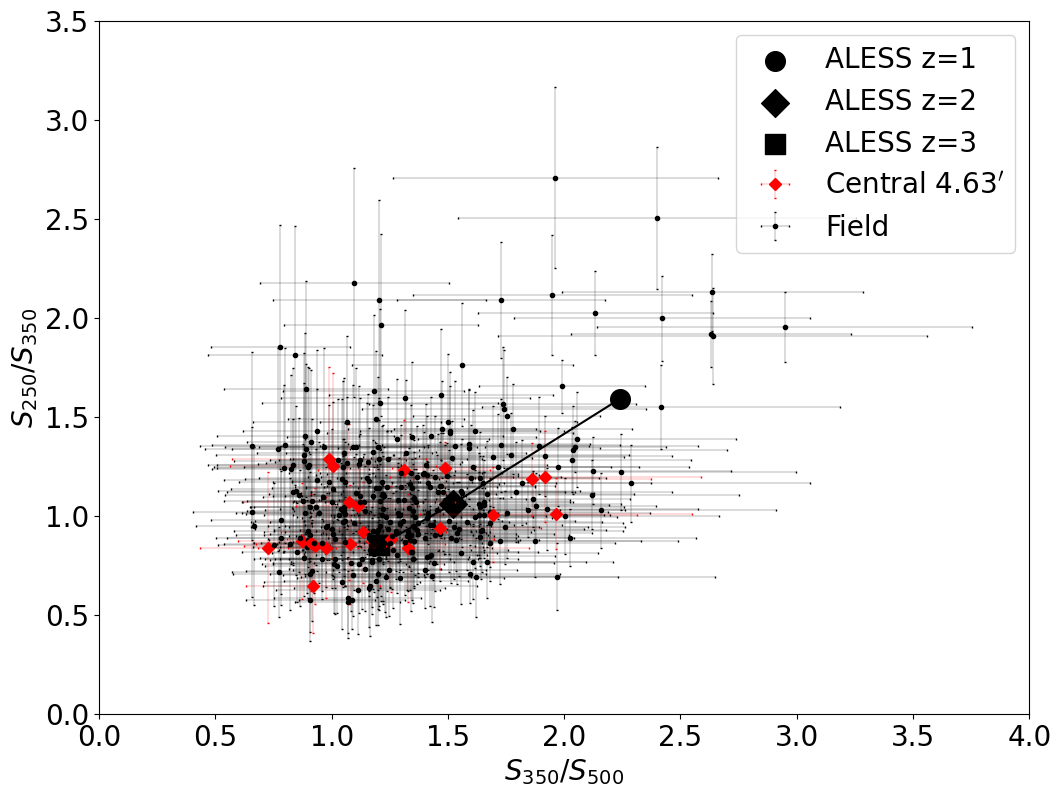}}
		\subfloat[]{\includegraphics[width=0.32\linewidth]{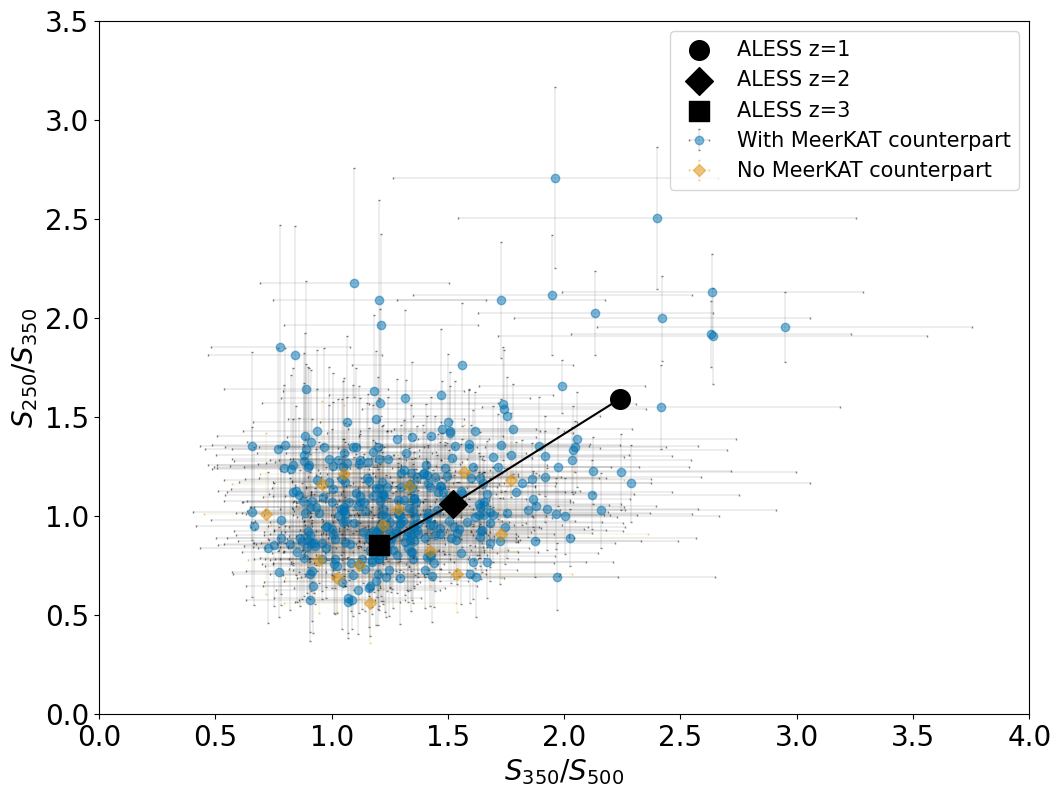}}
		\subfloat[]{\includegraphics[width=0.32\linewidth]{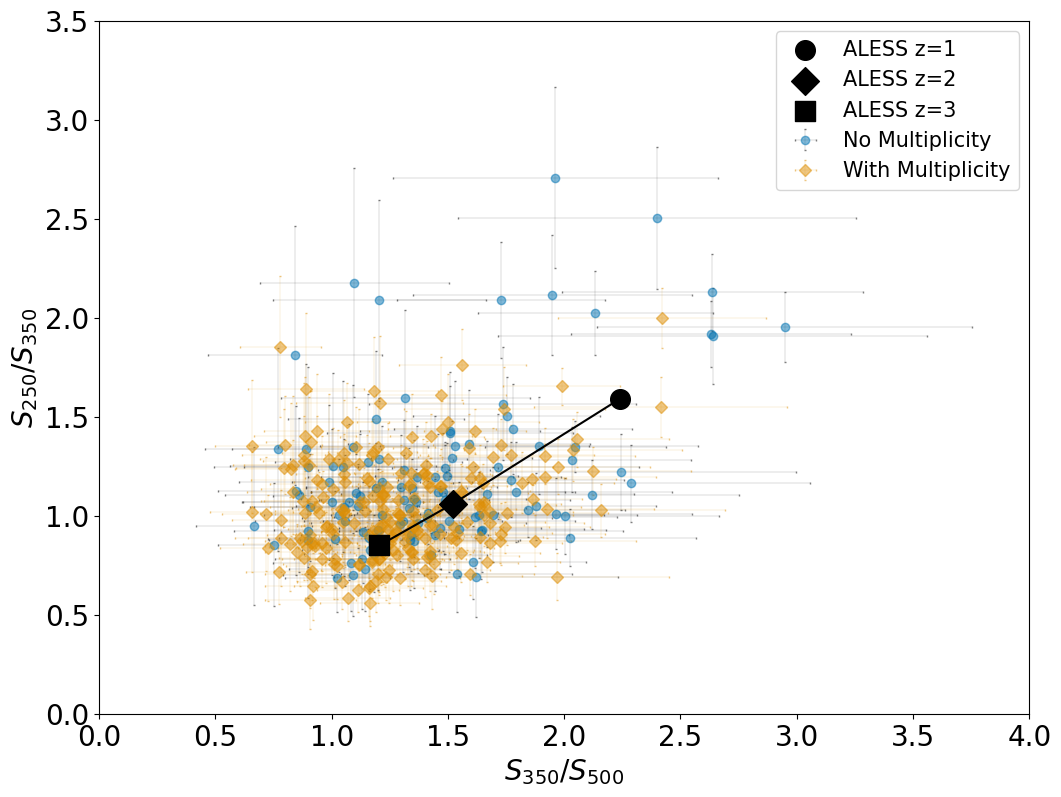}}
		\caption{Similar figure to Figure \ref{fig.colour}, but the sample is divided into (a) Sources within the central $4.63'$ \textit{Planck} beam, marked as red diamonds, and the other sources in the fields as black dots; (b) \textit{Herschel} sources with and without MeerKAT counterparts, blue dots and orange diamonds respectively; and (c) \textit{Herschel} sources with and without MeerKAT multiplicity, blue dots and orange diamonds respectively.}
		\label{fig.colour.centre}
	\end{figure*}

	We next examine the \textit{Herschel} SPIRE colours of our sources. These are commonly used as redshift indicators for submillimeter sources (e.g. \citealt{clements2014herschel}, \citealt{dowell2014hermes}, \citealt{ivison2016space}, \citealt{greenslade2018candidate}). It is widely accepted that sources whose SEDs peak at longer wavelengths tend to lie at higher redshifts (\citealt{casey2014dusty}). Therefore sources with SEDs peaking at $250$, $350$ and $500~\mu$m are likely to have progressively higher redshifts. We investigate the \textit{Herschel} SPIRE colours of our MeerKAT sources to obtain a rough estimation of their redshifts, and to see if their \textit{Herschel} colours have any correlation with their 1.28 GHz fluxes. Again,  \textit{Herschel} detections below $3\sigma$ in any of the 3 SPIRE bands are removed.\par 
	
	In Figure \ref{fig.colour}a we plot $S_{250}/S_{350}$ against $S_{350}/S_{500}$ for our sources, colour-coded by their corresponding 1.28 GHz MeerKAT fluxes. The ALESS SED for a template starburst galaxy (\citealt{da2015alma}) is used as a benchmark, and its \textit{Herschel} colours at various redshifts are displayed in the figure. As can be seen, most sources lie in regions suggesting redshifts between 2 and 3, while there is no obvious correlation between the 1.28 GHz flux and \textit{Herschel} colours. Figure \ref{fig.colour}b shows a similar diagram, but colour-coded with the monochromatic FIRC coefficient $q_{250\mu \text{m}}$. Again, there is no apparent correlation between the $q_{250\mu \text{m}}$ and the \textit{Herschel} colours. 
	
	In addition, we differentiate our sample into two sub-samples by (a) sources inside/outside the $4.63'$ \textit{Planck} beam at the field centres (b) whether the \textit{Herschel} sources have MeerKAT counterparts and (c) whether the \textit{Herschel} sources have multiple or single MeerKAT counterparts. The \textit{Herschel} colours of these sub-samples are shown in Figure \ref{fig.colour.centre}. Note that we use the multiplicity catalogue in Figure \ref{fig.colour.centre}c.
		
	The average SPIRE colours of the various subsamples in Figure \ref{fig.colour.centre} are as follows: (a) the mean $S_{250}/S_{350}$ and $S_{350}/S_{500}$ are $1.17\pm 0.09$ and $1.34\pm 0.18$ for sources in the centre, while the mean $S_{250}/S_{350}$ and $S_{350}/S_{500}$ are $1.26\pm 0.02$ and $1.37\pm 0.04$ for the other sources in the field. Most central sources locate at $z\approx 2-3$ region with a slightly lower $S_{250}/S_{350}$; (b) The mean $S_{250}/S_{350}$ and $S_{350}/S_{500}$ are $1.27\pm 0.01$ and $1.33\pm 0.02$ for sources with MeerKAT ID and $1.10\pm 0.05$ and $1.23\pm 0.10$ for sources with no MeerKAT ID. Both colours for sources with no MeerKAT ID are lower, the difference in $S_{250}/S_{350}$ is over $3\sigma$; (c) The mean $S_{250}/S_{350}$ and $S_{350}/S_{500}$ are $1.40\pm 0.02$ and $1.44\pm 0.04$ for sources with single MeerKAT ID and $1.17\pm 0.01$ and $1.28\pm 0.02$ for sources with multiple MeerKAT IDs. In this case both colours for sources with multiple MeerKAT IDs are significantly lower. \par 
	We further apply K-S tests to examine if there is a statistical difference between these sub-samples. The p-values of the K-S tests between \textit{Herschel} sources inside the $4.63'$ \textit{Planck} beam in field centres and the others in the fields are $0.238$ and $0.995$ for $S_{250}/S_{350}$ and $S_{350}/S_{500}$ respectively, meaning that there is likely no difference between these two sub-samples. Similarly for sources with and without MeerKAT ID, the p-values are $0.556$ and $0.999$ for $S_{250}/S_{350}$ and $S_{350}/S_{500}$ respectively, with no statistical difference between the SPIRE colours of the two population. Meanwhile, the p-values of the K-S tests between \textit{Herschel} sources with MeerKAT multiplicity and those without multiplicity are $4.27\times 10^{-14}$ and $9.67\times 10^{-5}$ for $S_{250}/S_{350}$ and $S_{350}/S_{500}$ respectively, indicating a difference between these two populations. However, this is expected since the \textit{Herschel} flux density distributions between these two populations are different as discussed in Section \ref{sec.id.1}. Therefore as the difference in \textit{Herschel} colour distributions is just a direct result of the different flux density distributions, no further conclusion can be drawn.\par 
	
	Despite there being no strong correlation between \textit{Herschel} colours and MeerKAT fluxes or $q_{250\mu \text{m}}$, the \textit{Herschel} colours suggest that most of our sources have redshift larger than 1 which is as expected from \citet{greenslade2018candidate}. Nevertheless, estimating the redshift of sources from their \textit{Herschel} colours can only provide rough indications with large uncertainties. Later in Section \ref{sec.optical.z}, we carry out photometric redshift estimations using accurate optical to IR cross-IDs, to further investigate the properties of our sample.\par

	\subsection{Using the FIRC for \textit{Herschel} cross-IDs}\label{sec.dis}
	
	\begin{figure}
		\centering
		\includegraphics[width=\columnwidth]{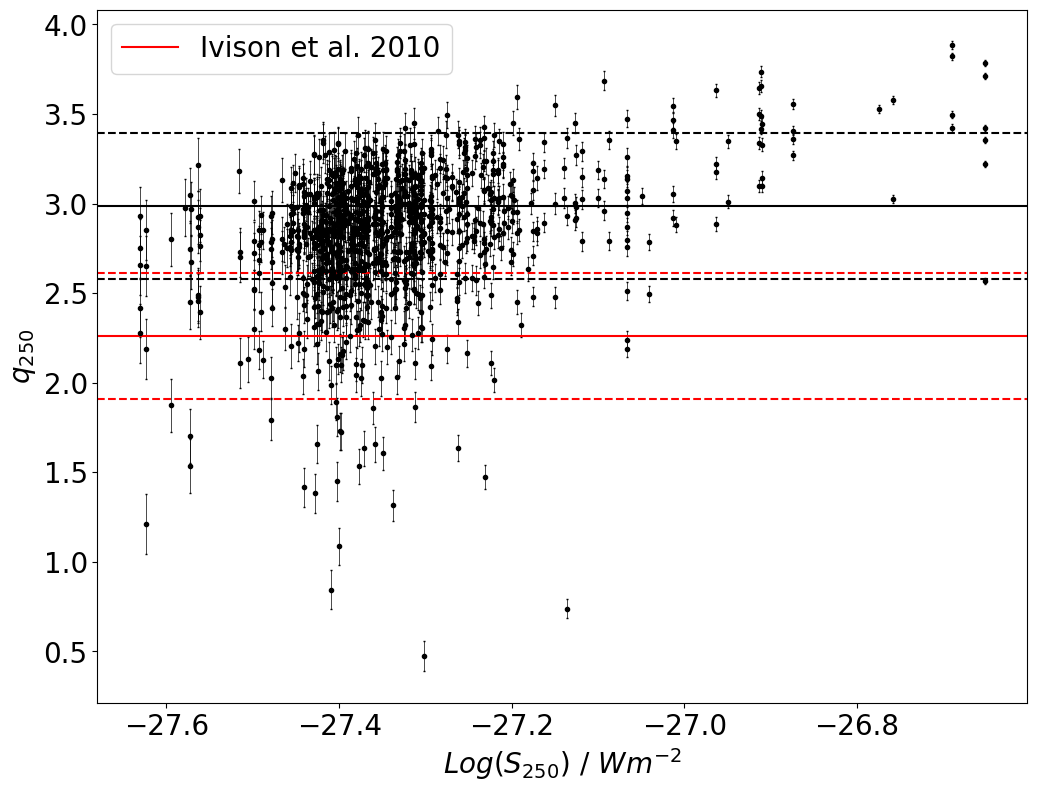}
		\caption{The $q_{250\mu \text{m}}$ against the $250~\mu$m flux density for \textit{Herschel} sources matched to simulated MeerKAT sources, with a matching radius of $7.5''$. The average $q_{250\mu \text{m}}$ of these sources is $2.99\pm 0.41$, the black solid line shows the mean $q_{250\mu \text{m}}$ of these sources and the dashed lines show their $1\sigma$ scatter. The mean and $1\sigma$ scatter of $q_{250\mu \text{m}}$ in \citet{ivison2010blast} are shown as red solid line and dashed lines. }
		\label{fig.firc.fake}
	\end{figure}
	
	\begin{figure}
		\centering
		\includegraphics[width=\columnwidth]{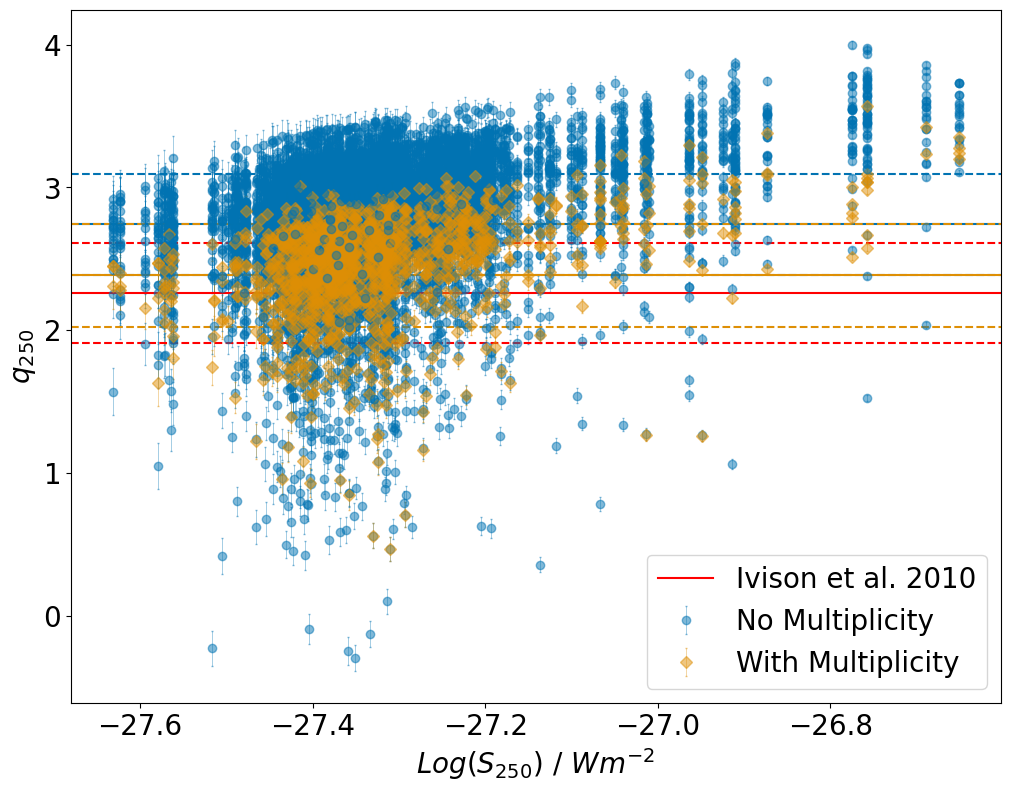}
		\caption{The $q_{250\mu \text{m}}$ against the $250~\mu$m flux density for \textit{Herschel} sources matched to simulated MeerKAT sources, with a matching radius of $17.6''$. The average $q_{250\mu \text{m}}$ for sources with multiple counterparts (orange diamonds) and without multiple counterparts (blue dots) are $2.38\pm 0.36$ and $2.74\pm 0.35$ respectively. The solid line shows the mean value and the dashed lines show the $1\sigma$ scatter. The mean and $1\sigma$ scatter of $q_{250\mu \text{m}}$ in \citet{ivison2010blast} are shown as red solid line and dashed lines.}
		\label{fig.firc.fake.multi}
	\end{figure}

	We have performed cross-identifications of \textit{Herschel} sources with our MeerKAT 1.28 GHz catalogue as presented in Section \ref{sec.id}. However, we only use the Poisson probability of chance alignment as the measure of the cross-identification's robustness. There is still the possibility of contamination in our cross-IDs due to random 1.28 GHz sources aligned with \textit{Herschel} objects by chance, especially for the cross-IDs used for the multiplicity study where a larger matching radius of $17.6''$ is used. Here, we explore the feasibility of using the FIRC to improve the quality of our cross-IDs without introducing other more complicated methods.  \par 
	We create a simulated catalogue in the G014 map, reproducing the \citet{hale2023mightee} blank field number counts in flux density bins ranging from $0.016$~mJy to $630$~mJy. In each flux bin, the flux density of a simulated source is randomly assigned from the flux density range of this bin. To eliminate the possibility of simulated sources coinciding with real MeerKAT sources, all simulated sources are at least $6''$ (roughly the synthesised beam size of MeerKAT) away from any real sources. The simulated catalogue is then matched to the H-ATLAS catalogue with matching radius of $7.5''$ and $17.6''$, corresponding to our cross-ID catalogue and multiplicity catalogue respectively. Since the simulated sources are randomly located on the sky, any source that matches a \textit{Herschel} source should be a purely chance alignment, and the FIRC coefficient $q_{250\mu \text{m}}$ of it should be very different to the \citet{ivison2010blast} result. The same procedure is then carried out 100 times to ensure robust statistics.\par 
	There are 438 \textit{Herschel} sources inside the G014 map, on average $11 \pm 3$ of them have counterparts within the $7.5''$ search radius, leading to a matching percentage of $2.60\pm 0.70\%$. The $q_{250\mu \text{m}}$ of those \textit{Herschel} sources with identifications in the simulated catalogue are calculated and displayed in Figure \ref{fig.firc.fake}. The average $q_{250\mu \text{m}}$ of these sources is $2.99\pm 0.41$. The mean $q_{250\mu \text{m}}$ of these chance coincidences is higher than both $2.26\pm 0.35$ in  \citet{ivison2010blast} and $2.33\pm 0.26$ from our MeerKAT images, but only $18.0\pm 12.0\%$ of these source have $q_{250\mu \text{m}}$ values inside the $1\sigma$ scatter of the $q_{250\mu \text{m}}$ in \citet{ivison2010blast}. Since all these sources are randomly distributed on the sky and should not have any physical association with real \textit{Herschel} sources, this much higher $q_{250\mu \text{m}}$ is likely a reflection of the blank field source counts. There are many more faint radio sources than bright ones. These faint radio sources have a higher possibility to coincide with a \textit{Herschel} source, leading to a higher $q_{250\mu \text{m}}$. This implies that only $18.0\pm 12.0\%$ of chance-alignment sources will have $q_{250\mu \text{m}}$ values comparable to true cross-IDs. Given the $7.33\pm 0.04\%$ Poisson probability of chance alignments, the probability of random MeerKAT sources matched to \textit{Herschel} sources having comparable $q_{250\mu \text{m}}$ is only $1.35\pm 0.86\%$, which is reduced by a factor of $\sim 6$. \par	
	
	Similar procedures are carried out for the $17.6''$ matching radius catalogue. There are now $26.4\pm 2.50\%$ of the 438 \textit{Herschel} sources in G014 that have matched simulated sources, $11.2\pm 2.90\%$ of these matched sources have multiple counterparts. $92.7\pm 3.72\%$ of \textit{Herschel} sources with multiple radio counterparts have 2 components, $6.80\pm 0.75\%$ sources have 3 components and $0.46\pm 0.19\%$ have four. In total, the percentage of \textit{Herschel} sources that have multiple aligned-by-chance counterparts is only $3.00\pm 0.70\%$, which implies that our identifications of \textit{Herschel} sources with multiple counterparts are reliable. The $q_{250\mu \text{m}}$ of \textit{Herschel} sources with matched simulated sources are presented in Figure \ref{fig.firc.fake.multi}. The average $q_{250\mu \text{m}}$ for sources with multiple counterparts and without multiple counterparts are $2.38\pm 0.36$ and $2.74\pm 0.35$ respectively. Sources without multiple counterparts have higher $q_{250\mu \text{m}}$ than both our results from real MeerKAT sources and the results in \citet{ivison2010blast}, while sources with multiplicity have comparable values. The fact that sources without multiplicity here still have higher mean $q_{250\mu \text{m}}$ supports our hypothesis that the difference between $q_{250\mu \text{m}}$ of \textit{Herschel} sources with and without multiplicity is a bias due to our calculation method as discussed in Section \ref{sec.id.2}, which may also explain why the $q_{250\mu \text{m}}$ of sources with multiplicity are comparable to real values.  $25.0\pm 4.0\%$ of matched sources have $q_{250\mu \text{m}}$ consistent to the $q_{250\mu \text{m}}$ in \citet{ivison2010blast}, indicating this fraction of random radio sources matched to \textit{Herschel} sources by chance will have $q_{250\mu \text{m}}$ comparable to real cross-IDs. Combining this with the $34.2\pm 0.2\%$ Poisson probability of chance alignment at $17.6''$ matching radius, the probability of finding a chance-associated radio counterpart will be increased to $8.47\pm 1.52\%$.\par 
	It is thus fair to say that using the FIRC coefficient $q_{250\mu \text{m}}$ as a constraint will indeed improve the reliability of \textit{Herschel}-MeerKAT cross-identification. It can also be used to decide the most likely counterpart(s) of a \textit{Herschel} source if there are multiple radio sources coinciding with it. However, it should be noted again that the FIRC coefficient $q_{250\mu \text{m}}$ used here is not K-corrected due to the lack of redshift information. Therefore it is dependent on redshifts of the sample and the significance of using $q_{250\mu \text{m}}$ to improve cross-IDs will depend on redshifts as well. Moreover, using a $q_{250\mu \text{m}}$ cut will raise the risk of discarding true identifications with unusual $q$, for example radio-loud sources as shown in Figure \ref{fig.cutout}i. Without more data, it is difficult to determine whether these sources with unusual $q_{250}$ are correct identifications or not. \par  
	Overall, using the FIRC coefficient as a constraint to \textit{Herschel}-radio cross identifications potentially provides a relatively simple method to improve the reliability of the cross-IDs. However, care must be taken as it will also exclude potential good identifications of radio-loud sources (or radio-quiet sources with large $q$), combining this method with other techniques such as visual examination of excluded sources can be a more reliable way of using it. More investigations should be done before one can safely use the FIRC to improve \textit{Herschel} identification.
	
	\subsection{Accurate Optical/IR Cross-IDs of \textit{Herschel} sources}\label{sec.optical}
	
	\begin{figure*}
		\centering
		\subfloat{\includegraphics[width=0.5\linewidth]{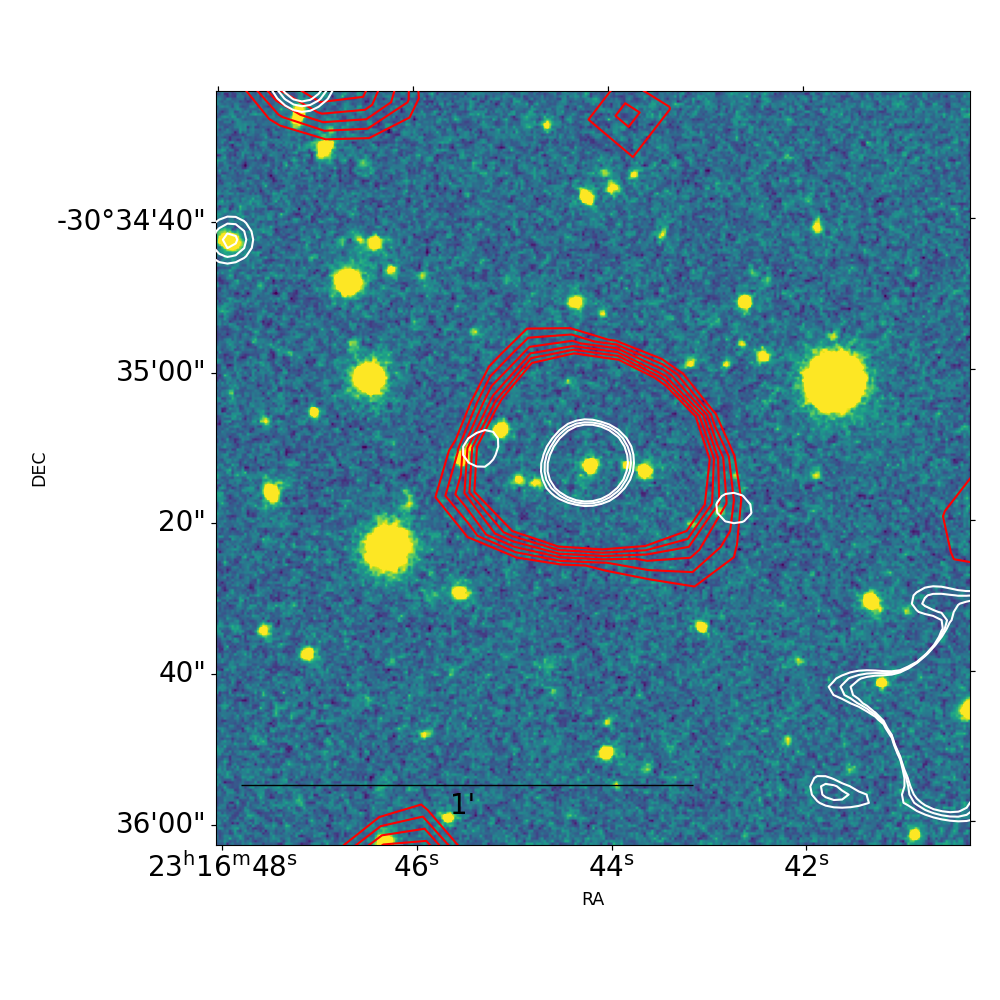}}
		\subfloat{\includegraphics[width=0.5\linewidth]{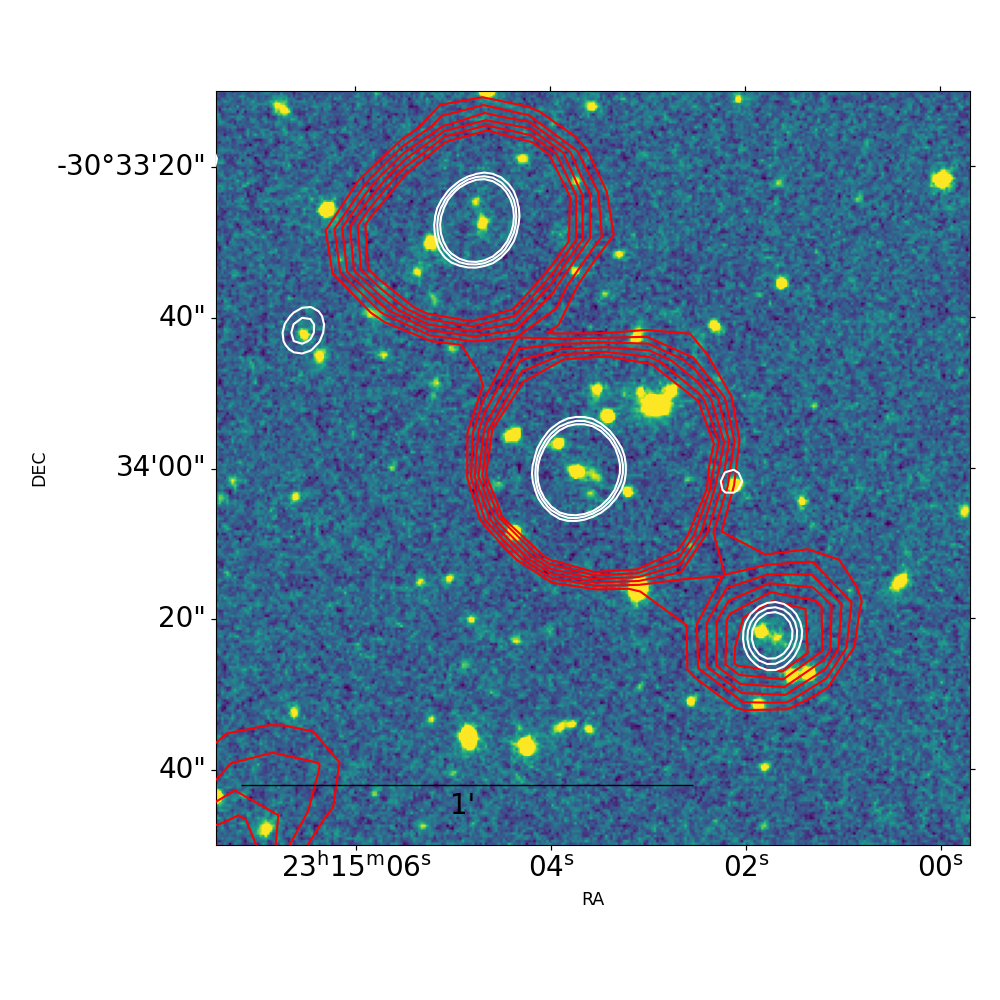}}
		\caption{Two examples of MeerKAT data helping \textit{Herschel} identifications with SHARKS sources. The background images are the SHARKS Ks data, the red contours show the $5-10\sigma$ \textit{Herschel} 250 $\mu$m flux density, the white contour shows the $3-5\sigma$ MeerKAT 1.28 GHz flux density. In the left panel, the large \textit{Herschel} source has overlapped with multiple faint SHARKS sources, which prevents direct Ks identification of this source. However, with the help of MeerKAT data, where there is only one source coincides with the \textit{Herschel} source, the SHARKS identification is pinpointed. Similarly in the right panel, 3 \textit{Herschel} sources are connected to each other and cover a lot of faint SHARKS objects, but there is one MeerKAT counterpart for each of the \textit{Herschel} sources, which greatly helps locating the potential true SHARKS identifications.}
		\label{fig.id}
	\end{figure*}

	The identification of \textit{Herschel} sources in optical/NIR data is difficult to implement, as optical/NIR surveys such as DES and SHARKS generally have much better angular resolution and higher source number densities, resulting in multiple optical/NIR sources overlapping with one \textit{Herschel} source. Moreover, \textit{Herschel} sources are generally dusty objects which means they may be very faint in the optical/NIR, making their optical/NIR counterparts harder to find. Conventionally, \textit{Herschel} cross-identifications with optical/NIR data are carried out using methods such as likelihood ratio to determine the most likely counterpart between several optical/NIR objects overlapping with one \textit{Herschel} source (e.g., \citealt{smith2011herschel}). However, all these difficulties no longer exist for \textit{Herschel}-radio cross-identification,  since dust is transparent to radio emission and the FIRC can be used to determine the most likely counterpart of a \textit{Herschel} object. With the help of high-quality radio data such as the MeerKAT data presented in this work, we can obtain accurate positions of \textit{Herschel} sources which will significantly help the cross-identification of these \textit{Herschel} sources in optical/NIR data. Here, we make use of the recently released SHARKS Ks data to demonstrate the power of radio data in \textit{Herschel} identification. Two examples are illustrated in Figure \ref{fig.id}. As can be seen, \textit{Herschel} sources generally coincide with many faint SHARKS Ks sources. With the help of MeerKAT data, the true SHARKS identification of \textit{Herschel} sources can be either directly pinpointed or limited to a small number of potential counterparts. Precise identifications will enable more multi-wavelength analysis, such as accurate photometric redshift estimates to search for evidence of protoclusters. \par

	Despite its power, there are still some problems in using MeerKAT data for \textit{Herschel} identifications in the optical/NIR. First, the average synthesised beam size of MeerKAT is significantly larger than those in optical/NIR surveys. Thus a MeerKAT source may encompass several faint optical/NIR objects, as shown in Figure \ref{fig.id}. Secondly, some faint \textit{Herschel} sources match no MeerKAT counterpart or only a tentative MeerKAT detection, and hence cannot be identified in optical/NIR data using this method. The next-generation radio telescopes such as the SKA will improve these issues with their better resolutions and sensitivities. 
	
	\subsubsection{Photometric Redshift Estimations} \label{sec.optical.z}
	Using the accurate positions of \textit{Herschel} sources obtained from MeerKAT IDs, we cross-match our sources to several optical/NIR catalogues and then estimate their photometric redshifts, followed by analysis to further examine if the three DSFG protocluster candidates are real. \par
\begin{figure}
		\centering
		\includegraphics[width=\columnwidth]{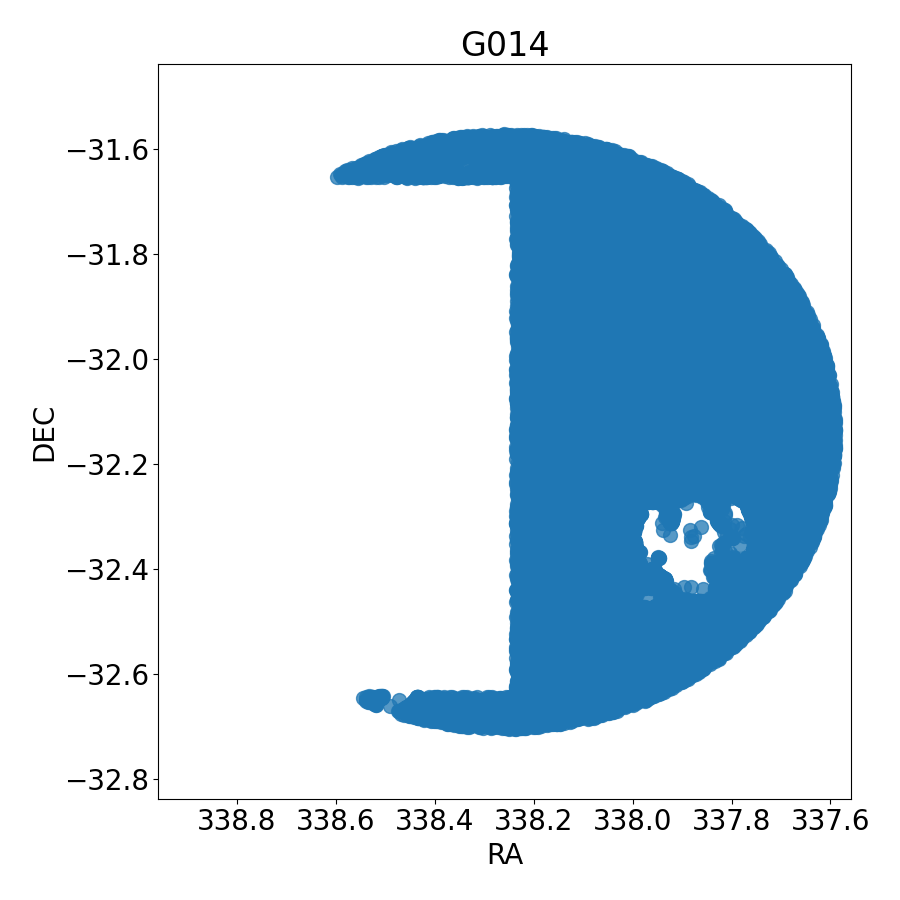}
		\caption{The KiDS DR4 catalogue footprint in G014. About half of the central region is covered, a tile in the east is not covered. There is also a hole in the KiDS catalogue in this region, which is the result of a very bright star.}
		\label{fig.g014.kids}
	\end{figure}

 
	We cross-match our sources to the KiDS data release 4 (DR4) catalogue (\citealt{2019A&A...625A...2K}) which is an optical survey in $ugri$ bands. KiDS DR4 additionally includes aperture-matched $ZYJHK_{s}$ photometry from their partner NIR survey VIKING (\citealt{edge2013vista}), resulting in a 9-band optical-NIR photometric catalogue. KiDS DR4 covers the entire areas of G017 and G257. About half of G014 including part of the central area is covered, with a tile in the eastern edge not included. Figure \ref{fig.g014.kids} illustrates the KiDS DR4 footprint in G014.
	
\begin{figure*}		
		\centering
		\subfloat{\includegraphics[width=0.33\linewidth]{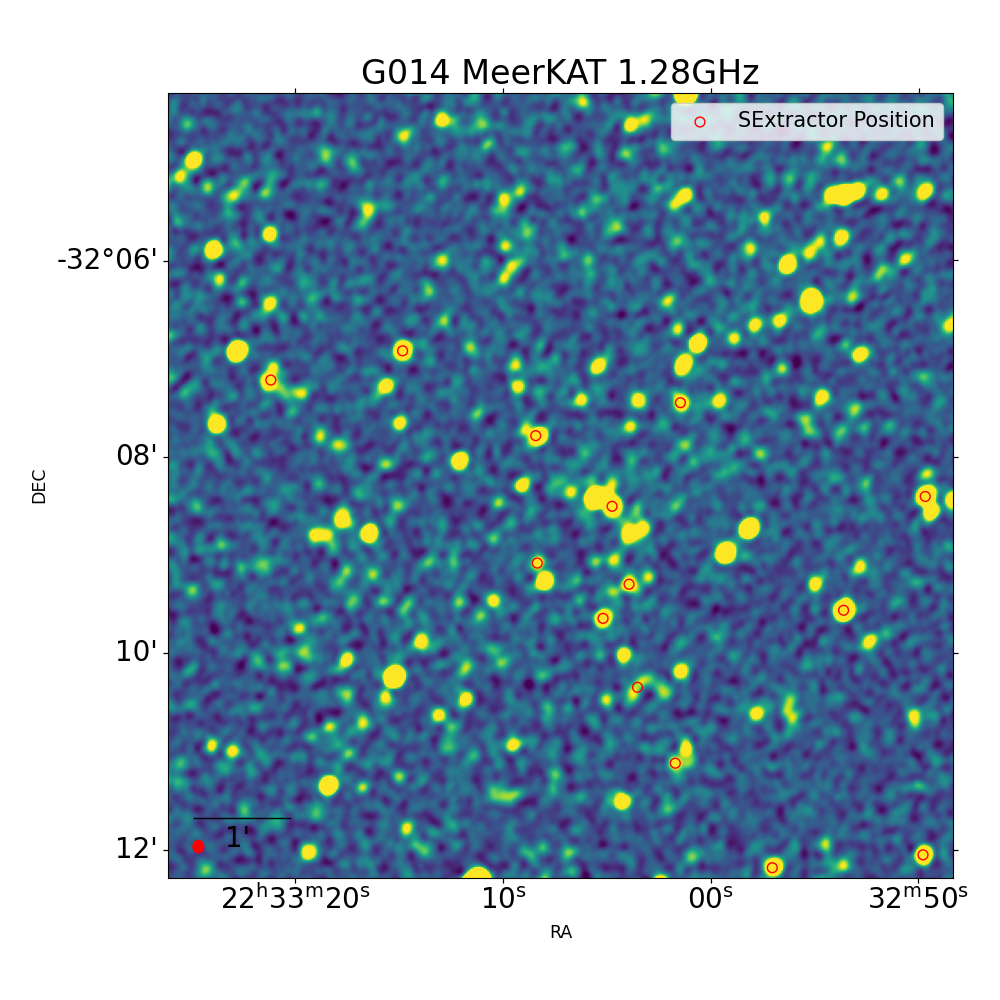}}
		\subfloat{\includegraphics[width=0.33\linewidth]{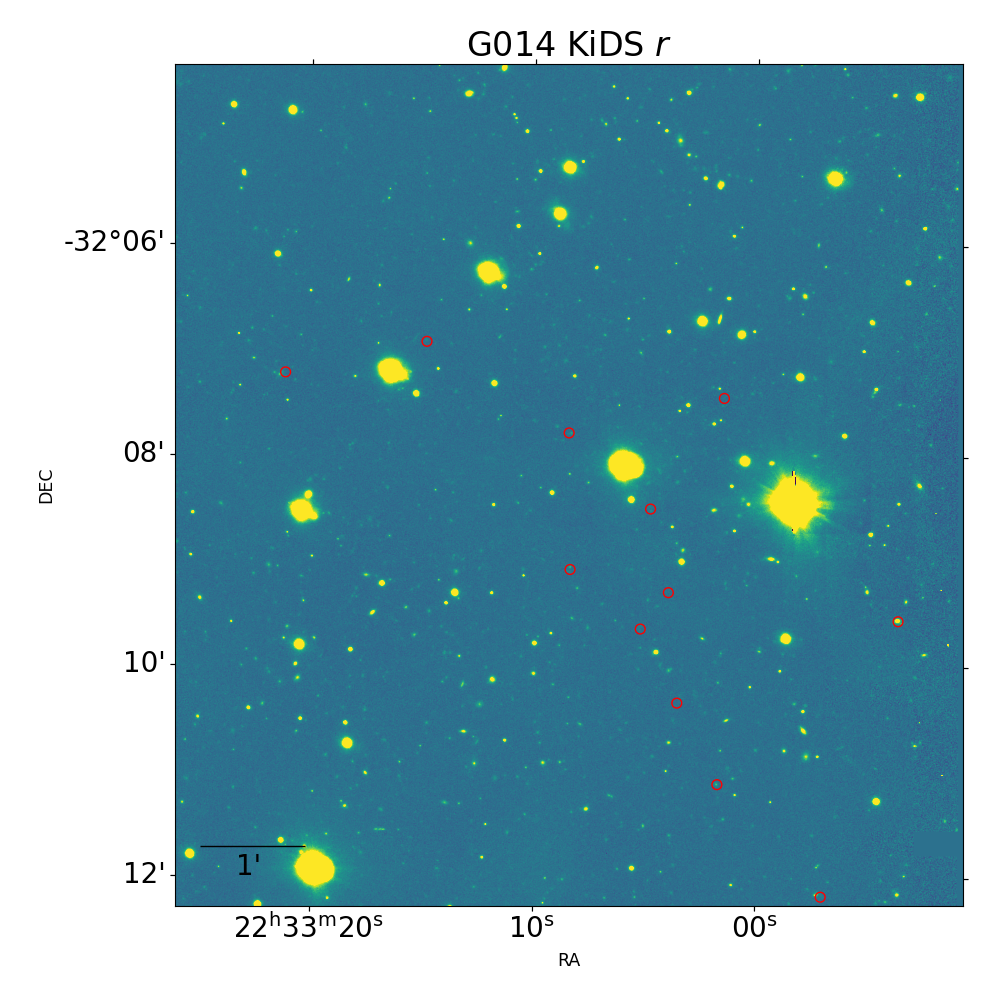}}
		\subfloat{\includegraphics[width=0.33\linewidth]{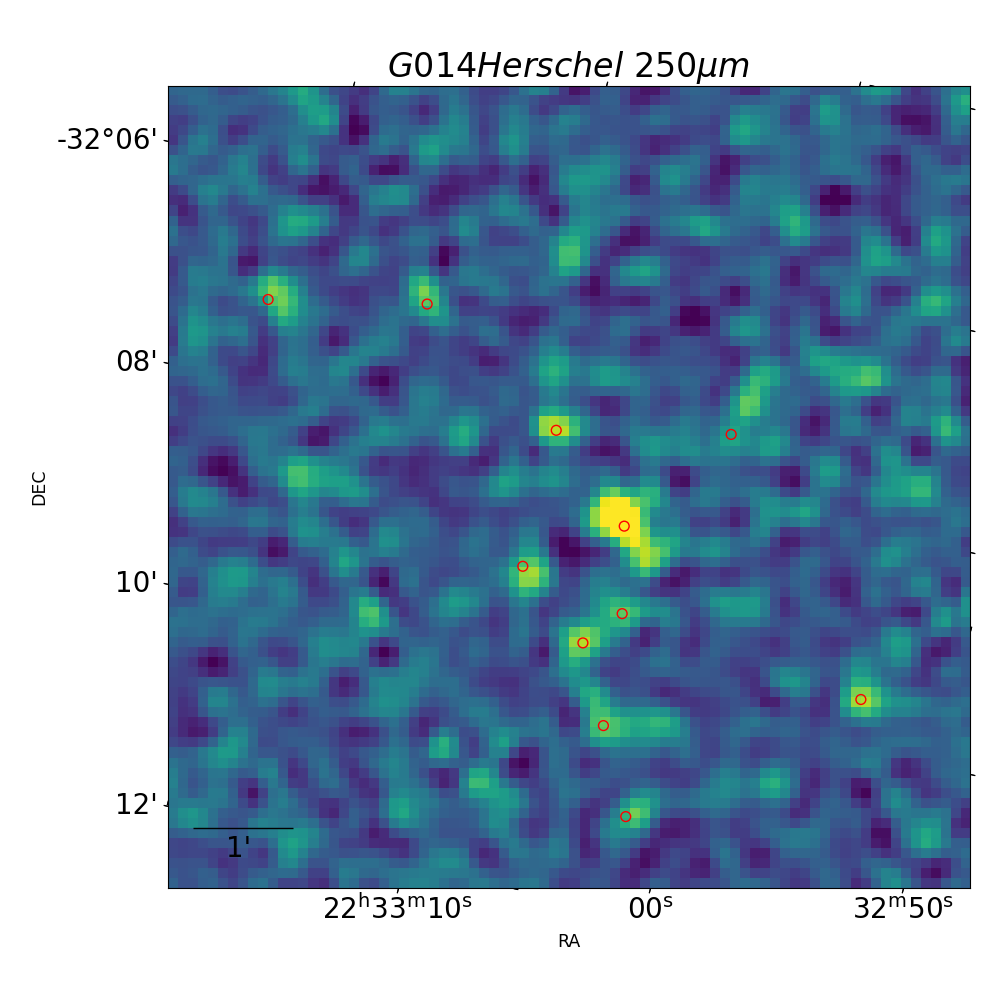}}\\
		\subfloat{\includegraphics[width=0.33\linewidth]{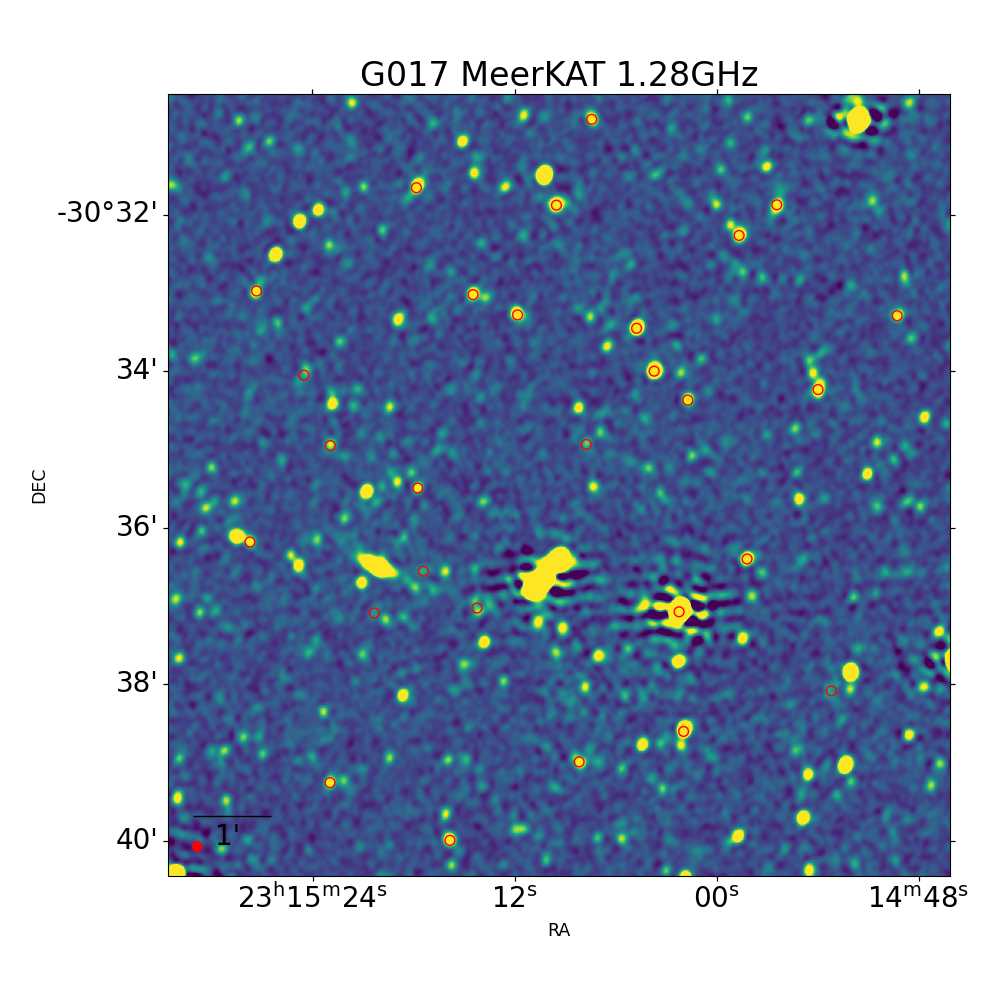}}
		\subfloat{\includegraphics[width=0.33\linewidth]{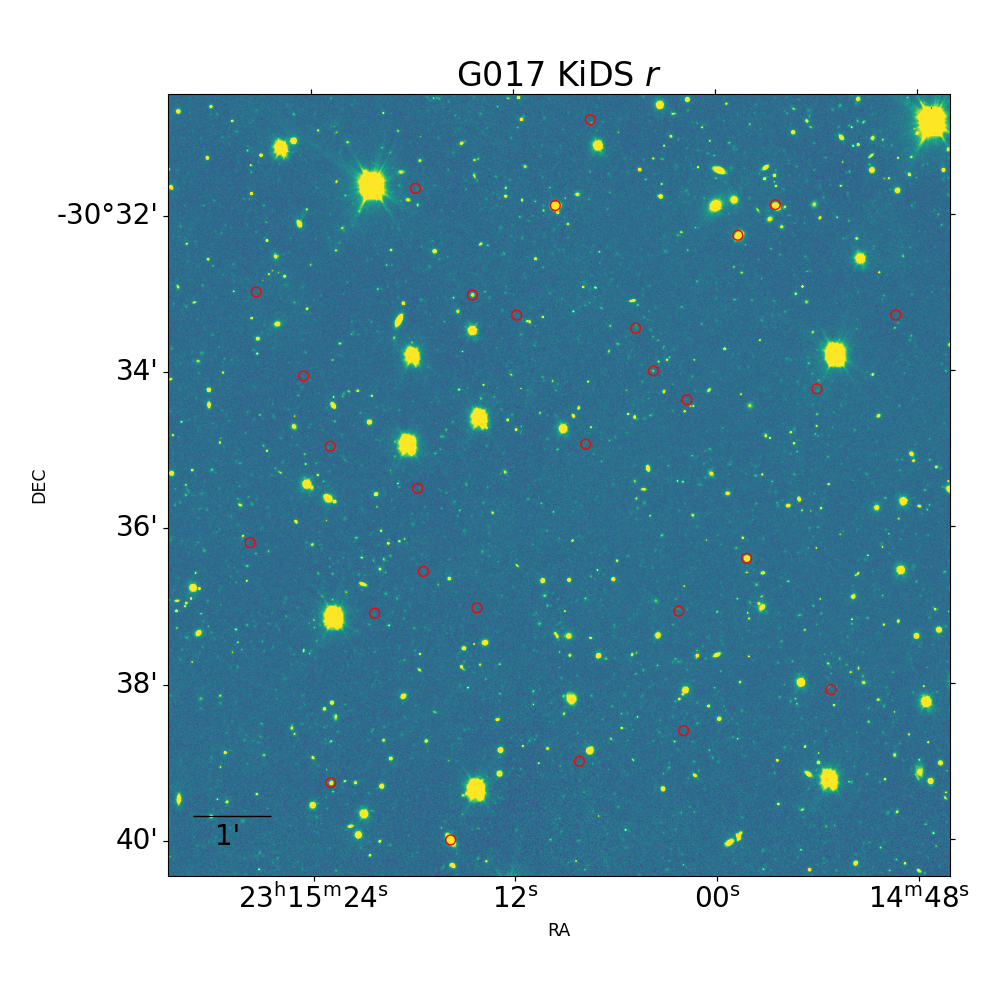}}
		\subfloat{\includegraphics[width=0.33\linewidth]{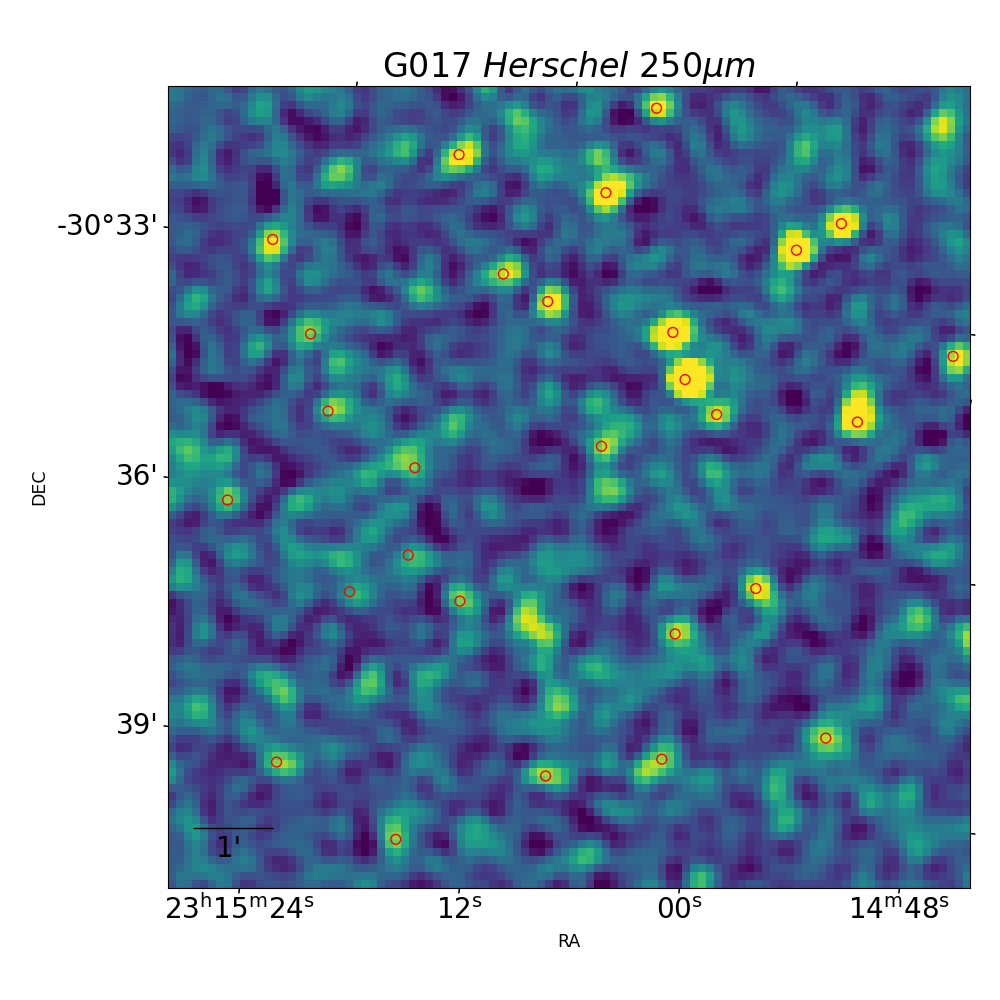}}\\
		\subfloat{\includegraphics[width=0.33\linewidth]{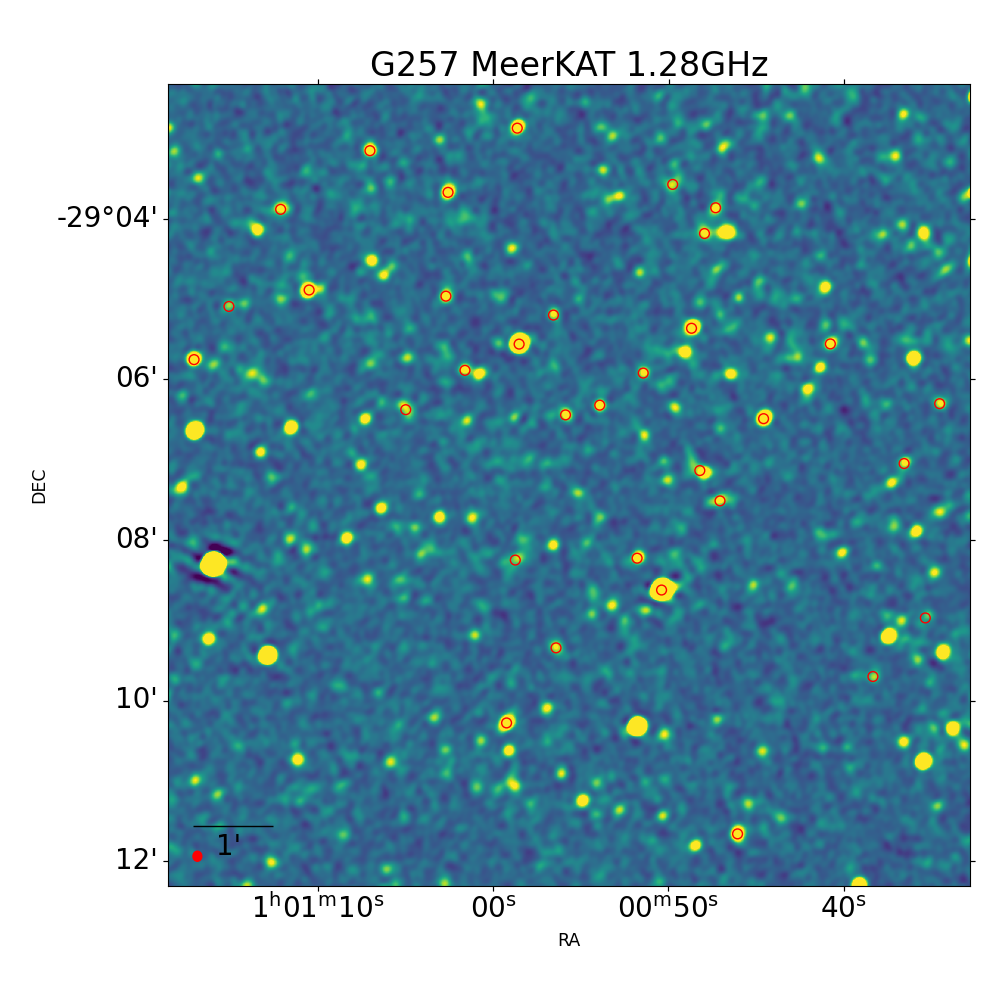}}
		\subfloat{\includegraphics[width=0.33\linewidth]{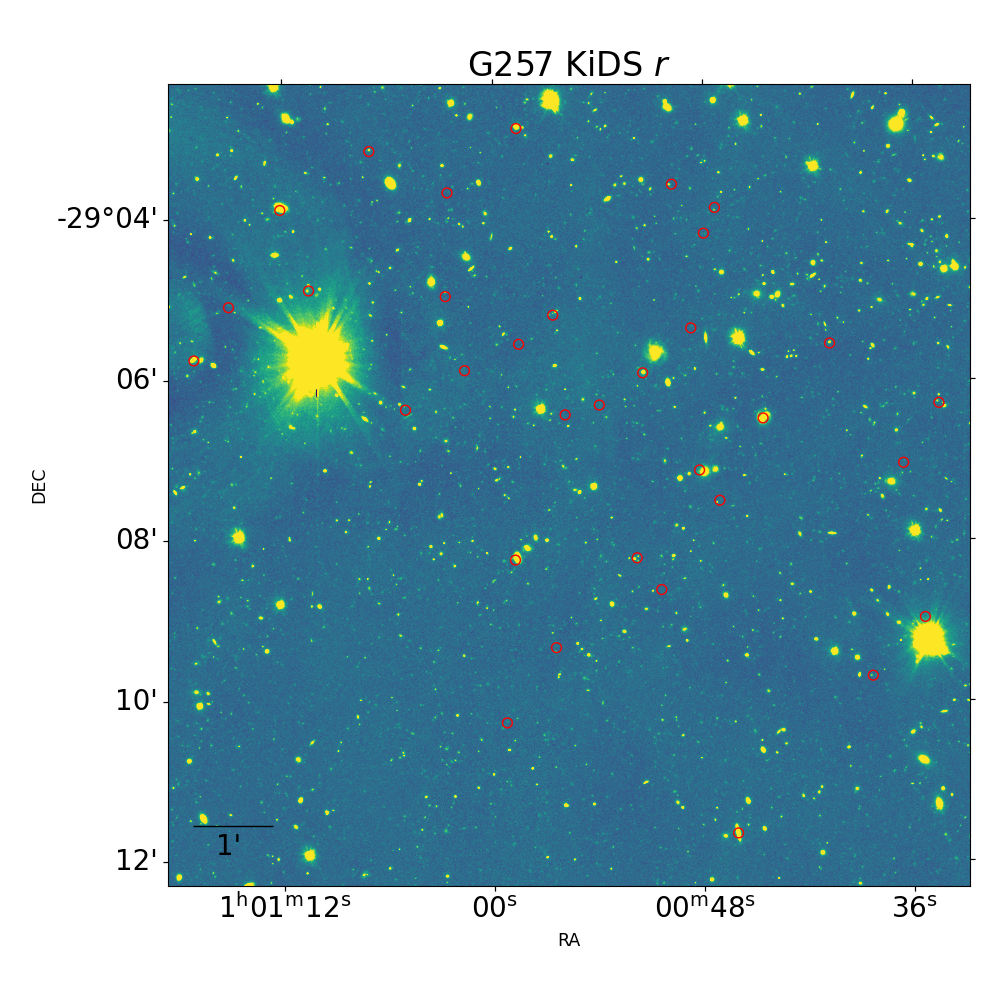}}
		\subfloat{\includegraphics[width=0.33\linewidth]{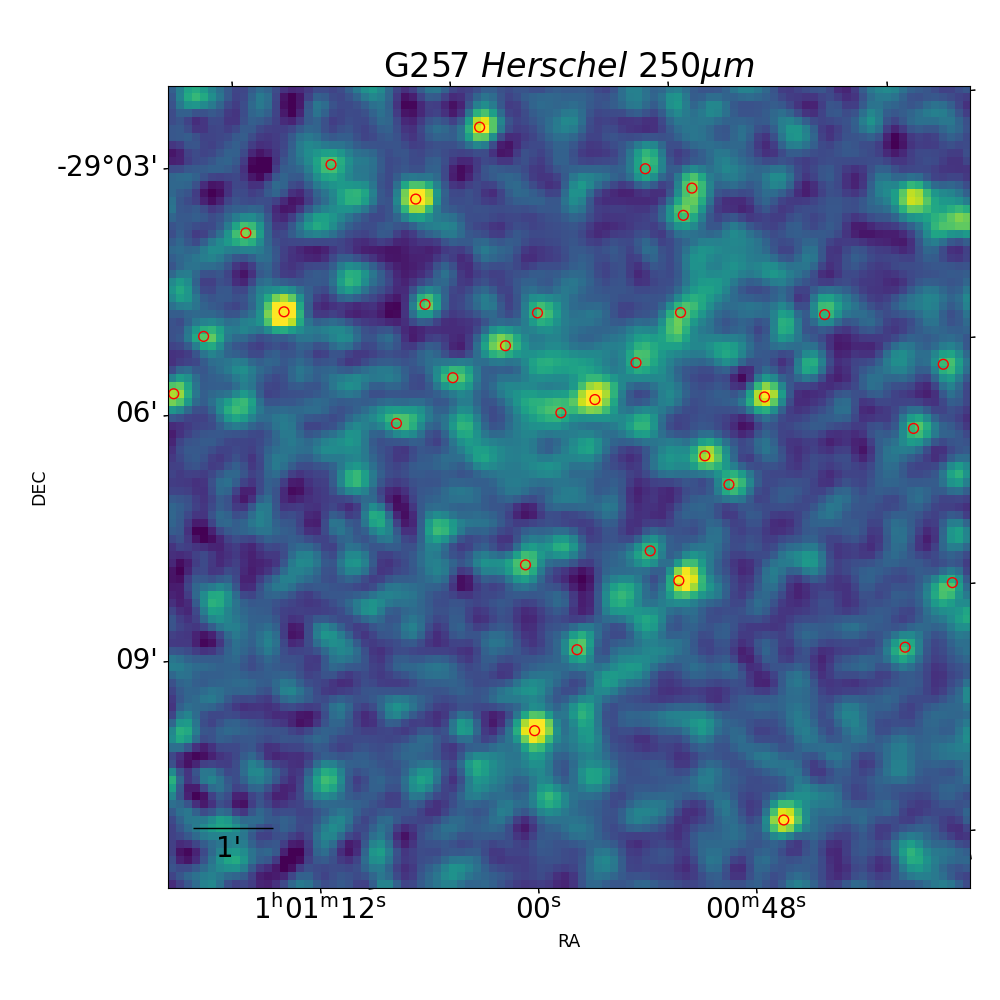}}\\
		\caption{$10'\times 10'$ cutouts of the centres of our MeerKAT 1.28 GHz images (left column), and positionally matched KiDS DR4 $r$ band images (middle column) and H-ATLAS \textit{Herschel} $250~\mu$m images (right column). From top to bottom is G014, G017 and G257 respectively. The red circles show the SExtractor positions of MeerKAT-\textit{Herschel} matched sources. The MeerKAT synthesised beam shape in each field is shown as red solid ellipse in the bottom left of each MeerKAT cutout. A $1'$ scale bar is also shown in the bottom left corner of each cutout. Cutouts of G014 are slightly smaller with a size of $10'\times 8'$ as G014 is not fully covered by KiDS DR4. It can be seen that while most \textit{Herschel} sources have clear MeerKAT counterparts, a significant fraction of MeerKAT-\textit{Herschel} matched sources have no KiDS $r$ band counterpart. }
		\label{fig.centre}
	\end{figure*}

	Following a similar procedure in Section \ref{sec.id}, we choose a matching radius of $2''$ to cross-match our \textit{Herschel}-matched MeerKAT sources to the KiDS sources, which leads to 105, 350, and 498 cross-matched sources in G014, G017, and G257 respectively.  Excluding G014 since it is not fully covered by KiDS and VIKING, about $58.4\%$ and $62.3\%$ of \textit{Herschel}-MeerKAT matched sources in G017 and G257 are identified, leading to a total matching rate of $\sim60.7\%$. Figure \ref{fig.centre} shows the $10'\times 10'$ centres of our MeerKAT images, together with matched KiDS $r$ band images and \textit{Herschel} $250~\mu$m images. As shown by the cutout images, a significant fraction of MeerKAT-\textit{Herschel} matched sources have no KiDS $r$ band counterpart which demonstrates the $\sim 60.7\%$ matching rate found.\par
	
	With the 9-band optical-NIR photometry, we then use the SED template fitting code {\sc EAZY} (\citealt{brammer2008eazy}) to estimate the photometric redshifts of our sample. The default EAZY\_v1.3 galaxy template set is used. This template set is designed to incorporate nebular emission lines and also includes a dusty galaxy SED template and an extremely blue SED with strong line emission. {\sc EAZY} allows all the templates to be included in the final results using fits of N-linear combinations. No magnitude-dependent redshift prior is applied to our fit. The average 68\% confidence interval of our {\sc EAZY} redshift estimation is 0.174. By cross-matching our KiDS-matched sample to the 2dF spectroscopic catalogues, we are able to evaluate the accuracy of our {\sc EAZY} redshifts. There are in total 119 of our sample with 2dF redshifts. We calculate the photometric redshift deviation as $|\Delta z/(1+z_{\text{spec}})|$ where $\Delta z=z_{\text{phot}}-z_{\text{spec}}$. An average of $0.0383$ is found. Following \citet{brammer2008eazy}, we further calculate the normalised median absolute deviation $\sigma_{\text{NMAD}}$, which is less sensitive to outliers, as:
	\begin{equation}\label{eq.5}
		\sigma_{\text{NMAD}}=1.48\times \text{median}\left( \left| \frac{\Delta z-\text{median}(\Delta z)}{1+z_{\text{spec}}} \right| \right)
	\end{equation}
	The average $\sigma_{\text{NMAD}}$ of the 119 sources is $0.178$. This shows that the accuracy of our {\sc EAZY} redshifts are reasonably good.\par 
	
\begin{figure*}
		\centering
		\includegraphics[width=\linewidth]{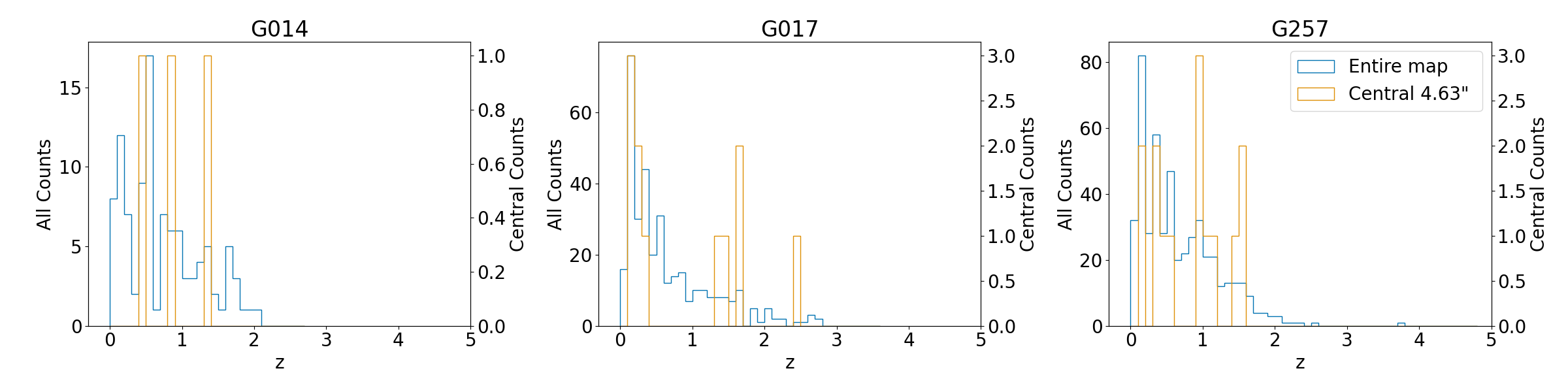}
		\caption{Photometric redshift distributions of \textit{Herschel} sources identified in both MeerKAT and KiDS/VIKING in the three observed fields. The left y-axis shows the total number of sources in the field, while the right y-axis shows the number of sources in the $4.63'$ radius centre of the field. A bin size of 0.1 is used. The blue histogram shows the redshift distribution of sources in the entire map, while the orange histogram shows the distribution of sources inside the $4.63'$ radius circle in the field centre. The fraction of $z\geq1$ sources is significantly larger in the centres in comparison to the overall distribution for G017 and G257. In G257, there are 5 sources at $0.9<z<1.2$, leading to a density contrast of $2.6\pm 1.7$ at this redshift range which is potentially a sign of overdensity.}
		\label{fig.photoz}
	\end{figure*}

	We then examine the redshift distributions of sources within the $4.63'$ radius circles at the centres of the three observed fields where the candidate protoclusters are. Figure \ref{fig.photoz} shows the redshift distributions of sources in the centres as well as the redshift distributions of all sources in the fields. There are 3, 11, and 14 sources with optical-IR IDs in the centres of G014, G017 and G257, while the numbers of all \textit{Herschel}-MeerKAT matched sources in the same areas are 17, 24, and 26. The fraction of $z\geq1$ sources is significantly larger in the centres in comparison to the overall distribution for G017 and G257, which is $45.0\%$ and $57.1\%$ respectively, while the same fraction for the entire field of G017 and G257 is $24.0\%$ and $24.5\%$. In G257, there are 5 sources at $0.9<z<1.2$, leading to a density contrast of $2.6\pm 1.7$ at this redshift range. This is potentially a sign of a $z\approx1$ overdensity associated with the candidate protocluster. However, the numbers of sources in the centres are too small to draw any statistically robust conclusions directly from them. Excluding G014, about $50.0\%$ of \textit{Herschel}-MeerKAT matched sources are not seen in KiDS and VIKING optical/NIR data, implying these sources have very faint optical/NIR counterparts and potentially lie at high redshifts. Given the $\sim 60.7\%$ identification rate in the entire fields of G017 and G257, the lower identification rate ($\sim 50\%$) in the field centres suggests that there may be more high redshift \textit{Herschel} sources in the centres than in the whole fields with their optical/NIR counterparts too faint to be detected in KiDS/VIKING.\par 
	
	Since a DSFG protocluster may also have underlying non-starbursting members associated with it, overdensities seen in optical/NIR data along the line of sight of our candidate protoclusters will also be evidence of real protoclusters. We additionally check the redshift distributions of all KiDS sources in the three fields, without matching them to either \textit{Herschel} or MeerKAT objects. The redshift distribution of central sources are basically identical to the overall distributions, showing no apparent sign of overdensity of optical sources. Nonetheless, as there are a large ratio of \textit{Herschel} sources not being detected in both KiDS/VIKING, deeper optical/NIR observations may reveal more traces of overdensities in the observed fields at higher redshifts and provide further evidence of the candidate protoclusters. \par 

\begin{figure*}
		\centering
		\includegraphics[width=\linewidth]{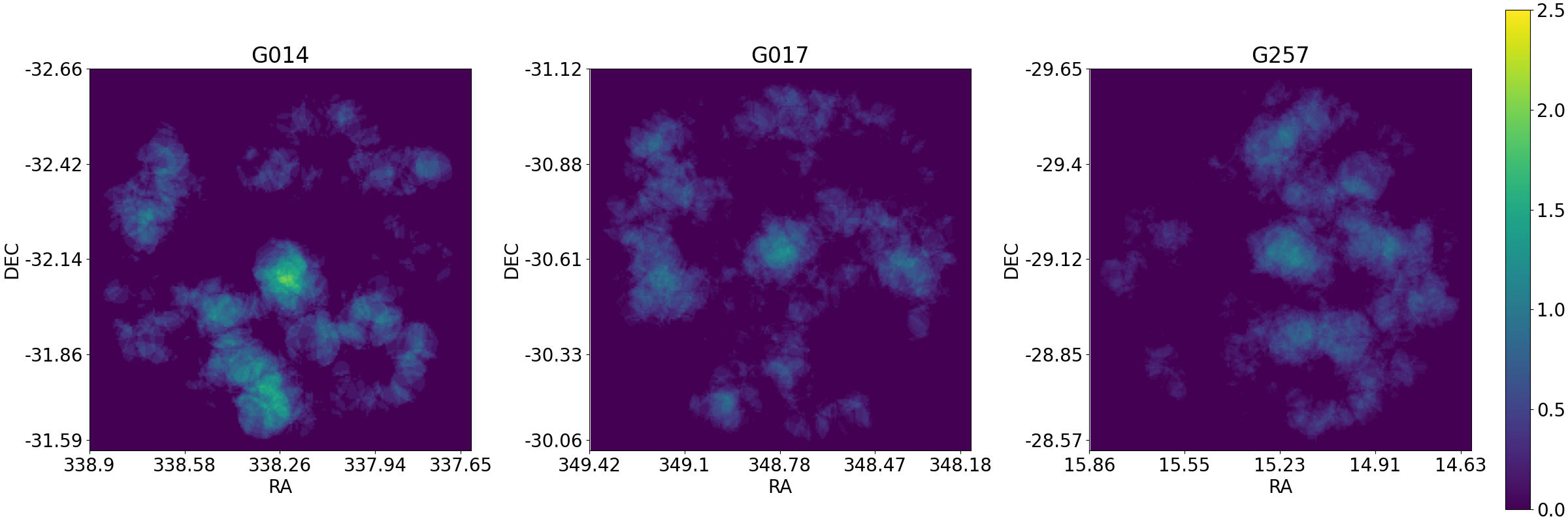} \\
		\includegraphics[width=\linewidth]{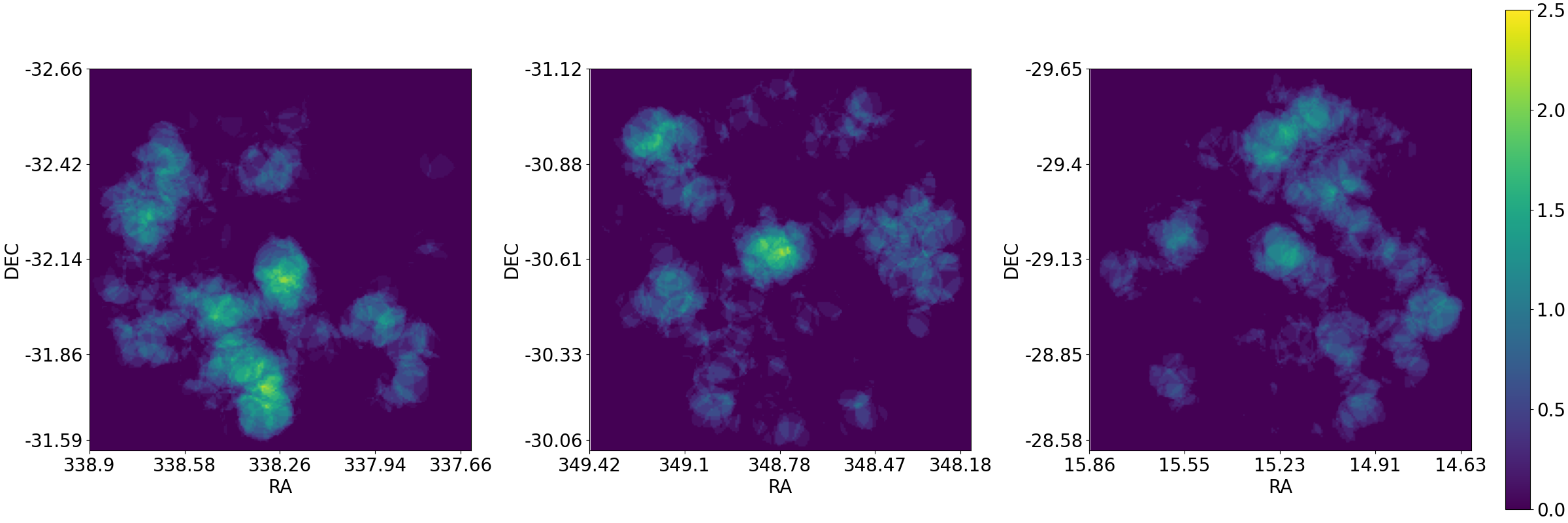}
		\caption{Density contrast maps of each observed field, with RA and DEC shown on x and y axes. For every pixel of these maps, the colour represents the local density contrast of KiDS-MeerKAT-\textit{Herschel} matched sources within a $4.63'$ radius circle centred at this pixel. The colour scale is the same for all panels. The top row shows the three fields before removing sources with {\sc EAZY} redshift lower than 1, while the bottom row shows the results after removing $z<1$ sources. The density contrasts $\delta$ are $1.1\pm 0.5$, $1.0\pm 0.4$ and $0.64\pm 0.34$ for G014, G017 and G257 before removing any sources, whereas $\delta (z>1)$ become $1.2\pm 0.6$, $1.8\pm 0.7$ and $1.1\pm 0.5$ after the removal. The density contrasts in the centres of G017 and G257 have significantly increased after removing low redshift sources, suggestive of $z>1$ protoclusters in the centres of observed fields. The change in G014 is less significant which is likely due to the incomplete coverage of optical/NIR photometry in this field.}
		\label{fig.ndensity}
	\end{figure*}

	\citet{greenslade2018candidate} estimates that most member galaxies of the candidate protoclusters should lie at $z>1$ and the optical/IR survey data used here is probably not deep enough to detect most sources at $z>1$ (over $50\%$ KiDS sources have $z_{\text{phot}}<1$, see Figure 11. in \citealt{2019A&A...625A...2K}). This means that the majority of \textit{Herschel} sources identified in KiDS/VIKING are unlikely to be associated with our candidate protoclusters. By excluding \textit{Herschel} sources with photometric redshifts lower than 1, we can reduce low-redshift contaminations and better isolate sources in higher redshift overdensities. Figure \ref{fig.ndensity} shows the density contrast maps before and after the removal. The maps are created by calculating the local density contrast of KiDS-\textit{Herschel}-MeerKAT matched sources within a $4.63'$ radius circle centred at every location across the image. At the centre of each field, the density contrast $\delta$ is $1.1\pm 0.5$, $1.0\pm 0.4$, and $0.64\pm 0.34$ for G014, G017 and G257 before removing any source, whereas $\delta (z>1)$ becomes $1.2\pm 0.6$, $1.8\pm 0.7$, and $1.1\pm 0.5$ after removing sources with {\sc EAZY} redshift lower than 1. It can be seen that at $z>1$ the central regions appear to have at least twice the number density as the surrounding fields and the significance of the central density contrasts increase after the removal, which supports the existence of $z>1$ protoclusters in the centres of observed fields. Note that the change of density contrast in the centre of G014 after removing low redshift sources is less significant. It is likely because of the incomplete coverage of KiDS/VIKING in this field, thus the number of sources with photometric redshift estimations in this field is smaller and the effect of removing $z<1$ sources is less obvious. \par 
	
	With the accurate positions of \textit{Herschel} sources provided by MeerKAT, we will be able to obtain more IDs from deeper optical/NIR data of high redshift \textit{Herschel} sources when it becomes available, and hence identify any overdensity associated with our candidate protoclusters. As the next step of this work, we have recently carried out deep optical/NIR observations on a number of DSFG protocluster candidates including the 3 studied here. Future work using our MeerKAT data to identify optical/NIR \textit{Herschel} counterparts in this new data is underway.

	\section{Conclusion}\label{sec.con}
	In this paper, we present MeerKAT 1.28 GHz radio imaging observations of 3 candidate DSFG protoclusters selected from \citet{greenslade2018candidate}. Sources are extracted using SExtractor, and then matched to the H-ATLAS DR2 in the SGP field. We do not find direct evidence of \textit{Herschel} protoclusters from our MeerKAT data alone, but with the help of accurate MeerKAT positions, the optical/NIR IDs and photometric redshifts of our \textit{Herschel} sources show evidence of real protoclusters in G017 and G257 at $z>1$ . More detailed results and conclusions are summarised as followed:\par 
	(a) Our 1.28 GHz source counts are comparable to other literature results. There is an excess in our G257 counts at $S\approx0.1$~mJy which is suggestive of an overdensity. By directly extracting sources from the MIGHTEE data (\citealt{heywood2022mightee}) using the same extraction procedures as in this work, we find that this excess disappears. It is likely caused by a difference in source extraction and source count corrections applied. Source counts just in the central $4.63'$ radius regions where the candidate protoclusters lie do not reveal any evidence of overdensity either. \par 
	
	(b) \textit{Herschel} sources are cross-matched to our MeerKAT sources using a matching radius of $7.5''$. The Poisson probability of chance association contaminants is $7.33\%$ at this searching radius. Approximately $95\%$ of the \textit{Herschel} sources in our observed regions have a MeerKAT counterpart. Visual inspection of \textit{Herschel} sources without MeerKAT counterparts reveal that most of them overlap with an extended 1.28 GHz source, but the positions of these extended radio sources as assigned by SExtractor are $>7.5''$ away from the \textit{Herschel} sources. Correcting for these sources with extended morphologies leads to a near 100 percent matching rate.\par 
	
	(c) To study the multiplicity of \textit{Herschel} sources in 1.28 GHz data, we also carry out another cross-identification of \textit{Herschel} sources to our MeerKAT sources, using a larger matching radius of $17.6''$, which is the beam size of the \textit{Herschel} SPIRE 250 $\mu$m band. We find 1885 out of the 1914 \textit{Herschel} sources have MeerKAT counterparts at this matching radius with 1038 out of the 1885 \textit{Herschel} sources having 2 or more radio components. The 250 $\mu$m flux distributions of \textit{Herschel} sources with and without multiplicity are statistically different, but the same difference is not shown in the 350 and 500 $\mu$m flux distributions. We suspect that this may be a result of insufficient faint sources in the 350 and 500 $\mu$m bands due to worse sensitivities. The distribution of $q_{250\mu \text{m}}$ of \textit{Herschel} sources with and without multiplicity are also different, but this is likely a bias induced by our method of calculating $q_{250\mu \text{m}}$ of \textit{Herschel} sources with multiple 1.28 GHz components.\par 
	
	(d) The FIRC is investigated by calculating the monochromatic FIRC coefficient $q_{250\mu \text{m}}$ of our sample. The average $q_{250\mu \text{m}}$ of our sources is $2.33\pm 0.26$, which is comparable to that in \citet{ivison2010blast}. Using $q_{250\mu \text{m}}$ as a constraint to the \textit{Herschel}-Radio identification can greatly reduce the probability of contaminations due to chance alignments by a factor of $\sim 6$ for the $7.5''$ matching radius catalogue, but with the risk of ignoring potential good identifications of unusual sources such as radio-loud AGNs.\par 
	
	(e) The \textit{Herschel} SPIRE colours of sources in the central regions, which are potential members of the candidate protoclusters, indicate that their redshifts should be at least larger than 1. No significant correlation between the \textit{Herschel} colours and the 1.28 GHz flux density or $q_{250\mu \text{m}}$ is found.\par 
	
	(f) We demonstrate that MeerKAT cross-IDs can significantly help \textit{Herschel} source identifications in optical/NIR data. We cross-match of our \textit{Herschel} sources with the public KiDS/VIKING data using MeerKAT positions, then carry out photometric redshift estimations with the SED fitting code {\sc EAZY}. About $60.7\%$ of \textit{Herschel} sources are identified but only $\sim50.0\%$ of \textit{Herschel} sources in the central $4.63'$ radius regions have optical/NIR IDs, which implies that there are more higher redshift \textit{Herschel} sources with very faint optical/NIR counterparts in the centres of the fields. The redshift distribution of sources in the centre of G257 shows a unique redshift peak at $0.9<z<1.2$, corresponding to a group of 5 sources with a density contrast of $\delta =2.6 \pm 1.7$. This could be evidence of the candidate protocluster in G257 being a real protocluster, but the robustness of this conclusion is hindered by small number statistics. By removing sources with their {\sc EAZY} redshifts lower than 1, the density contrasts of the $4.63'$ centres of G017 and G257 increase from $\delta = 1.0\pm 0.4$ and $0.64\pm 0.34$ to $\delta= 1.6\pm 0.6$ and $1.0\pm 0.5$, implying there are overdensities in these regions at $z>1$. With deeper MeerKAT observations or more powerful radio telescopes such as the SKA coming in the future, radio observations will be very powerful in identifying FIR/submm sources with optical/NIR data. This will be the future of optical/NIR IDs in FIR studies. Future work using our MeerKAT data to identify optical/NIR \textit{Herschel} counterparts is underway and will be discussed elsewhere.\par

\section*{Acknowledgements}
We thank the anonymous referee for detailed comments which significantly improved the paper and especially made it more understandable for the reader.
The MeerKAT telescope is operated by the South African Radio Astronomy Observatory, which is a facility of the National Research Foundation, an agency of the Department of Science and Innovation. The Herschel-ATLAS is a project with Herschel, which is an ESA space observatory with science instruments provided by European-led Principal Investigator consortia and with important participation from NASA. The H-ATLAS website is http://www.h-atlas.org/. \par 

Based on data products created from observations collected at the European Organisation for Astronomical Research in the Southern Hemisphere under ESO programme 198.A-2006. \par 

For the creation of the data used in this work, the SHARKS team at the Instituto de Astrofísica de Canarias has been financial supported by the Spanish Ministry of Science, Innovation and Universities (MICIU) under grant AYA2017-84061-P, co-financed by FEDER (European Regional Development Funds), by the Spanish Space Research Program “Participation in the NISP instrument and preparation for the science of EUCLID” (ESP2017-84272-C2-1-R) and by the ACIISI, Consejería de Economía, Conocimiento y Empleo del Gobierno de Canarias and the European Regional Development Fund (ERDF) under grant with reference PROID2020010107. \par 

We thank the support of the Wide-Field Astronomy Unit for testing and parallelising the mosaic process and preparing the releases. The work of the Wide-Field Astronomy Unit is funded by the UK Science and Technology Facilities Council through grant ST/T002956/1. \par

HD acknowledges financial support from the Agencia Estatal de Investigación del Ministerio de Ciencia e Innovación (AEI-MCINN) under grant (La evolución de los cúmulos de galaxias desde el amanecer hasta el mediodía cósmico) with reference (PID2019-105776GB-I00/DOI:10.13039/501100011033).

\section*{Data Availability}

Data used in this article can be shared on reasonable request to the corresponding author. The raw MeerKAT data can also be retrieved from the MeerKAT archive\footnote{\url{https://archive.sarao.ac.za}}.

\bibliographystyle{mnras}
\bibliography{refs.bib}

\bsp	
\label{lastpage}
\end{document}